\documentclass[a4paper,11pt]{scrartcl}
\usepackage[a4paper,left=2.54cm,right=2.54cm,top=2.54cm,bottom=2.54cm]{geometry}
\usepackage[english]{babel}
\usepackage{enumerate, comment}
\usepackage{authblk}
\usepackage[dvipsnames]{xcolor}
\usepackage{hyperref}
\usepackage{amsmath,amssymb,amsthm}
\usepackage[]{latexsym}
\usepackage[]{graphicx}
\usepackage{bbm}
\usepackage{subfigure} 
\usepackage{epstopdf}
\usepackage{lmodern}
\usepackage{caption}
\usepackage[titletoc,title]{appendix}

\usepackage{framed, xcolor}
 \definecolor{shadecolor}{gray}{0.9}

\usepackage{tikz-network}

\tikzstyle{mybox} = [draw=black, very thick, diamond, inner ysep=5pt, inner xsep=5pt]
\tikzstyle{myboxS} = [draw=black, thick, rectangle, rounded corners, inner ysep=5pt, inner xsep=5pt]

\tikzstyle{arrow}=[draw, -latex]
\usetikzlibrary{decorations.pathmorphing}

\usepackage[utf8]{inputenc}

\DeclareMathOperator*{\argmin}{arg\,min}

\newtheorem{lemma}{Lemma}[section]

\newtheorem{theorem}[lemma]{Theorem}

\newtheorem{algorithm}[lemma]{Algorithm}

\definecolor{salmon}{rgb}{1.0, 0.55, 0.41}

\usepackage{array}
\newcolumntype{M}[1]{>{\centering\arraybackslash}m{#1}}

\newtheorem*{theorem*}{Theorem}
\newtheorem*{algorithm*}{Algorithm}

\usepackage[style=bwl-FU, backend=bibtex8, doi=false,isbn=false, url=false,dashed=false, firstinits=true,maxbibnames=99]{biblatex}
\bibliography{ReferencesCyberProject.bib}

\begin{document}
	\title{Building Resilience in Cybersecurity\\ --\\ An Artificial Lab Approach}
	
	\author[b]{Kerstin Awiszus}
	\author[a]{Yannick Bell}
	\author[a]{Jan L\"uttringhaus}
	\author[a]{Gregor Svindland}
	\author[a]{Alexander Vo\ss}
	\author[a]{Stefan Weber}
	\affil[a]{\normalsize House of Insurance, Leibniz Universit\"at Hannover
	}
	\affil[b]{\normalsize University of Applied Sciences and Arts, Hannover}
	
	\date{\today}
	
\maketitle
\begin{abstract}
	Based on classical contagion models we introduce an \textit{artificial cyber lab}: the digital twin of a complex cyber system in which possible cyber resilience measures may be implemented and tested. Using the lab, in numerical case studies, we identify  two classes of measures to control systemic cyber risks: security- and topology-based interventions. We discuss the implications of our findings on selected real-world cybersecurity measures currently applied in the insurance and regulation practice or under discussion for future cyber risk control. To this end, we provide a brief overview of the current cybersecurity regulation and emphasize the role of insurance companies as private regulators.  Moreover, from an insurance point of view, we provide first attempts to design systemic cyber risk obligations and to measure the systemic risk contribution of individual policyholders.    
\end{abstract}\vspace{0.2cm}

\section{Introduction}

Cyber risks pose a major threat to societies, governments, businesses and individuals worldwide. For example, the annually published Allianz Risk Barometer, see \cite{allianz2022}, recently identified cyber incidents as the most important global business risks, ahead of business interruptions, natural disasters and pandemic outbreaks. In addition, cyber risk continues to increase, firstly due to the continued digitization of business processes, and secondly due to the COVID-19 pandemic and the associated increase in teleworking, see e.g. \cite{Lallie2021}, and thirdly in the context of current political conflicts and wars.

Regulatory and macro-prudential leaders are increasingly aware of the potentially catastrophic consequences of cyber risks. In particular, the systemic relevance of certain types of cyber threats, so-called systemic cyber risks, is highlighted, see e.g. \cite{Lagarde2021}.
Two illustrative systemic cyber incidents from the past are the WannaCry and NotPetya attacks\footnote{An in-depth risk analysis of these two incidents can be found, for example, in  \cite{Euro2020}.}:
\begin{itemize}
\item In May 2017, the WannaCry ransomware infected around 230,000 computer devices in more than 150 countries. It encrypted data on the infected systems and demanded a ransom payment of USD 300. The encryption resulted in data loss and rendered IT systems unusable in healthcare services and
in industry. It is estimated that the damage caused ranges from hundreds of millions to four billion US dollars. The discovery of a ``kill switch'' helped contain the incident.
\item 
In June 2017, the NotPetya malware was used for a global cyberattack that mainly targeted Ukraine. This version of the Petya malware was disguised as ransomware, but with the intention of causing maximum damage by encrypting data and disrupting IT systems. The encryption of data resulted in a permanent loss of its availability with immediate impact on institutions such as the Ukrainian Central Bank and a disruption of the country's major stock markets. In addition, the malware was able to infect other organizations outside the Ukrainian financial sector with offices  in Ukraine, compromising machines also elsewhere. For example, the global shipping company Maersk experienced widespread business disruptions at other locations around the world, which nearly destroyed the company.
\end{itemize}
{ This paper, in view of the previous examples, focuses on systemic cyber risks which are characterized by contagion effects in interconnected systems. Other instances of cyber accumulation scenarios are attacks based on a common risk factor such as the dependence on joint IT architecture or service providers, see for instance the infamous SolarWinds attack.\footnote{{\cite{Awiszus2021} propose to classify aggregate cyber risks which depend on a common risk factor as {\em systematic} while reserving the notion {\em systemic} for cyber risks caused by local or global contagion effects.}} For insurance stress testing of accumulation scenarios which may not follow a contagion pattern, like DoS attacks or cloud outage, see the discussion in \cite{EIOPA2022}.}

In light of the rapidly growing and evolving cyber threat landscape, cybersecurity approaches that focus solely on preventing attacks may be insufficient to manage and mitigate this class of systemic cyber risks. Therefore, building \textit{cyber resilience} requires taking a more expansive approach that targets the ``ability to \textit{anticipate}, \textit{withstand}, \textit{recover from}, and \textit{adapt to} adverse conditions, stresses, attacks, or compromises on systems that use or are enabled by cyber resources''\footnote{See the definition of ``cyber resiliency'' in \cite{NISTglossary}.}.

Legislators and regulators have enacted a variety of laws and policies governing cybersecurity and identified the need to enhance the resilience of cyber systems. 
In addition, private actors may also take a leading role in shaping and guiding cybersecurity standards. In particular, the idea of (re)insurance companies acting as \textit{private regulators} to fill existing regulatory gaps and mitigate residual risks has emerged. 

But how can private and government regulators ensure an adequate level of protection against cyber threats and implement appropriate measures to build cyber resilience? What characteristics of networked cyber systems are critical to managing and controlling cyber threats and, in particular, to preventing, managing, and responding to the onset of a systemic cyber risk event? And is regulatory intervention even necessary to build effective levels of cyber resilience?
In this paper, we address these questions. Our key contributions are:
\begin{enumerate}
       \item We design the \emph{artificial cyber lab}, the digital twin of a complex cyber system, to evaluate different types of cyber resilience measures. Digital twins consist of a ''physical entity, a virtual counterpart, and the data links between them,'' cf. \cite{Jones2020}.  The virtual counterpart of interconnected cyber-physical systems is based on network contagion models and is therefore tailored to the analysis of \emph{systemic cyber risks} such as the aforementioned WannaCry and NotPetya attacks.    
       \item In \emph{two exemplary case studies}, we leverage the lab to generate artificial data from virtual counterparts of real-world cyber systems to analyze specific types of cyber resilience interventions.
    \begin{enumerate}
         \item \emph{Security-related interventions}: Interconnected actors in a cyber network use security investments to protect themselves from cyber risk contagion. We
         \begin{itemize}
             \item study a \textit{security investment game} modeling network interaction and interdependence effects related to IT security standards; unlike the vast majority of game-theoretic models in the cyber insurance literature, our game is based on the underlying \textit{dynamic contagion} captured by stochastic Monte-Carlo simulations,
             \item rigorously \textit{prove} that there exists a steady state (Nash equilibrium) of security investment decisions which, however, generally does not minimize the overall cyber risk losses of the network, 
             \item develop and evaluate different \textit{regulatory allocation strategies} to further improve the overall system security in a steady state of security investment choices, 
             \item and analyze centrality measures to identify \textit{systemically relevant nodes} for the targeted allocation of cybersecurity obligations.  
         \end{itemize}
      	\item \emph{Topology-based interventions}:
      Network topology is important for both network functionality and the risk of cyber epidemic contagion. Therefore, we
        \begin{itemize}
            \item characterize the cyber contagion risk exposure of large-scale networks,
            \item study the effect of network heterogeneity on risk amplification,
            \item discuss possible \textit{efficient intervention strategies} that minimize the negative impact on network functionality,
            \item present a \textit{novel approach} to quantify contagious cyber risks and effectively allocate associated surcharges or insurance premiums based on the identification of critical network connections. 
        \end{itemize}
       \end{enumerate}
       Our digital twin approach provides an experimental framework for testing and evaluating different regulatory intervention strategies. This is particularly important due to the lack of data on historical cyber incidents and the non-stationarity of the cyber environment. Our results clearly indicate a \emph{need for regulation} in order to build an appropriate level of cyber resilience. 
        \item Based on the findings from the case studies, selected regulatory measures that are currently in use or under debate to strengthen resilience in real-world networked cyber systems are discussed. 
        \item To this end, we provide a brief overview of the current regulatory framework for cybersecurity in the European Union and the United States. In addition, we also discuss the role of private actors, particularly cyber insurance companies, in shaping security standards.
\end{enumerate}

\paragraph{Literature} In the following, we will only briefly review the relevant literature.
For a comprehensive overview of the various modeling and pricing approaches in the field of cyber risk and insurance, we refer the interested reader to the most recent survey  \cite{Awiszus2021}.  \cite{Dacorogna2023} provides another recent discussion on characteristics, models, and the management of cyber risks.

In the actuarial literature, cyber loss models are often based on classical frequency-severity approaches; see, e.g., \cite{zeller2021comprehensive} for an exemplary loss model and a comprehensive literature overview, and \cite{eling2020cyber} for a recent review of research in business and actuarial science. While at first glance such approaches appear to be the most feasible from an insurer's perspective, they suffer from insufficient  or inadequate data,  see also \cite{Zeller2023}. Furthermore, in the case of systemic cyber risks such as WannaCry or NotPetya, the structural importance of network effects for risk emergence and amplification cannot be adequately captured by these classical approaches.  The dynamics of incidents are similar to feedback mechanisms in financial systems such as the propagation of economic distress in a network of creditors or business partners. Interaction mechanisms of this type were, for example, studied in \cite{Giesecke2004} and \cite{Giesecke2006} using results from the theory of interacting particle systems. A similar approach was first introduced in microeconomics in the seminal work \cite{Follmer1974} in which actors interact on a grid.  

Regulatory aspects are also not considered in frequency-severity models for cyber claims. In a game-theoretic framework, by contrast, regulatory issues  as well as network interdependence of policyholders can be taken into consideration.  The existing literature on strategic interactions in cyber networks has focused mainly on the impact of cyber insurance on the self-protection efforts of interconnected actors, see, e.g. \cite{Ogut2005}, \cite{Bolot2009}, \cite{schwartz2014cyber}, and \cite{yang2014}. In most cases, market inefficiencies are observed and cyber insurance is not found to provide incentives for self-protection. However, in the absence of information asymmetries between insureds and insurer(s), simplified regulatory corrective actions and measures such as fines, rebates, or mandatory cyber insurance may increase incentives for self-protection, see, e.g. \cite{Pal2014} and \cite{liu2014}. For a detailed summary and comparative analysis of this literature, see \cite{Marotta2017}; see also \cite{Boehme2010}, and \cite{Boehme2018}.
However, the modeling framework adopted for risk contagion is often extremely simple and static, excluding risk amplification and the possibility of very high loss events.

In contrast, dynamic models of contagion processes provide a more realistic framework. Originally, such models were developed in the field of mathematical biology and epidemiology since the seminal work of \cite{Kermack1927}. In the last two decades, extensive efforts have been made to incorporate the underlying contact structure within populations into the modeling framework: Epidemic processes have been generalized to networks; see, for example, \cite{PastorSatorras2015} and \cite{Kiss2017} for detailed reviews. 
Because of their ability to capture interconnectedness, approaches to modeling epidemics in the context of cyber risk have also appeared recently.
For example, models of network contagion are utilized in \cite{fahrenwaldt2018pricing}, \cite{xu2019cybersecurity}, \cite{Jevtic2020}, \cite{Antonio2021}, and \cite{Chiaradonna2023} for the purpose of pricing cyber insurance policies. Furthermore, the impact of cyber risk contagion on insurance portfolios has been analyzed in \cite{Hillairet2021}, and more recently the network structure of interconnected industry sectors has been considered, see \cite{Hillairet2021b}. A dynamic contagion game was introduced in \cite{hayel2014} using the Markov-SIS model (but based on the easily tractable, albeit rough, NIMFA approximation). However, to our knowledge, the regulation, management, and control of contagious cyber risks have not yet been studied in a modeling framework based on dynamic contagion.

Beyond the field of cyber risk, applications of network models to insurance-related problems are less common in the literature. Existing works focus, for example, on the implementation of data science methods such as fraud detection techniques, see \cite{Tumminello2023},  or study the systemic risk in financial networks where insurance companies themselves are present as interdependent financial actors, such as in \cite{Chen20182} and \cite{Chen2020}.

Studies of network resilience and robustness can be found in the engineering and computer science literature. However, much of the work focuses exclusively on measurements of network topology properties (see \cite{Freitas2022} for a recent overview) or is based on models of lateral network movements that do not capture the infection and recovery dynamics of risk contagion (see \cite{Chen2018} or \cite{Freitas2020}). Moreover, resilience building is studied only from a network perspective and not in a regulatory framework. A specific attempt to build network resilience against self-propagating malware, and in particular the WannaCry worm, was recently presented in \cite{Chernikova2022}. The authors use synthetic WannaCry data to derive an adequate contagion model, similar to the classic SIR model, and appropriate parameter estimates. However, this model follows a deterministic top-down population-based approach, whereas our study is based on a stochastic bottom-up model of node-level interactions. Again, resilience is considered from an engineering perspective rather than a regulatory one, and issues of network economics and risk management are also not considered.

\paragraph{Outline} 
The paper is organized as follows. 
In Section~\ref{sec:legal}, we provide a brief overview of current cybersecurity legislation in the European Union (EU) and the United States of America (US), and also mention the regulatory role of private actors and insurance companies. Based on this, we present a selection of current approaches from the field to strengthen cybersecurity.
In Section~\ref{sec:mathematical}, we introduce the artificial cyber lab, and in the following two sections, we conduct the aforementioned illustrative case studies to analyze security- and topology-based cyber resilience measures. In light of these findings, we also revisit the selected real-world approaches from Section~\ref{sec:legal}. Section~\ref{sec:conclusion} concludes.

\section{The Real World: The Current State of Cybersecurity Regulation}\label{sec:legal}

In what follows, we briefly discuss the main characteristics of current cybersecurity legislation in the EU and the US, as well as the role of private actors such as insurance companies in shaping cybersecurity standards. 
This discussion will serve to identify and classify a set of real-world measures for improving resilience to cyberattacks, which we will then discuss in light of our findings from simulations conducted in the artificial cyber lab. 

\subsection{Current Government Regulations for Cybersecurity}

Due to the enormously increasing importance of cybersecurity to the functioning of modern societies, lawmakers have enacted several regulations, including a variety of legal norms. However, given the non-stationary nature of cyberspace, policymakers tend to future-proof their regulations by using indeterminate legal terms when formulating security requirements. Examples of such phrases include ``adequate security measures'' or ``adequate technical and organizational measures,'' see below. On the one hand, this can guarantee a high level of cybersecurity, even if a new technology or vulnerability is found. On the other hand, the indeterminacy of the legal terms introduces a significant degree of uncertainty as to the ``correct'' cybersecurity measures to be taken. 
In light of the latter problem, a growing number of technical standards and guidelines published by organizations such as the Cybersecurity \& Infrastructure Security Agency (CISA) and the National Institute of Standards and Technology (NIST) in the U.S., the International Organization for Standardization (ISO), the European Network and Information Security Agency (ENISA), or TeleTrusT - IT Security Association Germany and the Bundesamt f\"ur Sicherheit in der Informationstechnik (BSI) in Germany provide specific guidance, see for example \cite{Teletrust2021} and \cite{BSI2022}. While some of these standards actually serve as guidelines for government institutions, they are not legally binding for private companies, and furthermore they are usually characterized by a high degree of complexity, see for example \cite{BSI2022}. Both the non-legally binding nature and the complexity may prevent companies from implementing these standards in practice.

\begin{shaded}
\centering \textbf{Cyber Security Legislation in the EU and the US}
\bigskip

\centering \textit{Protection of Critical Infrastructure}
\begin{itemize}
\item The EU sets minimum standards for cybersecurity of critical infrastructure in the 2020 NIS Directive.\footnotemark ~The requirements include organizational provisions such as risk analysis and policies for information systems security, incident handling, business continuity and crisis management, supply chain security, and IT-related technical safeguards. 
In this context, critical infrastructure operators are required to implement ``appropriate security measures.'' However, the specific design of these measures is not specified in the directive.
\item In the US,  the Cybersecurity and Infrastructure Security Agency Act of 2018 entailed the establishment of the Cybersecurity \& Infrastructure Security Agency (CISA)  by the Department of Homeland Security. The CISA regularly publishes \textit{Binding Operational Directives} 
in which explicit actions improving the cybersecurity of federal civilian agencies are stated. 
For example, the recently published Directive BOD 22-01 requires all federal civilian agencies to remediate newly discovered exploits within a period of two weeks since disclosure, based on a regularly updated catalogue of known exploited vulnerabilities. Thereby, CISA sets a fixed threshold for software and service providers to roll out patches and updates for their respective end users. Although the BOD 22-01 targets federal civilian agencies only, CISA itself strongly recommends that private businesses review and monitor the catalogue 
to strengthen their cybersecurity.
\end{itemize}
\centering \textit{Data Protection}
\begin{itemize}
\item The \textit{General Data Protection Regulation (GDPR)} is the centerpiece of data protection legislation in the EU. 
It has been in force since May 25, 2018 and regulates the handling of personal data.
The central provision of data protection is addressed in Art. 32 GDPR, which requires the implementation of ``appropriate technical and organizational measures'', taking into account the ``state of the art, the implementation costs, and the nature, scope, circumstances, and purposes of data processing''. However, these terms are not further specified.
\item In the US, many federal states have introduced legislation on data protection. Again, indeterminate legal terms are used to define legislative requirements. For example, Section 1798.81.5 (b) and Section 1798.81.5 (e) of the \textit{California Consumer Privacy Act (CCPA)} state that ``a business that collects a consumer's personal information shall implement reasonable security procedures and practices appropriate to the nature of the personal information to protect the personal information from unauthorized or illegal access, destruction, use, modification, or disclosure'' -- without specifying which measures may be considered ``reasonable security procedures''.
\end{itemize}
\end{shaded}
\footnotetext{See the ``Directive (EU) 2016/1148 of the European Parliament and of the Council of 6 July 2016 concerning measures for a high common level of security of network and information systems across the Union''. Later, we will also discuss the newly proposed NIS2 Directive which is set to replace the existing regulatory framework for critical infrastructures in the EU.}

\subsection{Regulation by Private Actors and the Role of Insurance Companies}\label{sec:private:actors}
Against the backdrop of legal uncertainty associated with the presence of indeterminate terms under current legislation and the fact that recommended technical standards are typically not legally binding for business corporations, private actors may  play an essential role in cybersecurity governance by implementing and shaping security standards. For example, \cite{Hurel2018} discuss the role of private companies as entrepreneurs of cyber standards, with particular attention to Microsoft's efforts to influence global security standards and policies.

For insurance companies and financial institutions, cyber security is an increasingly important issue because of their significance to society and the sensitive data they hold. An empirical study on this issue and its growing relevance within the US banking and insurance industry has been presented in \cite{Gatzert2022}. Also, \cite{Sweetman2022} provides a first history of computer security and network protection within major institutions from the UK banking sector.

In this paper, in contrast,  we will focus to a greater extent on the particular role that cyber insurance companies can play in \textit{promoting} security standards \textit{among their policyholders}. This role has also been studied in, for instance, \cite{Trang2017}, \cite{talesh2018}, \cite{Woods2020}, \cite{Herr2021}, and \cite{Lemnitzer2021}. There, it is found that insurers may act as \textit{private regulators} in cybersecurity governance: 
Cyber insurance is an efficient way for companies to manage their cyber risk and seek assistance in implementing appropriate security measures. Hence, insurance companies can promote cybersecurity and resilience for their policyholders by setting certain standards in their contractual obligations.

\subsection{Selected Measures of Cyber Resilience}\label{sec:Add} 
In the previous sections, we discussed the current framework of cybersecurity regulation and emphasized the role of both governments and private actors such as insurance companies in implementing cyber security standards and strengthening resilience. In this section, we present a selection of concrete measures to improve cyber resilience focusing on systemic cyber risks that are either already part of current practice or currently under discussion. 
In particular, we include some measures which appear in the European Commission's proposal for replacing the existing  NIS legislation by a new NIS2 Directive\footnote{See the ``Proposal for a Directive of the European Parliament and of the Council on measures for a high common level of cybersecurity across the Union, repealing Directive (EU) 2016/1148—EU-doc. COM (2020) 823 final, dated 16 December 2020''. A discussion on the proposal is provided in \cite{Sievers2021}.}. Consistent with the previous discussion, we will distinguish between government regulation (GOV) and private regulation, particularly insurance-based regulation (INS). In addition, we will distinguish between measures targeting the IT-security (\textit{security-related interventions}) and those aiming at the structure of the network (\textit{topology-based interventions}).
To understand why we consider both, recall the infamous WannaCry and NotPetya attacks mentioned in the introduction, which can serve as models for studying systemic cyber risk. In both of these incidents, the risk propagation was due to the \emph{spread of malware} across a \emph{network} of interconnected actors and was characterized by the following two key aspects:
\begin{itemize}
    \item Both attacks resulted from an initial vulnerability of Windows-based computer systems: devices that had not applied the latest patches from Microsoft or were running outdated systems were affected. Improved IT-security -- in this case: regular software updates -- may have prevented these attacks.
    \item Both cyber epidemics spread through IT networks and affect many interconnected computers across different institutions at a global scale. Controlling the topology, especially the connections to critical parts of the network, might have reduced the damage caused. 
\end{itemize}

\paragraph{Security-Related Interventions}  We consider the following security-related interventions:
\begin{framed}
\begin{itemize}
\item[GOV]
\begin{itemize}
\item[$\diamond$] \textit{Size-cap rule:} Instead of covering all, the proposal for the new NIS2 Directive suggests limiting the scope of the Directive to medium-sized and large companies operating in the targeted sectors or providing services covered by the NIS2 Directive. In general, micro or small enterprises from critical infrastructure sectors should not be affected by the directive while exceptional cases are listed in Article 2, \S 2.
\item[$\diamond$] \textit{Supply chain protection:} Article 18, \S 2 of the NIS2 proposal contains a new catalog of cybersecurity risk management measures that are intended to reflect the state of the art. Specifically, supply chain security measures must be implemented by addressing ``security-related aspects concerning the relationships between each entity and its suppliers or service providers such as providers of data storage and processing services or managed security services''. Note the use of the indeterminate legal term ``state of the art''.  Nonetheless, we adopt the idea of supply chain protection as a concrete measure that can be analyzed.  
\end{itemize}
\item[INS]
\begin{itemize}
\item[$\diamond$] \textit{Assistance services}:  Depending on the policyholder's own (lack of) expertise, the policyholder's level of security can be significantly increased by providing or requiring investment in cyber assistance services. Cyber assistance services include implementation services, staff training, and external security testing for policyholders.  Some insurers also offer a 24/7 hotline with direct contact to technical experts, as well as public relations and legal experts at their own expense to minimize the potential damage from an ongoing cyberattack.
  Insurers could potentially mandate additional services for certain policyholders.
 \item[$\diamond$] \textit{Patch management and backup:} The use of a patch management procedure and the application of a backup process are already part of the current cyber insurance practice, see, for instance, Section A1-16 in \cite{GDV2017}. However, efficiently tailoring these obligations to the characteristics of the policyholder can further improve their effectiveness.
 \end{itemize}
\end{itemize}
\end{framed}
Intuitively, the requirements for individual cybersecurity investments should contribute to a higher level of security for the overall system. However, increasing the level of security comes at a cost. There is, of course, a trade-off between the cost of maintaining a high level of cybersecurity and potential losses from cyberattacks. The situation becomes particularly complex when one considers that networked actors imply interdependent levels of IT security. The question naturally arises whether individually rational security investment decisions by network actors already provide a sound level of security for the system as a whole, or whether interdependence calls for additional security commitments? And if such extra commitments are necessary, how should they be implemented within a cyber network?

\paragraph{Topology-Based Interventions}  The topological arrangement of the interconnected agents is critical to the extent of resulting cyber risk. We will consider the following topology-based arrangements:\begin{framed}
\begin{itemize}
\item[GOV]
\begin{itemize}
\item[$\diamond$] \textit{Incident response and reporting:} Computer security incident response teams (CSIRTs) shall be designated by each EU member state according to Article 9 of the NIS2 proposal. Specific requirements and tasks for CSIRTs are defined in Article 10, including the monitoring of cyber threats, the implementation of an early warning system, and the provision of proactive network scanning upon request of an entity. In addition, Article 20 obliges ``essential and important entities'' to report incidents with a significant impact on their functioning or the provision of their services to regulatory authorities or the CSIRT without undue delay.
\item[$\diamond$] \textit{Critical supply chains:} In addition to IT-security aspects, also the underlying pattern of connections between business partners (and their partners) and along production chains may play an important role in securing supply chains. Therefore, the risk assessment of network characteristics may help protect highly interconnected industries and infrastructures. Article 19 of the NIS2 proposal allows for EU coordinated assessments of critical supply chains, identified by the Commission in consultation with ENISA.
\end{itemize}
\item[INS]
\begin{itemize}
 \item[$\diamond$] \textit{Contact liability premiums}: A major concern are existing contagion channels for risk spreading and amplification. For instance, policyholders might have to provide so-called \textit{need-to-access} information when signing cyber insurance contracts, see for instance Kategorie B.4 in \cite{GDV2019}. 
 The idea is to monitor the number and type of access to a given IT facility and thus control potential contagion channels. To counteract possible accumulation scenarios, it might even be sensible to introduce additional risk premiums for systemic cyber events that depend on the existing contagion channels.
\item[$\diamond$] \textit{Insurance backstop mechanism}: \cite{Lemnitzer2021} argues for the necessity of a state-funded backstop mechanism for systemic cyber incidents to cover the losses of catastrophic events, similar to the Terrorism Risk Insurance Act (TRIA) which was established in the United States after 9/11. Here, a federal guarantee could be given to the insurance industry; after the occurrence of a systemic cyber event, mandatory surcharges could be imposed to the policyholders for the settlement of the costs incurred. Similar to the allocation of contact liability premiums, the size of these surcharges could correspond to the policyholders' individual contribution to the overall systemic risk.
\end{itemize}
\end{itemize}
\end{framed}
    Here too, of course, a trade-off exists between viewing the network links as contagion channels 
  and providing an \textit{effective infrastructure for data distribution}. Obligations should be implemented in a way that minimizes any negative impact on network functionality. 
    But how can the exposure to large cyber risk be assessed in complex network arrangements? What network characteristics do significantly increase the risk of large-scale cyber events? And how can effective topology-based measures be designed and implemented?
%

\section{The Artificial Cyber Lab - the Digital Twin of a Complex Cyber System}\label{sec:mathematical}
Important characteristics of cyber risk are the scarcity of data and the non-stationarity of the cyber environment due to the rapidly evolving IT-infrastructure.  
However, since classical statistical and actuarial models follow a frequency-severity approach and thus heavily rely on a sufficient amount of meaningful data, these standard methods are insufficient to evaluate the impact of cyber resilience interventions. To explore the questions from the previous section and assess the quality of proposed measures, we follow the digital twin paradigm and propose a novel approach based on models from network science and contagion theory; an experimental setup where cyber resilience measures can be implemented and tested through analysis and simulation - the \emph{artificial cyber lab}. 

 To build the  virtual counterparts of real-world cyber systems, a certain degree of abstraction is necessary to  provide a sufficiently complex but still tractable modeling framework. 
In general, network models for cyber risk contagion consist of three key components which we will sketch subsequently:
\begin{enumerate}
    \item[(i)] A \emph{network} representing interaction channels between agents or entities,
    \item[(ii)] a model for the \emph{spread} of a certain cyber threat through these interaction channels,
    \item[(iii)] and a \emph{loss model} determining the (monetary) losses occurring at the different agents due to the spread of the considered cyber threat.
\end{enumerate}

\subsection{Networks} Systems of interconnected agents, like  companies with data exchange, computer systems, or single devices, can mathematically be interpreted as {networks}. Agents are represented as \textit{nodes}, and the interaction channels (potential infection channels) between them as \textit{edges}. Exemplary network structures are depicted in Figure \ref{fig:exnetworks}.

\begin{figure}[h]
	\begin{center}
		\begin{minipage}[t]{0.3\linewidth}
			\centering
			\begin{tikzpicture}[scale=1.5] 
\tikzstyle{every node}=[draw, fill=White!85!gray, shape=circle];
\node (1) at (0, 1) {1};
\node (2) at ( 0.707, 0.707) {2};
\node (3) at (1, 0) {3};
\node (4) at (0.707, -0.707) {4};
\node (5) at (0, -1) {5};
\node (6) at (-0.707, -0.707) {6};
\node (7) at (-1, 0) {7};
\node (8) at (-0.707, 0.707) {8};
\draw (1) -- (2)
(1) -- (3)
(1) -- (4)
(1) -- (5)
(1) -- (6)
(1) -- (7)
(1) -- (8)
(2) -- (3)
(2) -- (4)
(2) -- (5)
(2) -- (6)
(2) -- (7)
(2) -- (8)
(3) -- (4)
(3) -- (5)
(3) -- (6)
(3) -- (7)
(3) -- (8)
(4) -- (5)
(4) -- (6)
(4) -- (7)
(4) -- (8)
(5) -- (6)
(5) -- (7)
(5) -- (8)
(6) -- (7)
(6) -- (8)
(7) -- (8);
\end{tikzpicture}
		\caption*{(a) fully connected}
			
		\end{minipage}
		\begin{minipage}[t]{0.3\linewidth}
			\centering
			\begin{tikzpicture}[scale=1.5] 
\tikzstyle{every node}=[draw, fill=White!85!gray, shape=circle];
\node (1) at (0, 0) {1};
\node (2) at ( 0, 1) {2};
\node (3) at (0.78183, 0.62349) {3};
\node (4) at (0.97493, -0.2225) {4};
\node (5) at (0.43388, -0.901) {5};
\node (6) at (-0.43388, -0.901) {6};
\node (7) at (-0.97493, -0.2225) {7};
\node (8) at (-0.78183, 0.62349) {8};

\draw 
(1) -- (2)
(1) -- (3)
(1) -- (4)
(1) -- (5)
(1) -- (6)
(1) -- (7)
(1) -- (8);
\end{tikzpicture}
			\caption*{(b) star-shaped}
			
		\end{minipage}
		\begin{minipage}[t]{0.3\linewidth}
			\centering
			\begin{tikzpicture}[scale=1.5] 
\tikzstyle{every node}=[draw, fill=White!85!gray, shape=circle];
\node (1) at (-2, 0) {1};
\node (2) at ( -1, 0) {2};
\node (3) at (-0.293, 0.707) {3};
\node (4) at (-0.293, -0.707) {4};
\node (5) at (0.647, 1.049) {5};
\node (6) at (0.647, 0.365) {6};
\node (7) at (0.647, -0.365) {7};
\node (8) at (0.647, -1.049) {8};
\draw 	(1) -- (2)
(2) -- (3)
(2) -- (4)
(3) -- (5)
(3) -- (6)
(4) -- (7)
(4) -- (8)
;
\end{tikzpicture}
			\caption*{(c) branching tree}
			
		\end{minipage}
	\end{center}
\caption{Exemplary network structures with $N=8$ nodes.}
\label{fig:exnetworks}
\end{figure}

A simple (unweighted) network connecting $N$ different agents can be represented by its adjacency matrix $A=(a_{ij})_{i,j\in\{1,\ldots,N\}}$ with $a_{ij}\in\{0,1\}$: here, $a_{ij}=1$ indicates that nodes $i$ and $j$ are directly connected, $a_{ij}=0$ indicates no direct connection.\footnote{Alternatively, weighted networks could be considered. Here, $a_{ij}>0$ represents the strength of the connection between nodes $i$ and $j$.} For example, in the case of the tree network depicted in Figure \ref{fig:exnetworks} (c), $A$ is given by
\begin{equation*}
A = \scriptsize\begin{pmatrix}
0 & 1 & 0 & 0 & 0 & 0 & 0 & 0  \\
1 & 0 & 1 & 1 & 0 & 0 & 0 & 0  \\
0 & 1 & 0 & 0 & 1 & 1 & 0 & 0  \\
0 & 1 & 0 & 0 & 0 & 0 & 1 & 1  \\
0 & 0 & 1 & 0 & 0 & 0 & 0 & 0 \\
0 & 0 & 1 & 0 & 0 & 0 & 0 & 0 \\
0 & 0 & 0 & 1 & 0 & 0 & 0 & 0 \\
0 & 0 & 0 & 1 & 0 & 0 & 0 & 0 \\
\end{pmatrix}_.
\end{equation*}

\subsubsection{Random Network Models}\label{sec:randnet} In applied network analysis, the exact network structure is often unknown. In this case, {random network models} enable sampling from a class of networks with given fixed topological characteristics (such as the overall number of nodes). In a random network, each possible edge in the network is present (or absent) with a given fixed probability.
We consider the following two standard classes of undirected random networks:\footnote{Pseudocode for both random network models is provided in Appendix \ref{app:networks}.}
\begin{itemize}
    \item \textit{Erd\H{o}s-R\'enyi networks:} The simplest random network model was introduced by \cite{Erdos1959}: The Erd\H{o}s-R\'enyi network $G_p (N)$ is constructed from a set of $N$ nodes in which each of the possible $N(N-1)/2$ edges is independently present with the same probability $p$, i.e., the expected number of edges is given by $(N(N-1)p)/2\approx (N^2p)/2$ for large $N$.
\item \textit{Barab\'asi-Albert networks:}  A phenomenon widely observed in the empirical analysis of networks, including the world wide web, IT networks, and social networks, is that newly formed connections tend to emerge at nodes with an already large degree. For example, newly created websites are more likely to link to an already existing popular website than to other websites. This principle is called \textit{preferential attachment}. Hence, real-world networks are usually more heterogeneous in terms of their topology than the Erd\H{o}s-R\'enyi model would suggest.  Often, a hierarchy of nodes is observable---with a few nodes of high degree (called \textit{hubs}), and a vast majority of less connected nodes.  A first model, motivated by the study of citation networks of academic papers, was introduced and discussed in \cite{Price1965} and \cite{Price1976}. The most commonly applied random graph model for networks which follow a preferential attachment principle is the one from \cite{Barabasi1999}. Different from the Erd\H{o}s-R\'enyi model, a Barab\'asi-Albert network $BA(N; m)$ with $N$ nodes is generated by a growing network algorithm: Starting from an initial core with $n_0$ nodes, $m\leq n_0$, and $\epsilon_0$ edges, a new node $i$ is added to the graph in each simulation step and $m$ edges for $i$ are randomly generated following a preferential attachment rule. The number of edges for the resulting network is given by $m(N-n_0) + \epsilon_0$, which, neglecting the initial core, can be approximated by $mN$.
\end{itemize}

\subsubsection{Measuring Centrality}\label{sec:centrmea} In network science, the structural importance of single nodes or edges within the network can be characterized using \textit{centrality measures} $\mathcal{C} $. However, centrality is not a rigorously defined term and a large variety of different concepts has been proposed.
\begin{enumerate}

\item For a network \textit{edge} $e$, a common way to measure centrality is to consider the fraction of shortest paths between any two nodes $i$ and $j$ that pass through $e$. The corresponding measure is then called \textit{edge (betweenness) centrality},\footnote{This centrality measure was introduced in \cite{Girvan2002}.} and, 
can be written as
\begin{equation}\label{eq:edgcen}
    \mathcal{C}^{edge}(e) = \sum_{i,j} \frac{\sigma_{ij}(e)}{\sigma_{ij}},
\end{equation}
where $\sigma_{ij}$ denotes the number of shortest paths between nodes $i$ and $j$, and $\sigma_{ij}(e)$ is the total number of these paths that go through edge $e$. 

\item For a network \textit{node} $i$, two of the most frequently used measures are\footnote{For an extensive overview we refer to Chapter 7 in \cite{Newman2010}.}
\begin{itemize}
        \item \textit{Degree centrality:} Here nodes are simply ranked by their number of network neighbors, i.e., 
        \begin{equation}\label{eq:degcen}
        \mathcal{C}^\text{deg}(i) = \sum_{j=1}^N a_{ij} = \sum_{j=1}^N a_{ji},\qquad i = 1,\ldots , N,
        \end{equation}
         for an undirected graph. It accounts for immediate network effects. 
        \item \textit{Betweenness centrality:} 
        In contrast to degree-based approaches, betweenness centrality focuses on the role nodes may play as connections or ``bridges'' between different network regions. In analogy to the concept of \textit{edge} betweenness centrality, the corresponding definition on the node level is given by 
		\begin{equation}   \label{eq:betnode}    
        \mathcal{C}^\text{bet}(i) = \sum_{j,h} \frac{\sigma_{jh}(i)}{\sigma_{jh}},\qquad i=1,\ldots, N,
		\end{equation}         
         where $\sigma_{jh}$ denotes the total number of shortest paths between nodes $j$ and $h$, and $\sigma_{jh}(i)$ is the particular number of these paths that go through node $i$.
\end{itemize}
\end{enumerate}
\subsection{Modeling Contagious Cyber Risks} Through the interaction channels described by the chosen network, a contagious cyber risk may spread. Mathematical models describing the spread of cyber epidemics on networks first divide the set of agents into distinct categories varying over time: e.g., individuals that are \emph{susceptible} to an infection, \emph{infected}, and \emph{recovered} individuals. The SIS \textit{(Susceptible-Infected-Susceptible)} and SIR \textit{(Susceptible-Infected-Recovered)} Markov models constitute frequently used epidemic spreading models on networks. The difference between them is the presence (SIR) or absence (SIS) of immunity: While reinfection events are possible in the SIS framework, in the SIR framework, recovered individuals gain permanent immunity. A rigorous discussion of the mathematical aspects is provided in Appendix \ref{app:SIR Dynamics}. 

\begin{figure}[h]
\centering
\begin{minipage}[t]{0.45\linewidth}
\centering
\begin{tikzpicture}[scale=1] 
\node[draw, shape=circle] (1)[fill=BrickRed!40!White] at (0, 0) {I};
\node[draw, shape=circle] (2) [fill=OliveGreen!40!White] at ( 1, 0) {S};
\node[draw, shape=circle] (3)[fill=BrickRed!40!White] at (4, 0) {I};
\node [draw, shape=circle](4) [fill=BrickRed!40!White] at ( 5, 0) {I};
\draw (1) -- (2)
(3) -- (4);
\draw[arrow, line width=0.5mm] (1.8,0) --  node[anchor=south] {$\tau$} (3.2,0) ;
\node[draw, shape=circle] (5) [fill=BrickRed!40!White] at (1,-1.5) {I};
\node[draw, shape=circle](6) [fill=OliveGreen!40!White] at (4,-1.5) {S};
\draw [->,decorate,decoration={snake,amplitude=.4mm,segment length=2mm,post length=1mm}, line width=0.5mm] (1.8,-1.5) -- node[anchor=south] {$\gamma$}(3.2,-1.5);
\end{tikzpicture}\\
\textbf{\sffamily (a) SIS Model}
\end{minipage}
\begin{minipage}[t]{0.45\linewidth}
\centering
\begin{tikzpicture}[scale=1] 
\node[draw, shape=circle] (1)[fill=BrickRed!40!White] at (0, 0) {I};
\node[draw, shape=circle] (2) [fill=OliveGreen!40!White] at ( 1, 0) {S};
\node[draw, shape=circle] (3)[fill=BrickRed!40!White] at (4, 0) {I};
\node [draw, shape=circle](4) [fill=BrickRed!40!White] at ( 5, 0) {I};
\draw (1) -- (2)
(3) -- (4);
\draw[arrow, line width=0.5mm] (1.8,0) --  node[anchor=south] {$\tau$} (3.2,0) ;
\node[draw, shape=circle] (5) [fill=BrickRed!40!White] at (1,-1.5) {I};
\node[draw, shape=circle](6) [fill=White!85!gray] at (4,-1.5) {R};
\draw [->,decorate,decoration={snake,amplitude=.4mm,segment length=2mm,post length=1mm}, line width=0.5mm] (1.8,-1.5) -- node[anchor=south] {$\gamma$}(3.2,-1.5);
\end{tikzpicture}\\
\textbf{\sffamily (b) SIR Model}
\end{minipage}
\caption{Infection and recovery for the SIS and SIR model in a network: A susceptible node is infected by its contagious neighbor with rate $\tau$. Independent from the state of its neighbors, an infected node recovers at rate $\gamma$. SIS and SIR differ in terms of immunity: In the SIS model, a recovered node is susceptible again such that multiple infections for the same node are possible. In contrast, recovery in the SIR model means that the node is immune and cannot be infected again. }
\label{fig:SISSIR}
\end{figure}
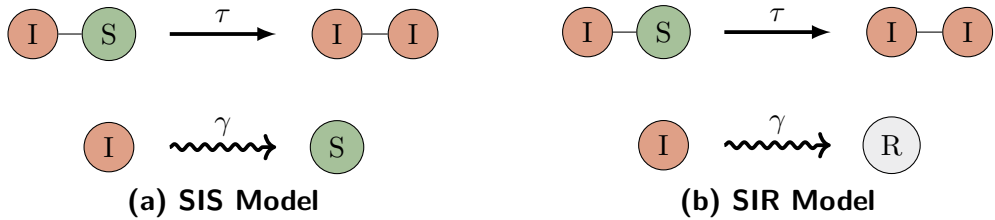

  The possible transitions in these two models as well as their two key parameters, the infection rate $\tau$ and the recovery rate $\gamma$, are illustrated in Figure \ref{fig:SISSIR}.\footnote{There also exist more nuanced models, e.g., containing only a limited immunity (SIRS) or an additional category of exposed individuals, i.e., individuals that are infected but not yet contagious (SEIR). For more details, see, e.g., \cite{PastorSatorras2015} and \cite{Kiss2017}.}

\subsection{Cyber Loss Models}  Finally, agents, i.e., nodes in the network, may experience \textit{losses} due to a cyber infection. Depending on the modeling purpose, the cyber loss model may emphasize  different aspects of an ongoing cyber incident, like the total number of affected network components, aggregate losses of network nodes, and the monetary losses of single entities. Typically, an adequate model should reflect on the stochastic nature of risk scenarios and capture key statistical aspects of cyber loss distributions, including loss expectations and tail risk properties. 

\subsection{Artificial Cyber Lab Setup}\label{sec:artificial:cyber:lab:setup}

For our design of the artificial cyber lab, a fundamental choice has to be made in terms of the contagious spread model, namely between an SIR and SIS approach (see Figure 2). Since we consider attacks similar to the WannaCry and NotPetya attacks, which were both based on the EternalBlue exploit, we assume that reinfections are rather unlikely because---once detected---the underlying security issues are easily solvable through the installation of the latest patches. Therefore, we will use the SIR model.\footnote{A similar choice has also been made in \cite{Hillairet2021} for modelling the WannaCry attack. Here, the authors use the population-based ODE system from \cite{Kermack1927} instead of a stochastic network model.}

The key \textit{in-} and \textit{output parameters} can be summarized as follows:

\begin{itemize}
    \item {\bfseries\sffamily Input:}
\begin{itemize}
\item \textbf{\sffamily Network:}
\begin{itemize}
    \item network size $N$ (number of agents)
    \item topological structure $A= (a_{ij})_{i,j =1,\ldots , N}$, i.e., the connectivity pattern between nodes (see Figure 1 for examples)
\item number and position of initially infected nodes
\end{itemize}

    \item {\bfseries\sffamily Epidemic Dynamics:}
\begin{itemize}
    \item infection rate $\tau = 0.1$ (determines the speed of the infection), assumed to be equal for all connections\footnote{Reasonable estimates of the infection speed in contagious cyber incidents cannot be derived due to insufficient data. We assume $\tau = 0.1$, which in a Markovian setting corresponds to the expected waiting time of 10 units of time for infectious transmission over a network edge. Our results can easily be adapted to a specific infection speed scenario by adequate interpretation of the time unit.}
\item individual recovery rates $\gamma_i$ for nodes $i=1,\ldots,N$ (influence the time needed for recovery---interpreted as IT security level, see Section \ref{subsec:CSI})
\end{itemize}

\item \textbf{\sffamily Loss Distribution:}
\begin{itemize}
    \item stochastic modeling framework for loss formation
\end{itemize}
\end{itemize}
\item {\bfseries \sffamily Output:}
\begin{itemize}
    \item \textbf{\sffamily Epidemic Dynamics:}
\begin{itemize}
    \item spread of cyber infection over time, total number of affected nodes, probability of infection for each node
\end{itemize}
\item \textbf{\sffamily Loss Distribution:}
\begin{itemize}
\item aggregate losses for single nodes or the entire network
\end{itemize}
\end{itemize}
\end{itemize}
 In the following, we use the lab to generate artificial data from our virtual model and evaluate two different types of cyber resilience interventions. Based on the results, we discuss the implications of our findings on the implementation of concrete cyber resilience measures for real-world cyber systems.

\section{Case Study I: Security-Related Interventions under Strategic Interaction}\label{subsec:CSI}
For both the WannaCry and NotPetya attacks, the vulnerability of systems was crucially dependent on the security efforts taken by individual network users.
Therefore, we firstly introduce a suitable model for security levels, benefits and costs within the framework of our artificial cyber lab. However, due to the interconnectedness of entities in cyber systems, the individual risk exposure is also influenced by the security choices of other network participants: Interdependence and strategic interaction of different actors constitute a key characteristic of systemic cyber risks. Therefore, we develop a \textit{security investment game} in order to study interdependence effects within the cyber network.
Finally, we evaluate if, and how, security-related interventions in the form of additional security obligations can efficiently be allocated among network nodes to improve the overall safety of the cyber system. 

\subsection{Security Investments and Strategic Interaction}
\begin{figure}[h]
\begin{minipage}[t]{0.45\linewidth}
\centering
\begin{tikzpicture}[scale=1] 
\node[draw, shape=circle] (1)[fill=BrickRed!40!White] at (0, 0) {I};
\node[draw, shape=circle] (2) [fill=OliveGreen!40!White] at ( 3, 1) {S};
\node[draw, shape=circle] (3)[fill=OliveGreen!40!White] at (3, -1) {S};
\node[draw=none] (4) at (-1.35, 0) {$\gamma_i$};
\draw[arrow, line width=0.7mm] (0.4,0.2) --   (2.6,0.8) ;
\draw[arrow, line width=0.7mm] (0.4,-0.2) --  (2.6,-0.8) ;
\draw[thick, ->,line width=.3mm] (-0.2,0.3) arc (30:320:0.5cm);
\end{tikzpicture}\\
{(a) Low Security Level}
\end{minipage}
\begin{minipage}[t]{0.45\linewidth}
\centering
\begin{tikzpicture}[scale=1] 
\node[draw, shape=circle] (1)[fill=BrickRed!40!White] at (0, 0) {I};
\node[draw, shape=circle] (2) [fill=OliveGreen!40!White] at ( 3, 1) {S};
\node[draw, shape=circle] (3)[fill=OliveGreen!40!White] at (3, -1) {S};
\node[draw=none] (4) at (-1.35, 0) {$\gamma_i$};
\draw[arrow, line width=0.3mm] (0.4,0.2) --   (2.6,0.8) ;
\draw[arrow, line width=0.3mm] (0.4,-0.2) --  (2.6,-0.8) ;
\draw[thick, ->,line width=.7mm] (-0.2,0.3) arc (30:320:0.5cm);
\end{tikzpicture}\\
{(b) High Security Level}
\end{minipage}
\caption{Contagious spreading for initially infected nodes with low and high recovery rates $\gamma_i$. The value of $\gamma_i$ reflects the IT security level and protection efforts of company $i$.}
\label{fig:UpdateFrequency}
\end{figure}
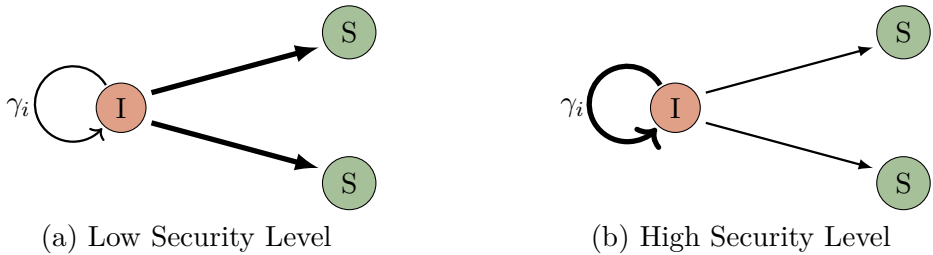

In our SIR model, the cyber risk exposure of network nodes depends on the epidemic infection and recovery rates. For tractability reasons, we assume a \textit{fixed} homogeneous infection rate $\tau$ (see Section~\ref{sec:artificial:cyber:lab:setup}) and vary the individual recovery rate $\gamma_i$ of node $i=1,\ldots,N$ which we will interpret as security level: 
The lower the security level $\gamma_i$, the longer it takes for firm $i$ to detect a cyber infection or an existing security gap. Consequently, this also affects the risk exposure of the firm's direct network neighbors, see Figure \ref{fig:UpdateFrequency}. 

From the perspective of the individual node $i$, the choice of security level $\gamma_i$ results from the \textit{trade-off} between the following two functions:
    \begin{enumerate}
    \item The \textit{cyber loss function} $L_i(\gamma_1,\gamma_2,\ldots , \gamma_N)$ describes the losses of node $i$ -- as a function of all nodes' security levels due to the interconnectedness of network agents. In general, a loss model may capture a variety of aspects, see the discussion in the previous section. Clearly, the amount of cyber losses should be related to the \textit{duration} of a cyber attack, which, for instance, may correspond to downtime of services\footnote{For example, this idea has been proposed in the loss model from \cite{xu2019cybersecurity}.} and business interruption costs.
    We choose a simple and tractable loss model by setting
    \begin{equation*}
        L_i := L_i(\gamma_1,\ldots , \gamma_N) := \mathbb{E}\Big[ \int_0^\infty I_i(t) dt\Big]
    \end{equation*}
    which represents the expected amount of time node $i$ will spend in the infectious state $I$, given the security levels $\gamma_1,\ldots , \gamma_N$. In particular, $L_i$ can be reduced by increasing the security level $\gamma_i$. For details, see Appendix \ref{sec:lossmodel}.
    
   SIR infection dynamics are described by an ordered system of equations; see Appendix \ref{app:SIR Dynamics} for details. Note that the order of SIR equations increases up to the network size $N$. 
    Hence, solving the exact system is intractable for complex networks due to the large number of system equations. We thus follow a \textit{stochastic simulation approach}.\footnote{Trajectories of the SIR dynamics can be generated using the well-known \textit{Gillespie algorithm}, cf.~\cite{Gillespie1976} and \cite{Gillespie1977}. Pseudocode is provided in Appendix \ref{app:Gillespie}.}
   Further details are given in Appendix \ref{sec:lossesgame}.
     \item The \textit{cost function} $C_i(\gamma_i)$ describes the cost of the implementation of security level $\gamma_i$ for node $i$. For simplicity, we let $C_i = C$ for all $i=1,\ldots, N$. Typically, such a cost function should be {strictly convex}, representing a rapidly increasing cost with increasing targeted security level. Further, $C$ should satisfy $C(0) = 0$. For simplicity and tractability, we choose an exponential function
    \begin{equation*}
        C(\gamma_i) = e^{k\gamma_i}-1,\quad k>0,
    \end{equation*}
    with growth constant $k$. In the following, we set $k=1/3$.
    \end{enumerate}
    
    A rational network agent $i$ will try to minimize her \textit{total expenses} 
    \begin{equation*}
        \mathcal{E}_i(\gamma_1,\ldots , \gamma_N) = C_i(\gamma_i) + L_i(\gamma_1,\ldots , \gamma_N),
    \end{equation*}
i.e., the competing sums of security costs and cyber losses, as a function of $\gamma_i$.

As noted in Section~\ref{sec:artificial:cyber:lab:setup}, we choose the fixed homogeneous rate $\tau = 0.1$ for the \textit{infection dynamics}. The contagion process is initialized at time $t=0$ by the random infection of a single node. We remark that there are, of course, many reasonable choices for the loss and cost function and thus the total expenses, and also the infection dynamics. However, since our studies are of a qualitative and not a quantitative nature, we believe that our choices are suited to gain a basic understanding of the problem.

\subsubsection{Individually Optimal Security Level} \label{subsec:CSIindividual}
 Under the assumption that for all nodes $j\neq i$ the security level $\gamma_j$ remains unchanged, a security level $\gamma_i$ is \textit{individually optimal} for node $i$, if it minimizes the total expenses $\mathcal{E}_i$, i.e., a rational agent will choose the individually optimal security level
\begin{equation*}
    \gamma_i^{\text{ind}}(\gamma_{-i}) := \underset{\gamma_i\in [0,\infty)}{\argmin}\;  \mathcal{E}_i(\gamma_1,\ldots , \gamma_N)\quad \mbox{where} \quad \gamma_{-i} :=(\gamma_1,\ldots , \gamma_{i-1},\gamma_{i+1},\ldots , \gamma_N).
\end{equation*}

\begin{figure}[h]
		\centering
		\includegraphics[width=0.7\textwidth]{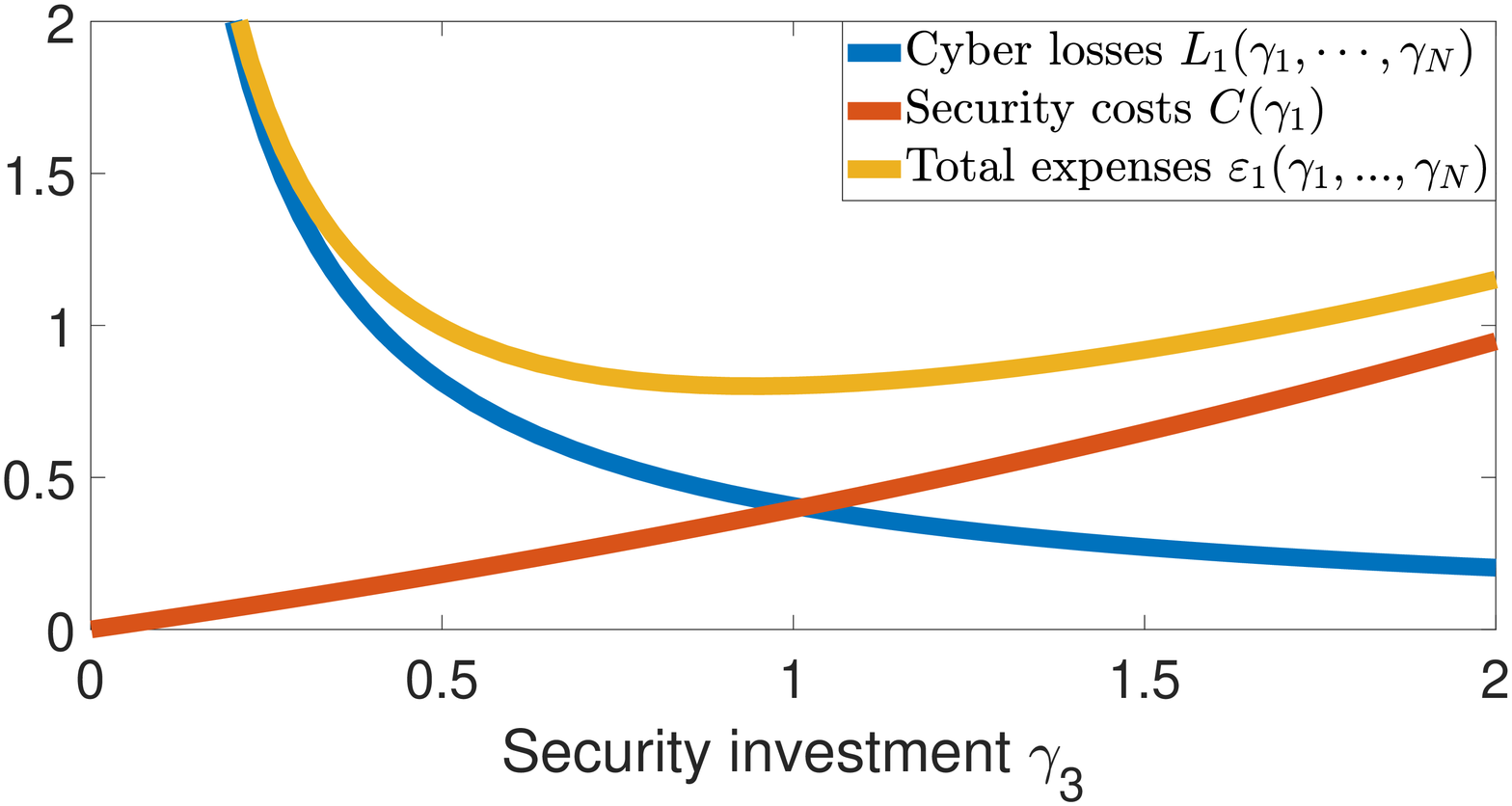}
		\caption{Cyber losses, security costs, and total expenses of node $3$ in the branching tree from Figure \ref{fig:exnetworks} as a function of the security level $\gamma_3$. Infection rates are assumed to be homogeneous, $\tau=0.1$, and security levels are set to $\gamma_j=0.1$ for $j\neq 3$. The value $\gamma_3^{\text{ind}} = 0.943$ is individually optimal. Cyber losses are calculated using the decomposition scheme from Appendix \ref{sec:lossmodel}: $\mathcal{T} = 10,000,000$ trajectories of the SIR process were generated to determine the probability $\mathbb{P}(A_3)$ where $A_3$ is the event that node $3$ becomes infected. For each simulation, the initially infected node was randomly chosen.}
		\label{fig:indoptimal}
	\end{figure}

An example for node 3 from the branching tree in Figure \ref{fig:exnetworks} is shown in Figure \ref{fig:indoptimal}.
	
\subsubsection{Strategic Interaction of Interdependent Actors} \label{subsec:CSIexternality}

The security level choices of network agents do not only affect their individual expenses $\mathcal{E}_i$ but also the cyber losses $L_j$, $j\neq i$, of other network nodes. Therefore, these nodes will in turn react to the new threat situation, initializing a cascade of \textit{strategic interactions}. We will call this the {\em security investment game}. A {\em steady state} of individually optimal security levels is a choice of security levels  $\gamma\in (0,\infty)^N$ such that  \begin{equation*}
\forall i=1,\ldots , N: \quad    \gamma^\text{ind}_i (\gamma_{-i}) = \gamma_{i}.
\end{equation*}
In other words, a steady state is a {\em Nash equilibrium} of the security investment game. The following theorem asserts the existence of steady states of individually optimal security levels. The proof of Theorem~\ref{theorem} is provided in Appendix~\ref{app:proof}.

\begin{theorem}\label{theorem} Steady states of individually optimal security levels exist.

\end{theorem}
{ Note that the theorem holds for basically any reasonable choices of cost functions $C_i$ and loss functions $L_i$ as long as the total expenses $\mathcal{E}_i$ remain strictly convex in $\gamma_i$ and admit a minimum point. In that case the proof would make use of Berge's maximum principle.}
We implement the security investment game as a dynamical game with several {rounds} $r=0, 1,\ldots, M$ where every round $r$ starts with a fixed vector of security levels $$\gamma(r) = (\gamma_1(r),\gamma_2(r),\ldots , \gamma_N(r)).$$  
\begin{algorithm}[The Security Investment Game]\label{alg:Security} $\,$\\ 
\emph{Input:} Initial configuration ${\gamma}(0)\in (0,\infty)^N$, number of rounds $M\in\mathbb{N}_{>0}$. 
 \begin{enumerate}
 \item \emph{(Initialization)} Set $r\to 0$.
    \item For every node $i$, $i=1,\ldots, N$, calculate 
    \begin{equation*}
    \gamma_i(r+1) = \underset{\gamma_i\in [0,\infty)}{\argmin}\;  \mathcal{E}_i(\gamma_1(r),\ldots ,\gamma_{i-1}(r), \gamma_i,\gamma_{i+1}(r),\ldots ,  \gamma_N(r)).
\end{equation*}
  More details are given in Appendix \ref{sec:lossesgame}. Set
  \begin{equation*}
\gamma({r+1}) = (\gamma_1(r+1),\gamma_2(r+1),\ldots , \gamma_N(r+1)).
\end{equation*}
    \item If $r<M$, set $r\to r+1$, and return to Step 2; otherwise end.
\end{enumerate}
\emph{Output:} Security configuration ${\gamma}(M)$ after $M$ rounds
\end{algorithm}

\subsubsection{Complex Network Interactions}
\begin{figure}[h!]
			\centering
	{\includegraphics[trim={2cm 3cm 0 0},clip,width=0.3\textwidth]{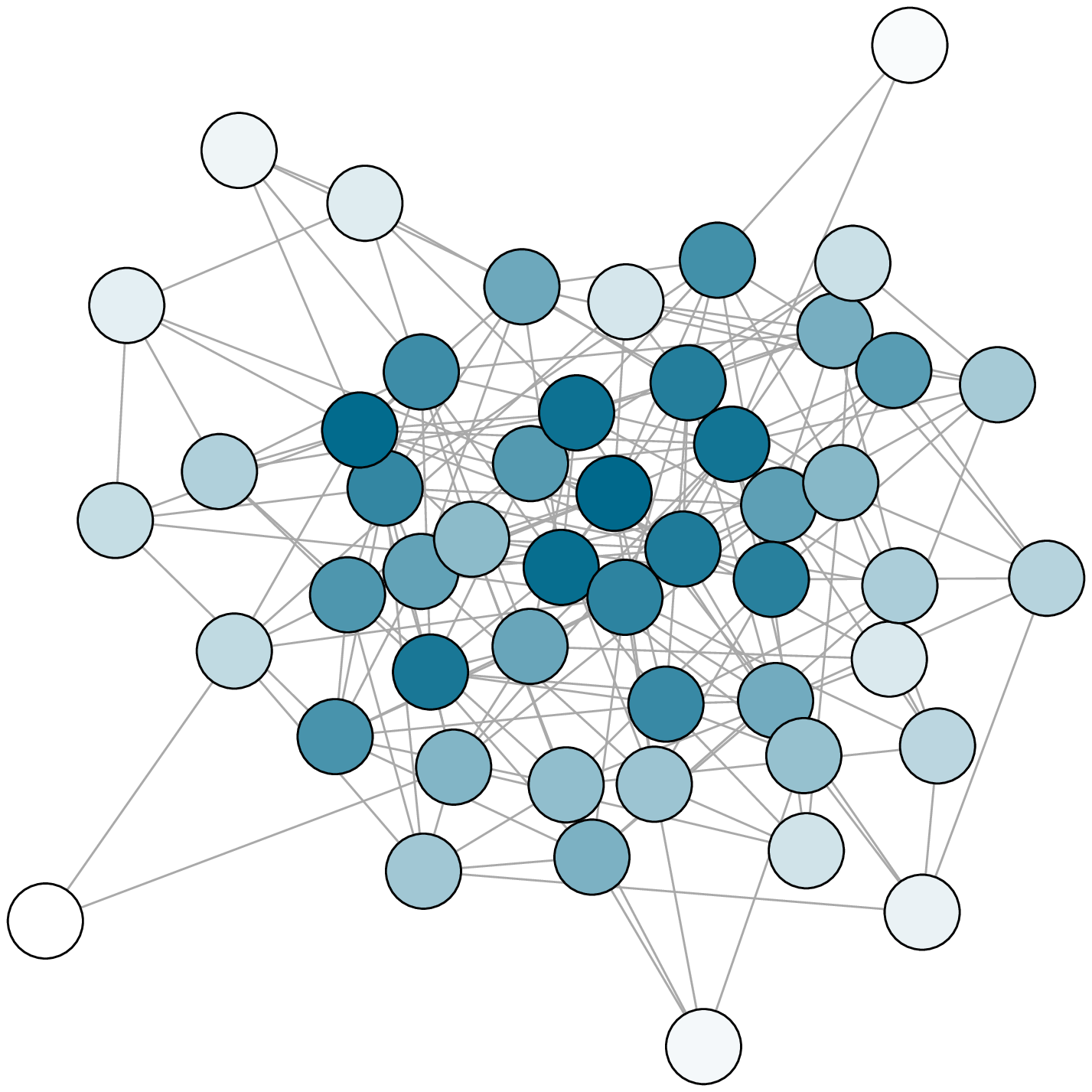}}
	{\includegraphics[trim={2cm 3cm 0 0},clip,width=0.3\textwidth]{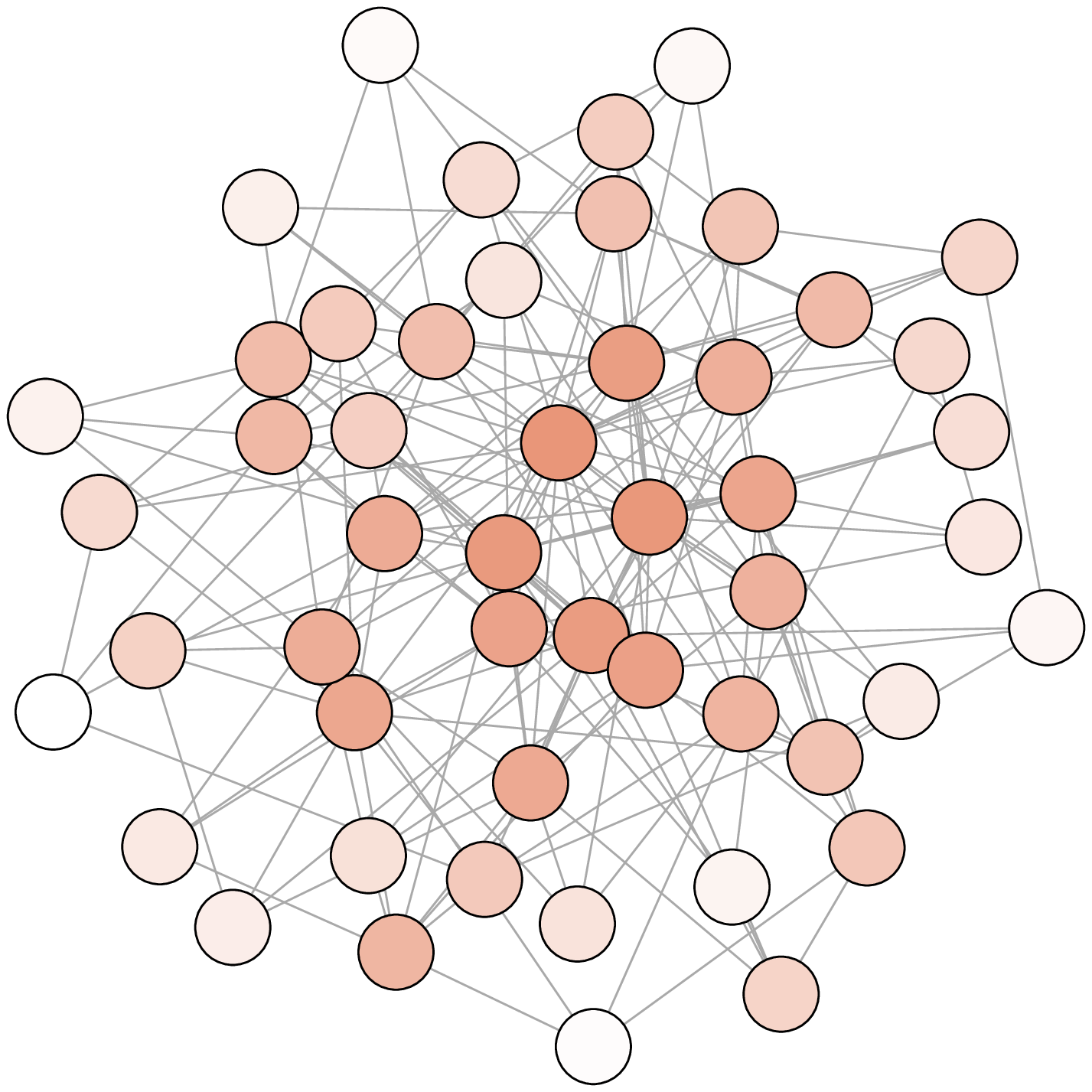}}
		\caption{Visualization of the considered exemplary networks drawn from the Erd\H{o}s-R\'enyi (left) and Barab\'asi-Albert (right) classes. Nodes are colored according to their chosen level of security after round 50 of the security investment game (Algorithm \ref{alg:Security}): the darker the color, the higher the chosen security level (\textit{for Erd\H{o}s-R\'enyi}: minimum: 0.3780, maximum: 0.6526; \textit{for Barab\'asi-Albert}: minimum: 0.4719, maximum: 0.7598). 		Data is based on $\mathcal{T}=10,000,000$ trajectories of the SIR epidemic process for each round of the security investment game.}
		\label{fig:SInetwork}
	\end{figure}

We study the strategic interaction in two particular \emph{fixed} networks: one generated from the Erd\H{o}s-R\'enyi class with parameters $N=50$ and $p=0.16$, and another one drawn from the Barab\'asi-Albert class with $N=50$ and $m = 4$. Note that these two exemplary networks are comparable with respect to their number of network connections, cf. Section \ref{sec:randnet}.
Visualizations are provided  in Figure \ref{fig:SInetwork}.

On both networks, we conduct the security investment game (Algorithm \ref{alg:Security}) with $M=50$ rounds and initial security level $\gamma_i(0) = 0.1$ for all nodes $i$. To generate values for the cyber losses $L_i$, in each round of the game, $\mathcal{T} =10,000,000$ trajectories of the SIR epidemic process are simulated; see Appendix \ref{sec:lossesgame} for details.

The results of the security investment game in the steady state, $\gamma^\text{stead} = (\gamma_1^\text{stead},\ldots, \gamma_N^\text{stead})$, are represented by node colors in Figure \ref{fig:SInetwork}.
For both network arrangements, we observe that more central nodes choose higher security levels than nodes in the periphery. The accumulated total expenses
\begin{equation*}
   \mathcal{E}(\gamma^\text{stead}) := \sum_{i=1}^N \mathcal{E}_i(\gamma_1^\text{stead},\ldots,\gamma_N^\text{stead})
\end{equation*}
are given by $\mathcal{E}(\gamma^\text{stead}) \approx 21.66$ for the Erd\H{o}s-R\'enyi-type, and $\mathcal{E}(\gamma^\text{stead}) \approx 21.92$ for the Barab\'asi-Albert-type network, respectively. 

\subsection{Demand for Regulation: Allocating Additional Security Investments}\label{subsec:allocatingMU}
In this section we address the question whether the individually optimal security choices given by a steady state $\gamma^\text{stead}$ are also favorable from an overall network perspective, i.e., do they minimize the accumulated total expenses $\mathcal{E}(\gamma)$, or can \textit{additional security investments} further improve the situation? In fact, the individually optimal security choices will in general \textit{not} lead to a minimization of the overall network expenses, see Appendix~\ref{app:2nodes} for a simple example. Indeed, it is well-known that Nash equilibria (steady states) do not in general minimize the social welfare function which in this case is the total expenses. In our case, a profound systematic characterization of the latter observation is, however, still an open challenge due to the lack of sufficient analytical tractability of the network dynamics, f.e., see Section 3.5.3 in \cite{Kiss2017} for a similar problem.


As indicated by the distribution of steady state security investments shown in Figure \ref{fig:SInetwork}, a key role in identifying good allocations of additional security investments may be played by the individual nodes' \textit{centrality}. To this end, recall the degree and betweenness centrality of network nodes introduced in Section \ref{sec:centrmea}. Note that these standard centrality measures from the literature are solely based on the underlying network topology. The security investment game, however, suggests yet another way of measuring centrality, namely by fixing a steady state $\gamma^\text{stead} $ and defining the centrality of node $i$ to be the individually optimal investment
\[\mathcal{C}^\text{inv}(i) = \gamma_i^\text{stead}.\] 
 In the following, we will refer to this latter centrality measure as the \textit{investment-based centrality}.

\subsubsection{Allocation Strategies}
In view of the previous discussion, in this section we proceed as follows: 
\begingroup\itshape
\begin{enumerate}
\item We start from a steady state $\gamma^\text{stead}$ of individually optimal security levels. Moreover, we fix an additional security budget $\beta>0$.
\item This extra amount of security is allocated amongst the nodes according to one of the following strategies:
\begin{enumerate}
    \item \textbf{Untargeted} allocation: $\beta$ is uniformly distributed among all network nodes, providing an additional security investment $\gamma_i^\text{all} = \beta /N$ for each node $i$.
    \item \textbf{Targeted} allocation: we choose a centrality measure $\mathcal{C}$ and determine the \textit{allocation weights}
   \[w_i := \frac{\mathcal{C}(i)}{\sum_{j=1}^N \mathcal{C}(j)},\qquad i = 1,\ldots , N.\] Based on these allocation weights we consider two opposing procedures:
   \begin{itemize}
       \item[i)] The \textbf{upper} allocation strategy allocates $\beta$ proportionally $\gamma_i^\text{all} := \beta\cdot w_i$. Here a higher amount of $\beta$ is assigned to nodes with a higher degree of centrality.
       \item[ii)] The \textbf{lower} allocation strategy does the opposite. To this end, we calculate the inverse allocation weights 
       \begin{equation*}
       \hat{w}_i := \begin{cases} w_i^{-1} & \text{if }w_i\neq 0 \\ 0 & \text{else}.\end{cases}
       \end{equation*}
       In this case the additional security investment $\gamma_i^\text{all}=\beta\cdot \Large(\hat{w}_i/ \sum_{j=1}^N \hat{w}_j\Large)$ for node $i$ assigns a higher amount of $\beta$ to nodes with a lower, yet positive, degree of centrality.
   \end{itemize}
 The proposed allocation procedures yield a new vector of security levels $\tilde{\gamma}$ with entries 
 \[\tilde{\gamma}_i =  \gamma^\text{stead}_i + \gamma^\text{all}_i,\qquad i=1,\ldots , N.\]
\end{enumerate}
\item Finally, we calculate the accumulated total network expenses $\mathcal{E}(\tilde{\gamma})$ under the new security configuration.
   \end{enumerate}
   \endgroup

\subsubsection{Allocation for Complex Networks}
\begin{figure}[h]
    \centering
\begin{minipage}[t]{0.32\linewidth}
\centering	
	{\includegraphics[trim={7.5cm 3cm 0 0},clip,width=1\textwidth]{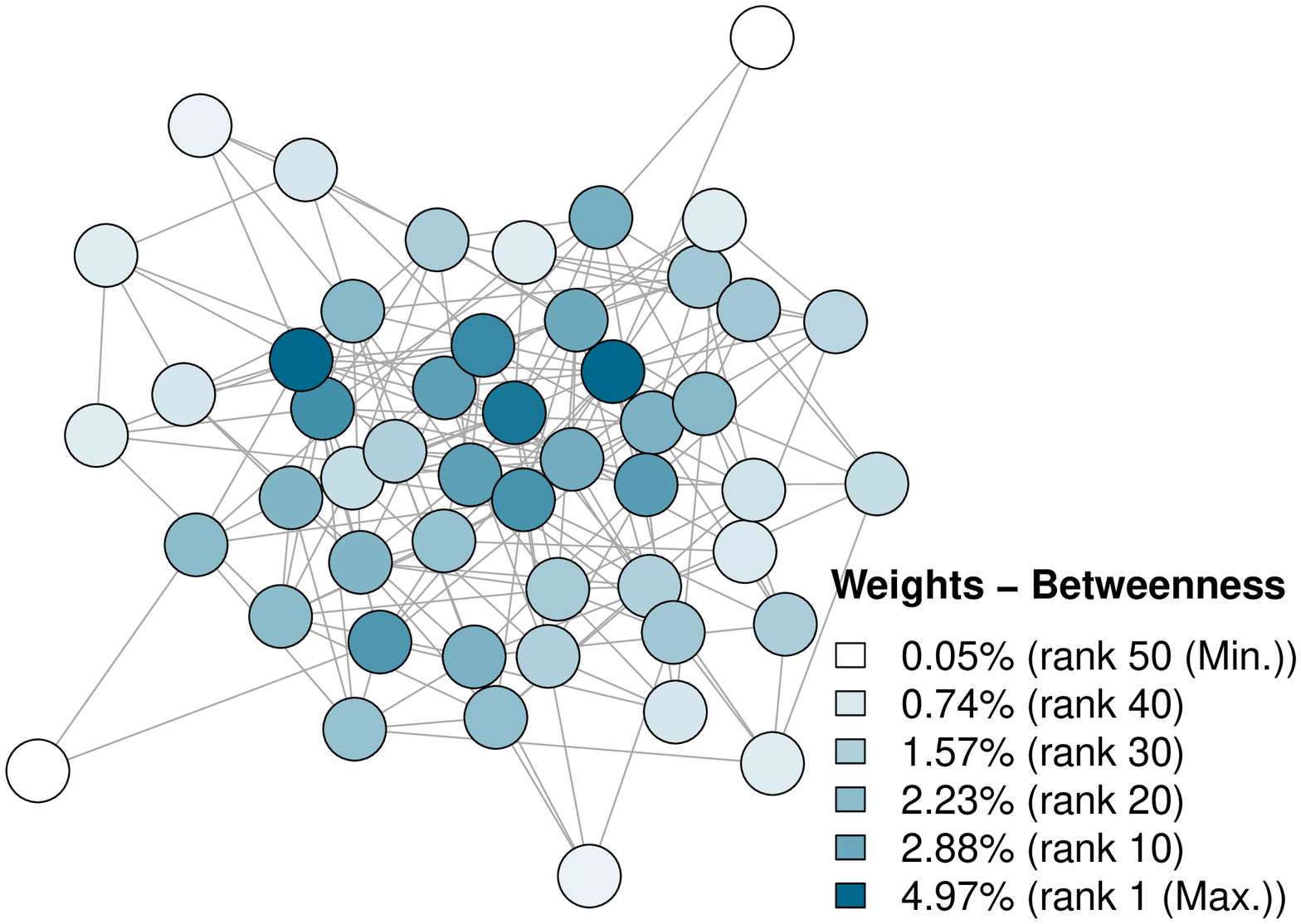}}
	\caption*{(a)}
\end{minipage}
\begin{minipage}[t]{0.32\linewidth}
\centering
	{\includegraphics[trim={7.5cm 3cm 0 0},clip,width=1\textwidth]{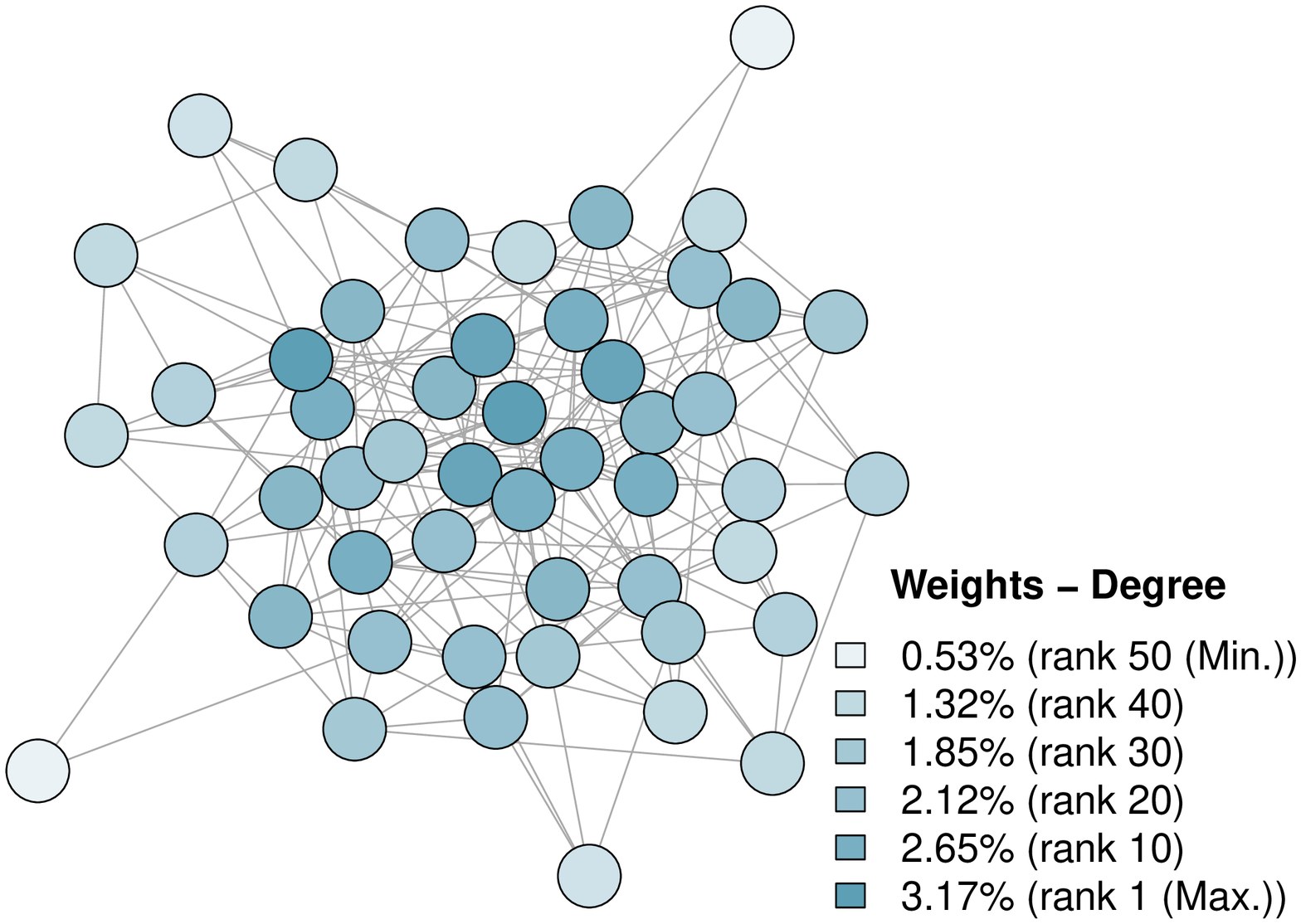}}
	\caption*{(b)}
	\end{minipage}
	\begin{minipage}[t]{0.32\linewidth}
\centering
	{\includegraphics[trim={7.5cm 3cm 0 0},clip,width=1\textwidth]{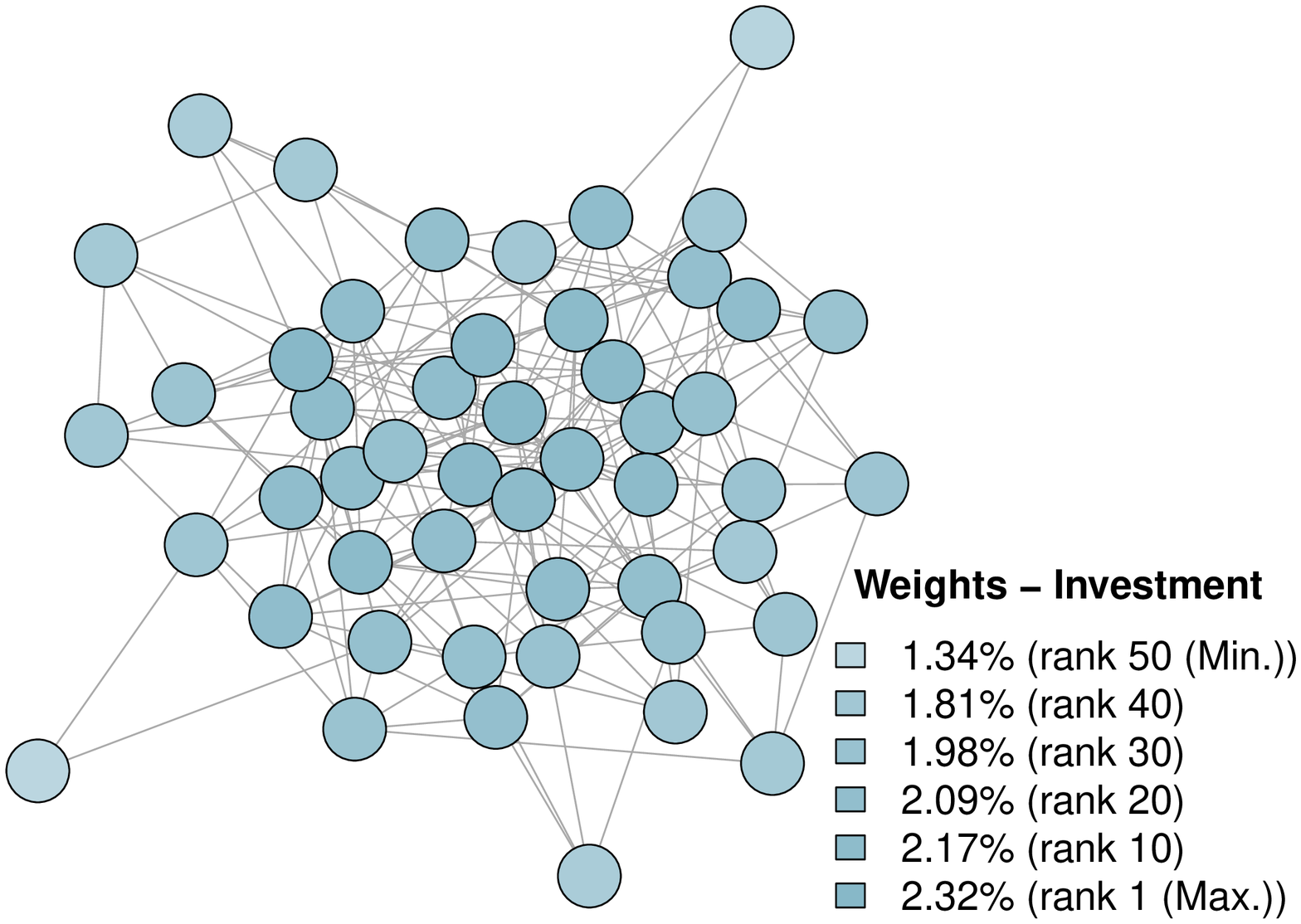}}
	\caption*{(c)}
	\end{minipage}
	
	\begin{minipage}[t]{0.32\linewidth}
\centering	
	{\includegraphics[trim={7.5cm 3cm 0 0},clip,width=1\textwidth]{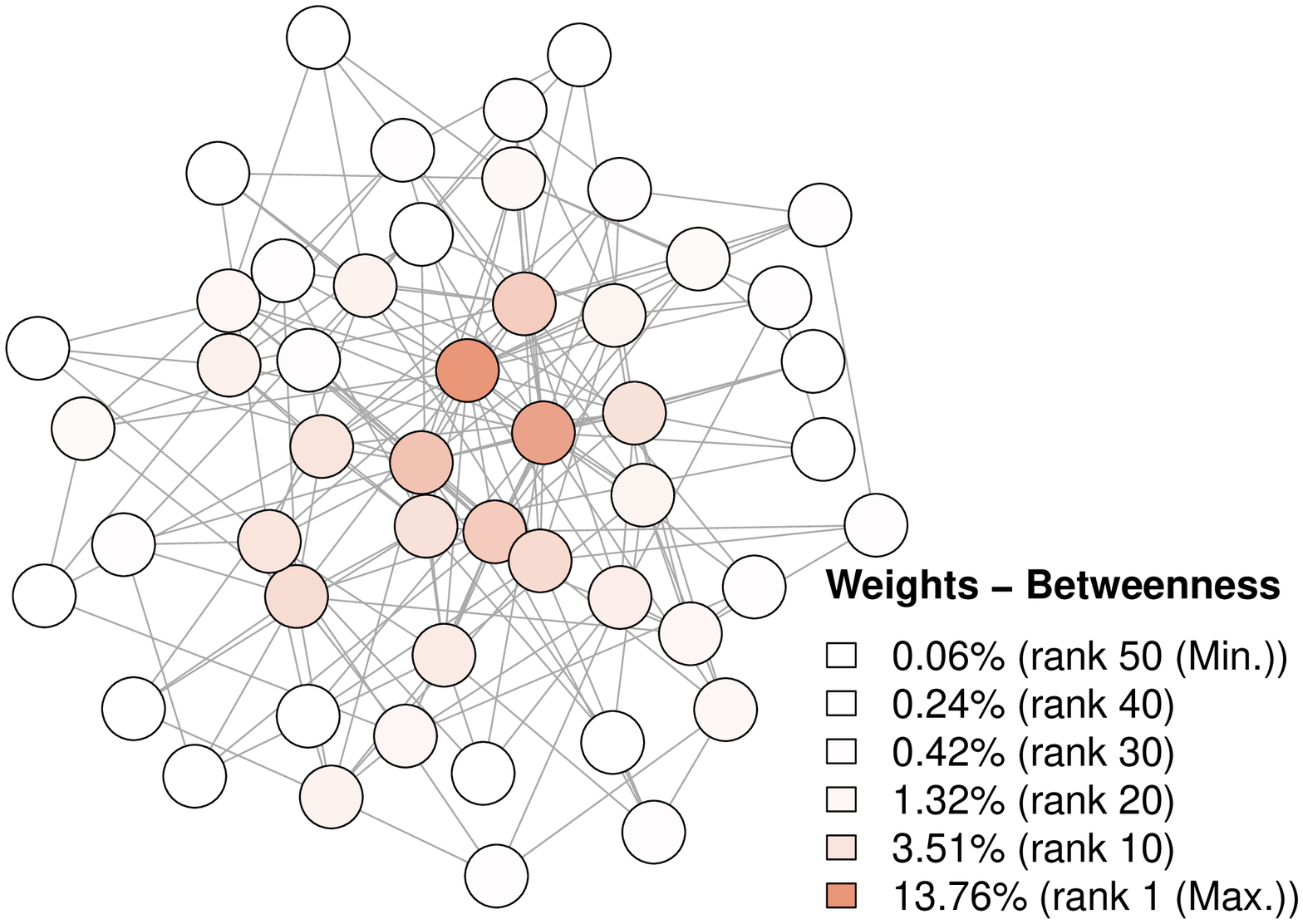}}
	\caption*{(a)}
\end{minipage}
\begin{minipage}[t]{0.32\linewidth}
\centering
	{\includegraphics[trim={7.5cm 3cm 0 0},clip,width=1\textwidth]{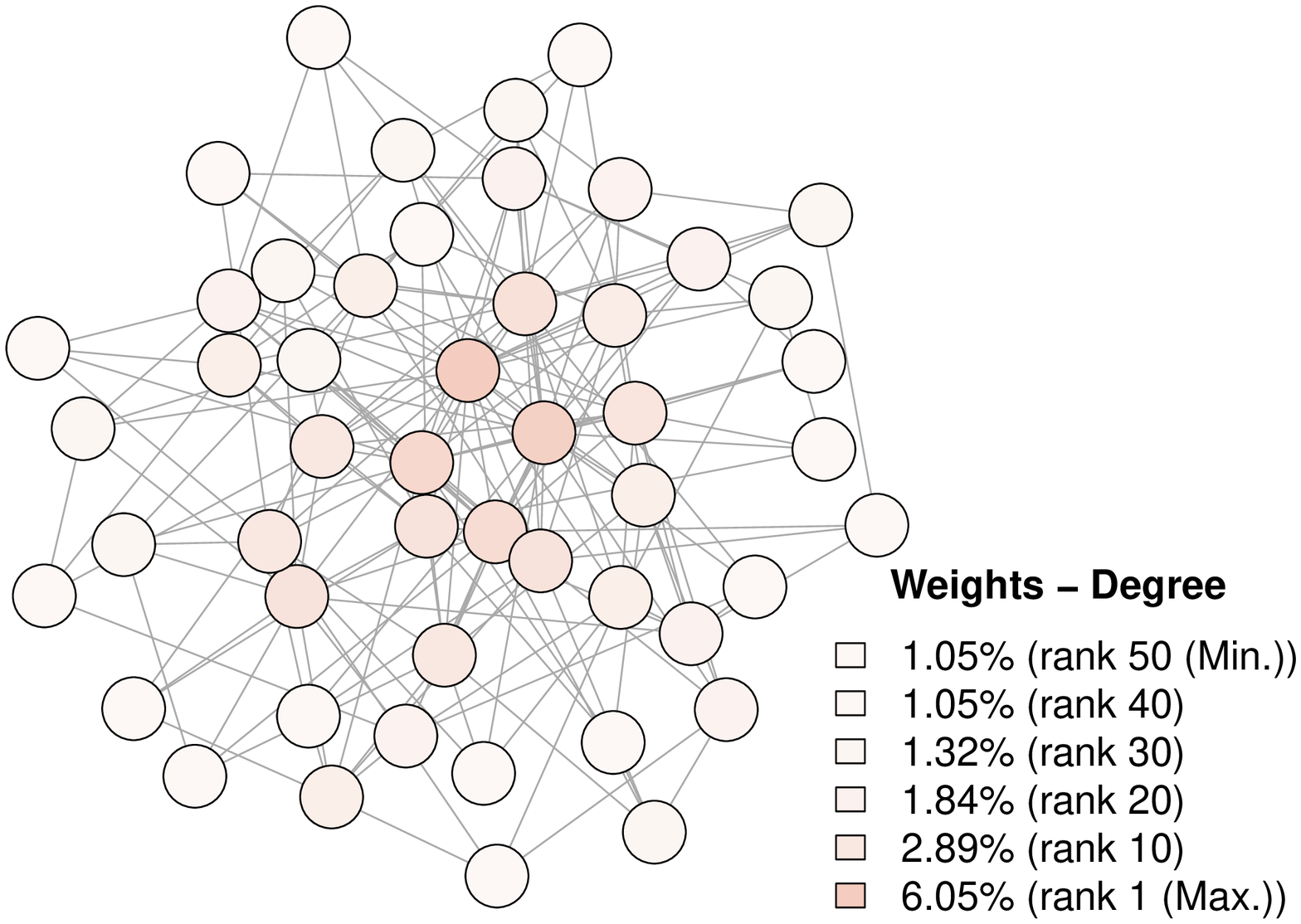}}
	\caption*{(b)}
	\end{minipage}
	\begin{minipage}[t]{0.32\linewidth}
\centering
	{\includegraphics[trim={7.5cm 3cm 0 0},clip,width=1\textwidth]{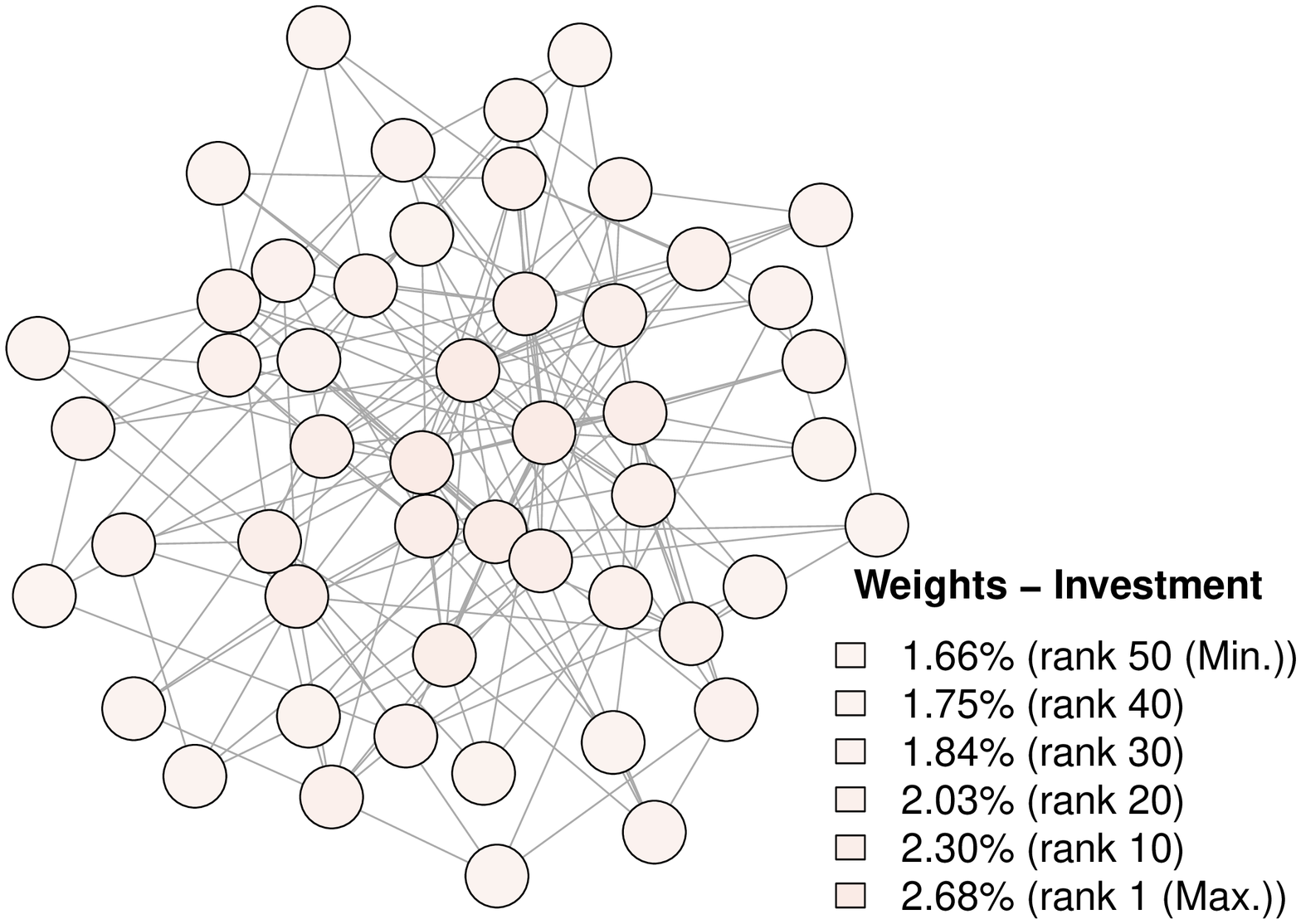}}
	\caption*{(c)}
	\end{minipage}
	
\caption{ Exemplary visualization of centrality weights $w_i$ in the Erd\H{o}s-R\'enyi (top) and Barab\'asi-Albert-type (bottom) network for (a) betweenness, (b) degree, and (c) investment centrality.}
	\label{fig:CSIweights}
\end{figure}

We compare the different allocation strategies and centrality measures for the  Erd\H{o}s-R\'enyi- and Barab\'asi-Albert-type networks from Figure \ref{fig:SInetwork} by allocating an additional budget of $\beta = 5$. The strategies are visualized in Figure \ref{fig:CSIweights} and the resulting  reductions of total network expenses are shown in Table \ref{tab:alloctable} on a percentage basis.

\begin{table}[h!]
\centering
     \begin{tabular}{|M{2cm} ||M{2.8cm}
     |M{2.8cm}|M{2.8cm}|}
     \hline
       & $c^\text{deg}$  
       & $c^\text{bet}$ & $c^\text{inv}$  \\
    \hline
    \hline
    
     upper  & {\color{NavyBlue!80!black} 10.6\%} {\color{salmon!80!black} 11.3\%}  
     & {\color{NavyBlue!80!black} 10.8\%} {\color{salmon!80!black} 12.3\%} & {\color{NavyBlue!80!black} 10.2\%} {\color{salmon!80!black} 9.6\%} \\
      \hline
    lower    & \:{\color{NavyBlue!80!black} ~8.2\%} \:{\color{salmon!80!black} ~6.7\%} 
    & \:\:{\color{NavyBlue!80!black} ~0.5\%} \:{\color{salmon!80!black} ~3.4\%}\: &  \:~{\color{NavyBlue!80!black} 9.5\%} {\color{salmon!80!black} 8.3\%} \\
    \hline
    untargeted  & \multicolumn{3}{|c|}{{\:\:\color{NavyBlue!80!black} ~9.9\%} \:{\color{salmon!80!black} ~9.0\%}\:} \\
    \hline
    \end{tabular}
    \caption{Percental reduction of accumulated total expenses $\mathcal{E}$ after the allocation of the additional budget $\beta = 5$ among all network nodes. The three proposed allocation strategies are evaluated for each of the suggested centrality measures. Entries for the Erd\H{o}s-R\'enyi network are colored in blue (left entries), and for the Barab\'asi-Albert network in salmon (right entries), respectively.  For each entry, cyber losses were generated from $\mathcal{T}=10,000,000$ simulations of the SIR epidemic process. Full data is given in Appendix \ref{app:upperalloc}.}
    \label{tab:alloctable}
\end{table}

In any case, we observe that the injection of additional network security \textit{clearly reduces} the accumulated total expenses.

Comparing the different allocation procedures, we see that the upper allocation strategy leads to lower overall losses than both the untargeted and lower allocation strategies -- regardless of the centrality measure chosen. 

Moreover, for both types of networks, we observe that the upper allocation strategy combined with topology-based centrality measures outperforms the investment-based approach. In particular, the upper allocation strategy based on betweenness centrality yields the best outcome.

A possible reason for this is that the proportion of budget which is allocated to periphery nodes is too large in both the untargeted and investment-based case: e.g., for the graph from the Erd\H{o}s-R\'enyi class, the investment-based centrality of periphery nodes is more than half the size of the maximum node centrality, see Figure \ref{fig:CSIweights} (c). In contrast, the betweenness centrality of the most isolated nodes is close to zero, 
and therefore, almost no additional security investment is allocated to these nodes, see Figure \ref{fig:CSIweights} (a).

\subsubsection{Further Centralization  of Upper Allocations}
Our previous observations suggest that additional security investments should not be distributed equally among nodes, but in accordance with their centrality following an upper allocation strategy. Introducing specific requirements for {\em every} network entity may be difficult or even impossible from a regulatory point of view. For example, the upcoming NIS2 Directive introduces a specific size-cap rule which solely targets medium-sized and large entities in sectors of critical infrastructure, see the discussion in Section \ref{sec:Add}. But does the exclusion of low-centrality nodes from the allocation procedure substantially reduce the beneficial effect of additional network security?   Or can we even improve the effectiveness of security obligations if only a certain fraction of highest-centrality nodes is considered? 

\begin{figure}[h]
    \centering
    \begin{minipage}[t]{0.495\linewidth}
    \centering
	{\includegraphics[width=1\textwidth]{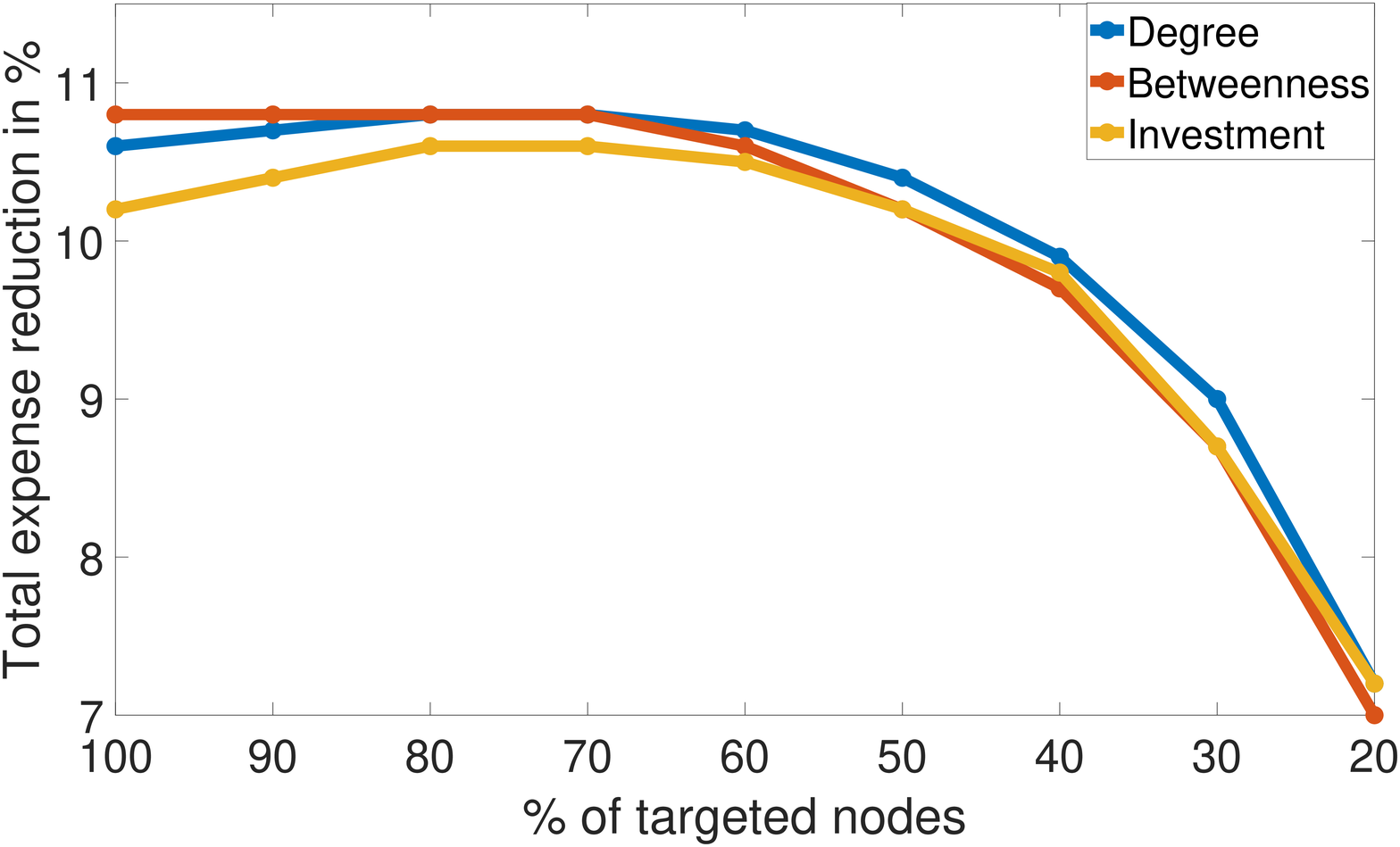}}
	\caption*{Erd\H{o}s-R\'enyi}
	\end{minipage}
	\begin{minipage}[t]{0.495\linewidth}
	\centering
	{\includegraphics[width=1\textwidth]{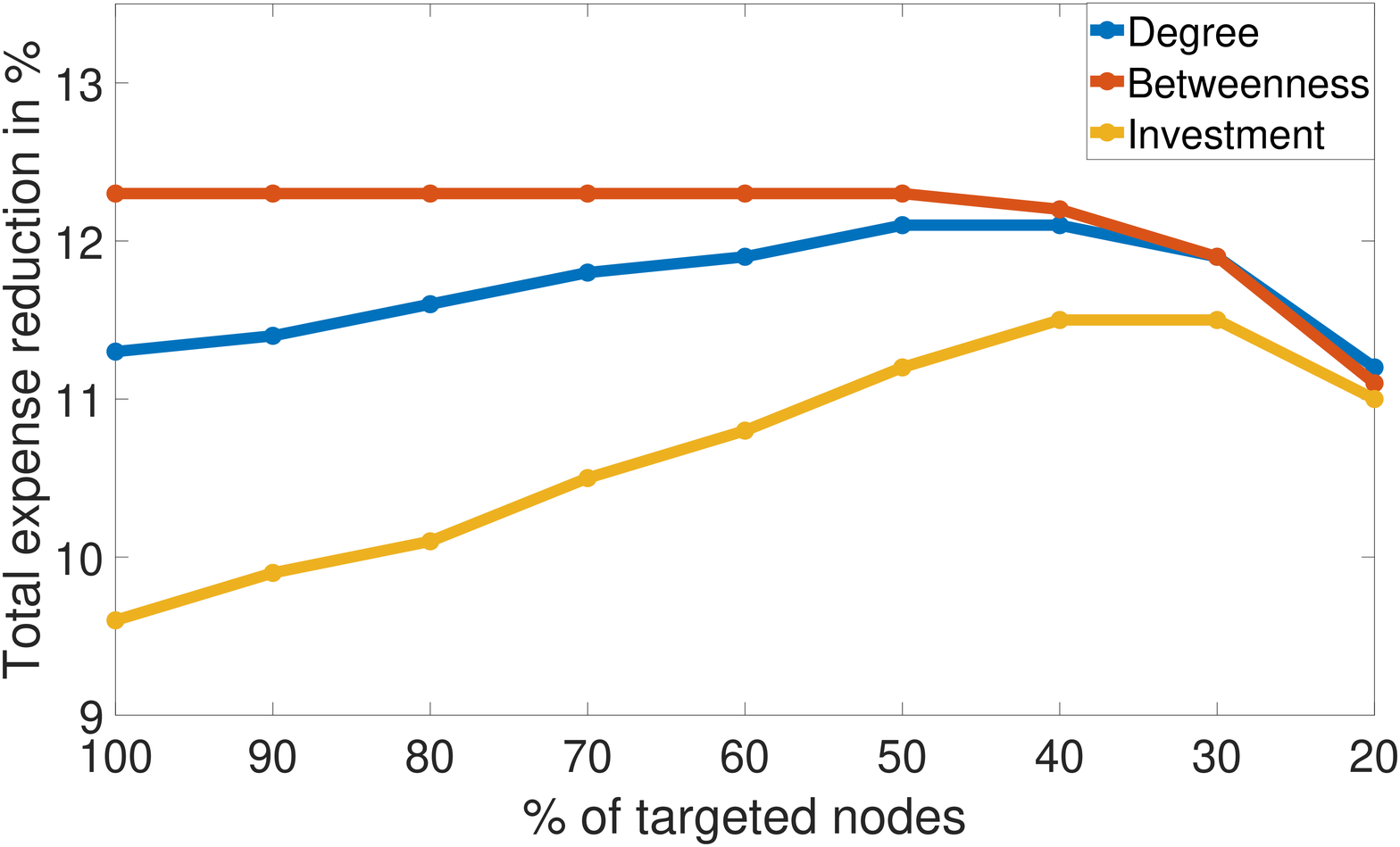}}
	\caption*{Barab\'asi-Albert}
	\end{minipage}
    \caption{Refinement of the upper allocation strategy for different percentages of targeted nodes. Again, the total additional security budget is $\beta = 5$. For each data point, $\mathcal{T}=10,000,000$ simulations of the epidemic process were generated. Full data is given in Appendix \ref{app:upperalloc}.}
    \label{fig:effalloc}
\end{figure}

To answer these questions, the upper allocation procedure is slightly modified: Suppose we want to restrict the budget allocation to a certain fraction $p$ of nodes with the highest centrality, and let $\mathcal{I}$ denote the set of corresponding node indices. Then, the amount of budget which is allocated to node $i$ is chosen as 
  \begin{equation*}
        \gamma_i^\text{all} = \begin{cases}  \beta\cdot \Large(\mathcal{C}(i)/\sum_{j\in\mathcal{I}} \mathcal{C}(j)\Large),  &\text{if } i\in\mathcal{I}, \\
        0, &\text{else}.
         \end{cases}
    \end{equation*}
Note that for $p=100\%$, this coincides with the previously studied upper allocation strategy on the full network. Results of the modified procedure for different percentages of targeted nodes are depicted in Figure \ref{fig:effalloc}.

For the Erd\H{o}s-R\'enyi network no  substantial change of expenses is found when excluding the most decentralized nodes from the allocation of the additional security budget. In contrast, for the Barab\'asi-Albert network and allocations based on investment and degree centrality, only targeting nodes with a medium to high degree of centrality is even beneficial. No substantial change, however, is observed in case of the betweenness-centrality based allocation. As noted before, this may be due to the fact that in the betweenness-centrality case, the allocation weights of periphery nodes are anyway close to zero. In sum, our observations provide evidence that budget allocations to periphery nodes are rather ineffective. 

Nevertheless, for all centrality measures and both types of networks under consideration, solely allocating the budget to a small fraction of nodes with the highest centrality does not prove to be optimal.  A reason for that might be the trade-off between costs and efficiency: Additional security investments for highly central nodes come with substantially increasing costs, since these nodes already invest a high amount in the individually optimal steady state (see Figure \ref{fig:SInetwork}) and the cost function $C_i(\gamma_i)$ is strictly convex.

For both types of networks, the overall best results are found for betweenness-based allocations. However, the corresponding optimal total expenses are only slightly below the total expenses corresponding to adequately targeted degree-based allocations. Determining the betweenness centrality of nodes requires information on the full network topology, and this information may not be available in practice. In contrast, node degrees, i.e., the number of IT contacts of an agent in the cyber network, are local quantities, and thus, they can more easily be determined, e.g., using questionnaires. Therefore, in view of the information gathering issue and given the comparable performance in our simulations, degree-based allocations targeting the upper 50\% of most central nodes may constitute a reasonable compromise.

\subsection{Evaluation of Security-Related Interventions}\label{subsec:CSIevaluation}
We find that mandatory security investments as a regulatory obligation can actually increase the overall cybersecurity in a system of interconnected agents. More precisely, our simulations suggest the following:
\begin{enumerate}[(i)]
    \item The strategic interaction of nodes in the cyber network leads to a \textit{steady state} of security investments as proven in Theorem \ref{theorem}. However, the self-regulation of interdependent actors does in general \textit{not} lead to an effective state of security configurations from an overall network perspective: A substantial improvement of this state is possible by the injection of additional security budget. Therefore, a  \textit{need for regulation} is found, and introducing adequate security-related obligations might be reasonable.
    \item Severe security requirements for weakly connected entities like private households or companies with a very small number of business partners do not seem to have any notable effect on reducing network vulnerability. However, solely focusing on the most central nodes does not produce the best results either. Provided that these central nodes at least make the significant investments given by some steady state of the security investment game, additional investments come with massively increasing costs.  Therefore, regulation should also focus on agents and companies with a medium to large number of IT or business contacts.     
    \item Centrality is not a rigorously defined concept. However, for both degree- and betweeness-based security allocations, good results are obtained. For practical reasons, degree-based allocations may be easier to implement: information on immediate network contacts can directly be obtained from agents, e.g., using questionnaires.
\end{enumerate}

As regards the selected cybersecurity measures given in Section~\ref{sec:legal} we find that: 
\begin{framed}
\begin{itemize}
\item[GOV] 
\begin{itemize} 
\item[$\diamond$] \textit{Size-cap rule:}  Remarkably, the approach proposed by the European Commission is in very good agreement with our findings: Security obligations for micro and small enterprises are ineffective, but both medium-sized entities as well as large businesses should be targeted.  Therefore, our results strongly suggest that the size-cap rule is an efficient tool for improving resilience in cyber systems.
\item[$\diamond$] \textit{Supply chain protection:} The observation that efficient security allocations cannot solely be restricted to a small fraction of nodes with the highest centrality illustrates the need for a strengthening of security levels further down along possible paths of contagious transmission. Therefore, similar to the size-cap rule, our study supports the implementation of security enhancing measures along supply chains.  
\end{itemize}
\item[INS]
\begin{itemize}
\item[$\diamond$] \textit{Assistance services:} Our study may help to identify companies for which assistance should be made mandatory in insurance contracts, and also give an estimate of the amount of services that should be made available to the specific policyholder. Further, in the case of an ongoing WannaCry- or NotPetya-type incident, the amount of resources for those assistance services may only be limited, see also the discussion in \cite{Hillairet2021}. Thus, our results may also be useful for an effective resource allocation in such situations.
\item[$\diamond$] \textit{Patch management and backup:} Our observations suggest that the effectiveness of mandatory obligations 
strongly depends on the systemic importance of the examined entity measured by a reasonable centrality criterion. Medium-sized as well as large businesses with respect to centrality should not only invest more in cybersecurity, and thus in particular in their back-up and patching procedures, than smaller entities, but they should even invest more than an individually optimal assessment would suggest. 

\end{itemize}
\end{itemize}
\end{framed}

\section{Case Study II: Topology-Based Interventions and Cyber Pandemic Risk} \label{subsec:CSII}

Due to the interconnectedness of modern IT systems, both the WannaCry and NotPetya incidents affected systems at a global scale, triggering large amounts of cyber losses. Clearly, a major regulatory concern is the prevention of such cyber pandemic incidents. Moreover, since risk pooling does not apply to systemic incidents, it is also important for insurance companies to reduce the risk of potential cyber accumulation scenarios within their portfolios. 

Digital information and technology networks often come at a size of several thousand nodes\footnote{{ The possibly largest existing network is the WWW with approximately $N=10^{12}$ nodes.}}, see the reference network data from Table 2.1 in \cite{Barabasi2016} and Table 10.1 in \cite{Newman2010}.
In this section we study the cyber pandemic risk exposure, first for homogeneous Erd\H{o}s-R\'enyi-type networks, and then for heterogeneous---probably more realistic---Barab\'asi-Albert-type networks { of large size}. We will observe that in order to control the cyber pandemic risk, regulatory approaches which solely focus on the improvement of individual cyber security are insufficient: interventions need to target the underlying \textit{topological} network structure. Thus a clear demand for the regulation of the network topology in large-scale cyber systems is found. 
We will also observe that network heterogeneity massively amplifies the cyber pandemic risk. 

\subsection{Demand for Regulation: Network Topology and Cyber Pandemic Risk}\label{subsec:CSIIconnectivity}

In large-scale networks, the frequency distribution of epidemic outbreak sizes in the SIR model can typically be characterized by the presence of two peaks,\footnote{Mathematical details are extensively discussed in Chapter 6 of \cite{Kiss2017}.} namely
\begin{itemize}
    \item  \textit{small outbreaks}, affecting only a very small fraction of network nodes, and
    \item  proper \textit{epidemic outbreaks} or \textit{pandemics}, where a large number of nodes becomes infected.
\end{itemize}
To assess the risk of cyber pandemics, simulation studies are conducted. Again, for all networks, we choose a global infection rate of $\tau = 0.1$. In contrast to the previous study, recovery rates are assumed to be fixed and homogeneous for all nodes, i.e., $\gamma_i = \gamma = 1$ for all $i=1,\ldots , N$. This parameter choice implies that detection of cyber incidents is expected to be 10 times faster than infectious transmission, i.e., we \textit{assume an overall high standard of IT security for the full network}.
\subsubsection{Cyber Pandemic Risk in Homogeneous Networks}
We firstly analyze the cyber epidemic risk exposure of homogeneous large networks drawn from the Erd\H{o}s-R\'enyi random graph model with a fixed size of $N=1,000$. The benefit of this model class is that the resulting networks are easily tractable due to the fact that their topology is entirely determined by the parameters $p$ and $N$. In particular, $p$ can be interpreted as the control parameter of network connectivity. 
Each simulation is performed in the following way:
\textit{\begin{enumerate}
    \item Randomly draw a network $G_p(1,000)$ from the Erd\H{o}s-R\'enyi class.
    \item Randomly choose a single node which is initially infected.
    \item Randomly generate an infection trajectory from our SIR model. We are interested in the total number of infected nodes, i.e., the \emph{outbreak size}.
\end{enumerate}}

	\begin{figure}[h]
		\centering
		{\includegraphics[width=0.7\textwidth]{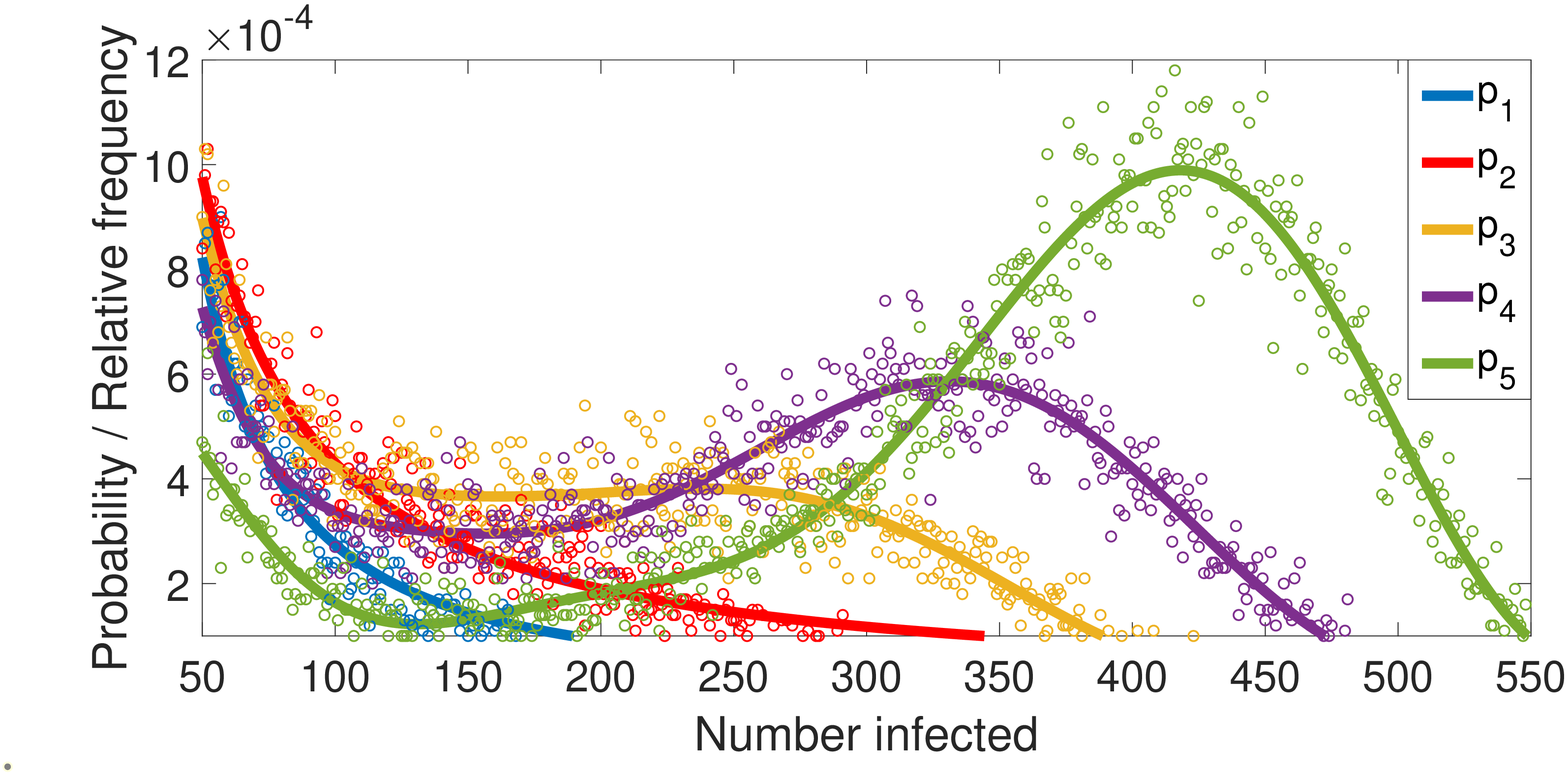}}
		\caption{Final outbreak size frequencies given an initial infection of a single network node, over 100,000 simulations for increasing values of $p$; values are $p_1 = 0.01< p_2 = 0.011 < {p_c} < p_3 = 0.012 < p_4 = 0.013 < p_5 = 0.014$. Exact data points from the simulation and appropriate regression curves (power law for $p_1$ and $p_2$, polynomial of degree 8 for $p_3, p_4, p_5$) are plotted. }
		\label{fig:ERoutbreaks}
	\end{figure}

The resulting frequency distribution of outbreak sizes is depicted in Figure \ref{fig:ERoutbreaks}. The following \textit{phase transition} can be observed:
\begin{itemize}
    \item For low connectivity probabilities $p$, only small outbreaks occur; the outbreak size frequency is exponentially decaying.
    \item {\itshape Tipping point behavior:} If a certain  
    critical edge probability $p_c$ is exceeded, the frequency distribution is characterized by a second peak around a characteristic large outbreak size.
    \end{itemize}

The apparent strong dependence between network connectivity and outbreak sizes suggest that a supervision and regulation of the network is beneficial to avoid large systemic outbreaks.  Naively speaking, in a homogeneous network, the regulator should aim at keeping the network connectivity \textit{below the critical threshold} $p_c$.

\subsubsection{The Heterogenous Case: Cyber Pandemic Risk in Scale-Free Networks}

On a larger scale, many real-world networks are characterized by a preferential attachment principle, see Chapter 4 in \cite{Barabasi2016}, and therefore, a more heterogeneous topology is often observed: Let $K$ be a random variable which represents the degree  $k_i$ of a randomly chosen network node  $i$. Then
\begin{itemize}
    \item the degree distribution of the Erd\H{o}s-R\'enyi random graph $G_p(N)$ is given by a binomial form, i.e., we have 
    \begin{equation*}
        \mathbb{P}(K=k) = \binom{N-1}{k} p^k (1-p)^{N-1-k},\qquad k= 0,\cdots, N-1,
    \end{equation*}
    \item whereas under preferential attachment, the distribution of node degrees typically follows a power-law, i.e., 
    \begin{equation*}
         \mathbb{P}(K=k) \sim k^{-\alpha},\qquad \text{with degree exponent }  \alpha\in\mathbb{R}_+.\footnote{For details and empirical examples, we refer to Chapters 3 and 4 in \cite{Barabasi2016}.}
    \end{equation*}
  Node arrangements with $\alpha = 3$ can be modeled using the Barab\'asi-Albert class introduced in Section \ref{sec:randnet}. These so-called \textit{scale-free} networks provide a hierarchy of nodes, with heavily connected high-degree hubs in their center and less connected nodes in their periphery.
\end{itemize}

\begin{figure}[h]
		\centering
		{\includegraphics[width=0.3\textwidth]{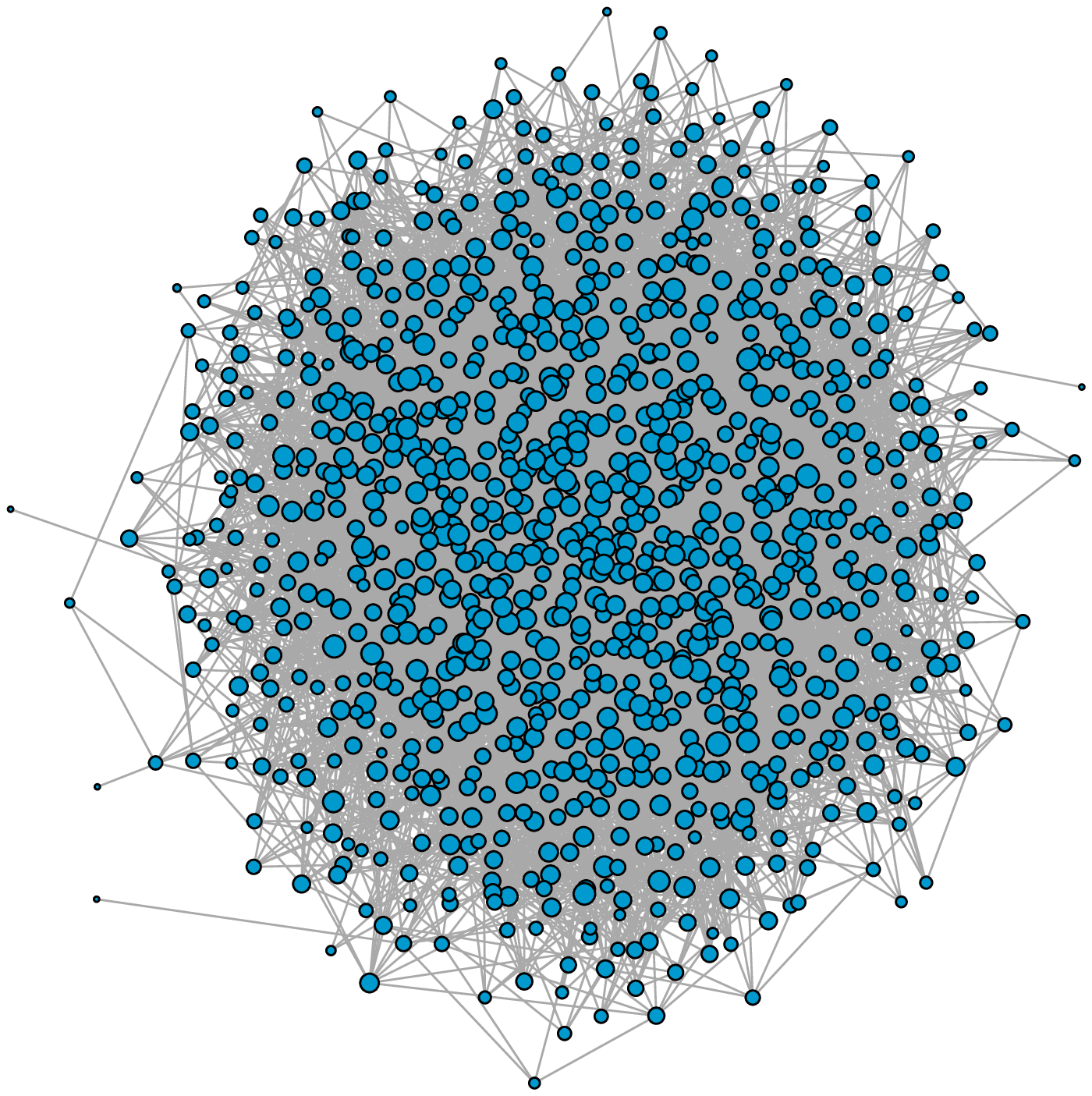}} 	{\includegraphics[width=0.3\textwidth]{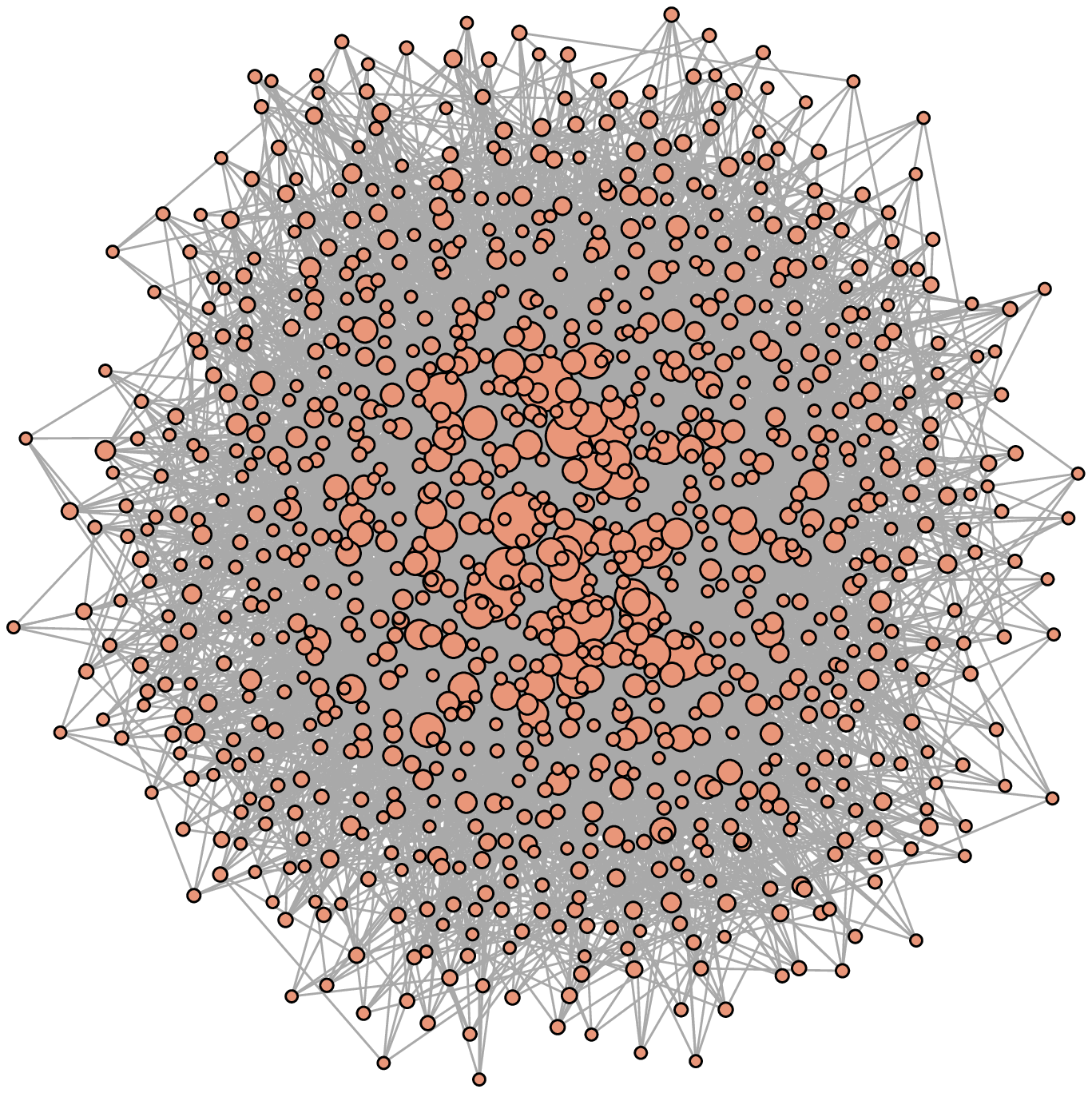}}\\
		\caption{Erd\H{o}s-R\'enyi $G_{0.01}(1,000)$ (left) and Barab\'asi-Albert $BA(1,000; 5)$ (right). In both cases the node size of node $i$ is given by $100\cdot\sqrt{k_i/\sum\nolimits_{j=1}^{1000} k_j}$, an increasing function of the node's relative degree $k_i/\sum\nolimits_{j=1}^{1000} k_j$}.
		\label{fig:Deg}
	\end{figure}

Figure \ref{fig:Deg} shows representative networks from both the Erd\H{o}s-R\'enyi and Barab\'asi-Albert class, highlighting the different degree distributions.

\begin{figure}[h]
		\centering
		{\includegraphics[width=0.7\textwidth]{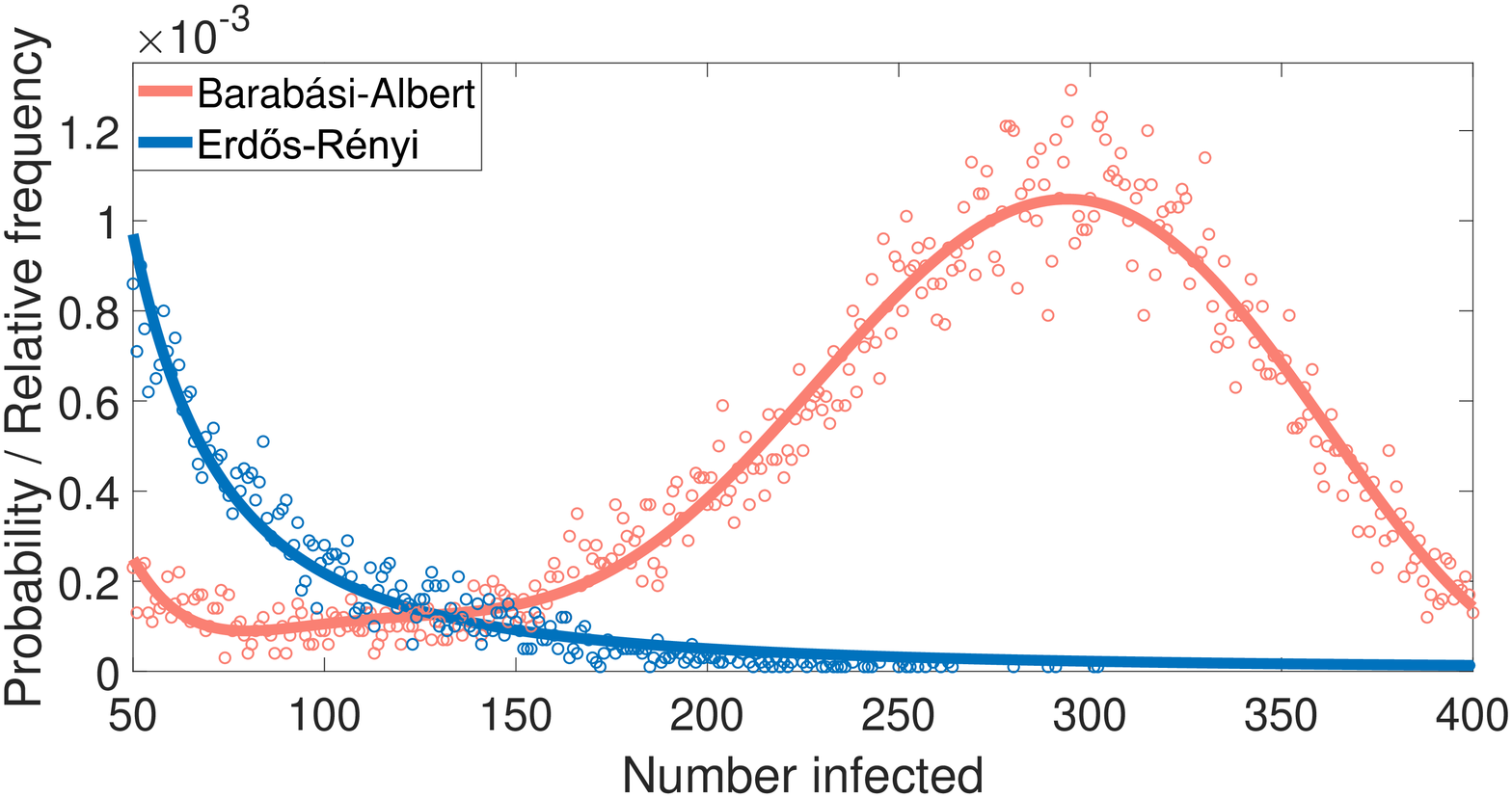}}
		\caption{Final outbreak size frequencies given an infection of a single network node for the Barab\'asi-Albert $BA(1,000; 5)$ and Erd\H{o}s-R\'enyi networks $G_{0.01}(1,000)$  from Figure \ref{fig:Deg} over 100,000 simulations. Exact data points from the simulation and a regression curve (power law for Erd\H{o}s-R\'enyi, polynomial of degree 8 for Barab\'asi-Albert) are plotted.}
		\label{fig:BA5vsER001}
	\end{figure}

This difference in the network topology has a strong impact on the epidemic vulnerability. Focusing on connectivity in terms of the sole number of edges, for networks of size $N=1,000$, the class of Barab\'asi-Albert networks with $m=5$ is comparable to Erd\H{o}s-R\'enyi graphs with $p=0.01$, since the resulting numbers of edges in both networks approximately coincide.\footnote{Approximately 5,000 edges should be present in both networks, see the discussion in Section \ref{sec:randnet}.} However, there exists a strong difference regarding their vulnerability to epidemic outbreaks, as shown by Figure \ref{fig:BA5vsER001}: In contrast to the Erd\H{o}s-R\'enyi graph,  a clear second peak in the frequency distribution of outbreak sizes is observed for the Barab\'asi-Albert network. Hence, the heterogeneity in the topology of Barab\'asi-Albert networks remarkably lowers the critical connectivity threshold for cyber pandemics, i.e., it {amplifies} the epidemic spread and {triggers} the emergence of large-scale outbreaks.

A profound characterization of this behavior { in relation to the distribution of node degrees} can be obtained in the limit of infinite network size $N\to\infty$: { Neglecting additional correlation effects}\footnote{{ The effect of degree correlations and clustering on the dynamics of spreading phenomena is difficult to quantify analytically due to the dimensionality of the system, see also the discussion in Appendix \ref{app:SIR Dynamics}. Findings on their impact on the epidemic threshold are surveyed in Sections B.1 and B.2 of \cite{PastorSatorras2015}.}} it is known that large-scale pandemic outbreaks are possible if and only if the threshold condition
\begin{equation}\label{eq:infthresh}
\frac{\tau}{\tau +\gamma}\frac{\mathbb{E}[K^2-K]}{\mathbb{E}[K]}>1
\end{equation}
is satisfied, see Equation 6.4 on p.221 in \cite{Kiss2017}\footnote{See also Equation 62 in \cite{PastorSatorras2015} for an equivalent expression of the threshold.}. Note:
\begin{itemize}
    \item For Erd\H{o}s-R\'enyi random graphs, in the limit     the degree distribution is Poisson with parameter $\lambda$ denoting the average degree, see Section 3.4 in \cite{Barabasi2016}. 
    Therefore, from \eqref{eq:infthresh}, it follows that cyber pandemics can be prevented in the infinite limit if the network security/recovery rate $\gamma$,  satisfies $\gamma \geq \tau (\lambda -1)$.
    \item In contrast, for scale-free networks with $\alpha\in (2,3]$ and a sufficiently high number of nodes, it may be difficult or even impossible to prevent cyber pandemics by solely improving the network security 
    or reducing the overall network connectivity. The reason for this is that in the infinite size limit, 
the second moment $\mathbb{E}[K^2]$ of the degree distribution diverges to $\infty$ while the first moment $\mathbb{E}[K]$ stays finite, see Section 10.4.2 in \cite{Newman2010} for more details. 
Hence, in view of \eqref{eq:infthresh}, with growing $N$, the security parameter $\gamma$ must be substantially increased to prevent the occurrence of cyber pandemics. This comes with massively increasing costs. In the limiting case $N\to\infty$, \eqref{eq:infthresh} is always satisfied, regardless of the infection and recovery parameters chosen, so cyber pandemics may always occur.
\end{itemize}

In scale-free networks with a degree exponent $\alpha$ in the range of $(2, 3]$ and a large number of entities, cyber pandemics are thus an \textit{inherent risk} of the underlying network topology. The risk of cyber pandemic outbreaks cannot be controlled by security-related interventions, i.e., by increasing the recovery rate $\gamma$, only, but requires a manipulation of the degree distribution, that is the topological network arrangement. This behavior is clearly relevant in the risk assessment of cyberspace, which consists of a very large number of entities and is characterized by a heterogeneous, possibly scale-free, structure of interconnections.\footnote{For example, the Internet's degree distribution is estimated to be scale-free with degree exponent $\alpha\approx 2.5$ in Table 10.1 of \cite{Newman2010}.}

\subsection{Implementing Suitable Interventions}\label{sec:ImpTOP}

In the previous subsection, we have seen how a network's vulnerability to large-scale cyber pandemic outbreaks depends on the topology of the underlying cyber network. 
The following approaches may be considered to limit or control critical network connections and nodes:
\begin{itemize}
    \item \textit{Edge removal:} Edge deletion comprises 
    \begin{itemize}
    \item  \textit{physical deletion of connections}, such as any unnecessary access to servers, or if not possible, \item  \textit{edge hardening}, which corresponds to strong protection of network connections via firewalls, the closing of open ports, or the monitoring of data flows using specific detection systems, see \cite{Chernikova2022}.
\end{itemize}

    \item \textit{Node splitting} to separate critical contagion channels and let them pass through two different nodes with the same operational task.
\end{itemize}
Since manipulating the network topology comes at a cost, probably reducing \textit{network functionality}, the aim in the following is to identify critical network connections and nodes in a way which reduces negative effects on the network functionality to a minimum. A classical measure for network functionality is the \textit{average shortest path length} $\langle l \rangle$: For nodes $i$ and $j$, $l_{ij}$ is the minimum number of edges connecting $i$ and $j$. The average shortest path length is the average over all these distances, i.e.,
		\begin{equation*}
			\langle l \rangle = \sum_{i,j, i\neq j} \frac{1}{N(N-1)} l_{ij}
		\end{equation*}
in case of a connected network. A small value of $\langle l\rangle$ is a measure for fast and efficient data flow, and hence, corresponds to a high network functionality.
  If a network consists of more than one component, then $l_{ij}$ is not well defined for any two nodes $i$ and $j$ which come from two different components. In this case, we follow \cite{Newman2010}, p. 311 and adopt the definition by only taking the average over those node pairs which are connected by an existing path.\footnote{ In particular, this modification is relevant for the random edge deletions in Figure \ref{fig:EdgeDel}, where larger amounts of links are removed. The networks in Figures \ref{fig:EdgeDel}, \ref{fig:NodeSplit}, and \ref{fig:NodeSplit_Bad} which are generated by targeted edge deletions and node splittings are not fragmented into disconnected components.}

\subsubsection{Edge Removal and Node Splitting}
\paragraph{Edge Removal}
To identify epidemically critical edges, we utilize the edge centrality given in  \eqref{eq:edgcen} in Section~\ref{sec:centrmea} and propose the following procedure: 
\textit{\begin{enumerate}
\item Consider a network $G$. Determine the centrality of G's edges. 
\item Consecutively delete the most central network edges. Stop the deletion process, if the resulting network does not exhibit a cyber pandemic outbreak any more.
\end{enumerate}}

The procedure thus ends when 
pandemic outbreaks are not any longer observed in the resulting network $G_c$. Let $\mathcal{E}_c$ denote the set of edges which are deleted from $G$ to obtain $G_c$, and let $\vert \mathcal{E}_c\vert$ be its number.

\begin{figure}[h]
		\centering
\begin{minipage}[t]{0.495\linewidth}
\centering	
	{\includegraphics[width=1\textwidth]{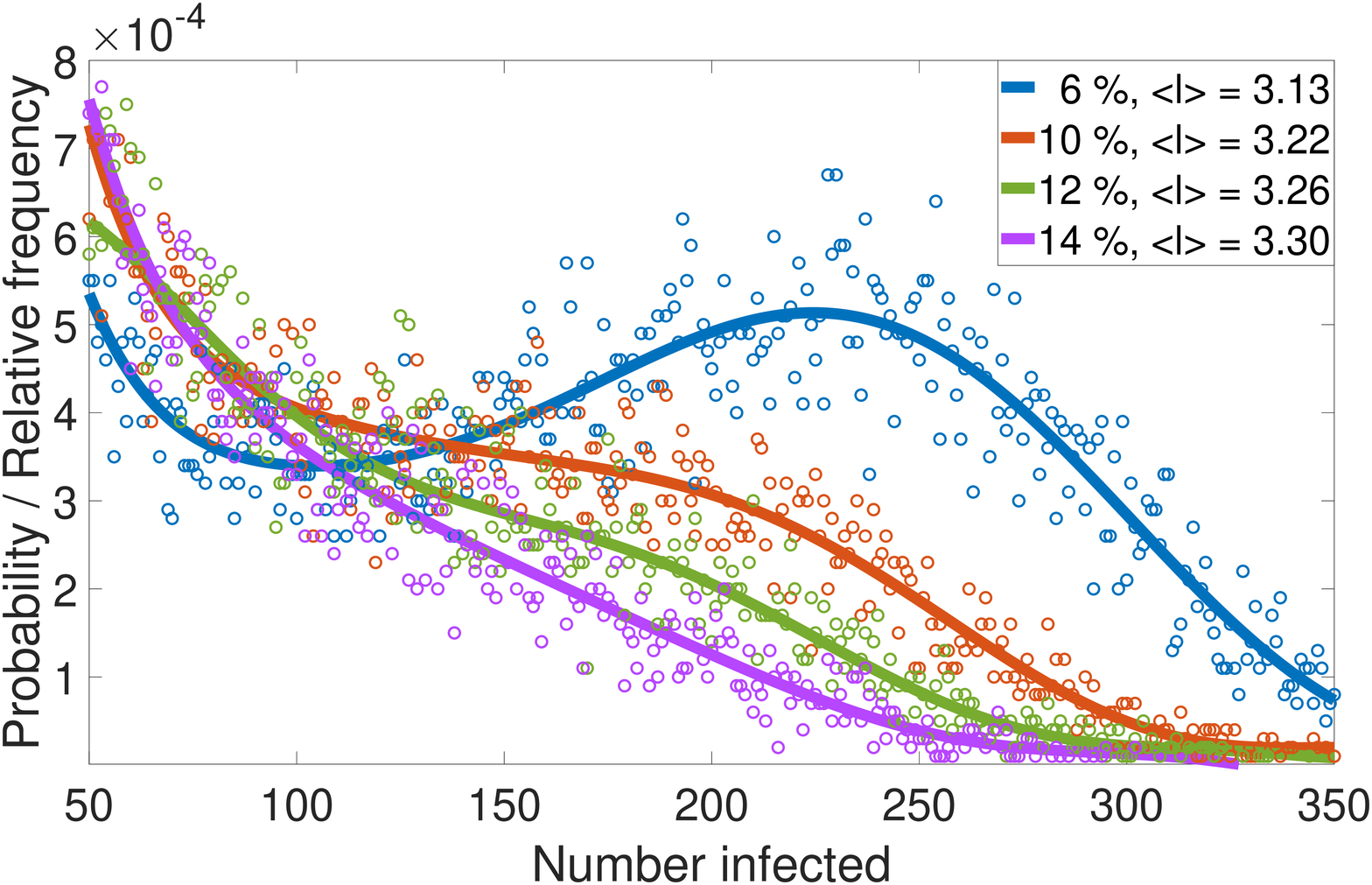}}
	\caption*{edge centrality}
\end{minipage}
\begin{minipage}[t]{0.495\linewidth}
\centering
	{\includegraphics[width=1\textwidth]{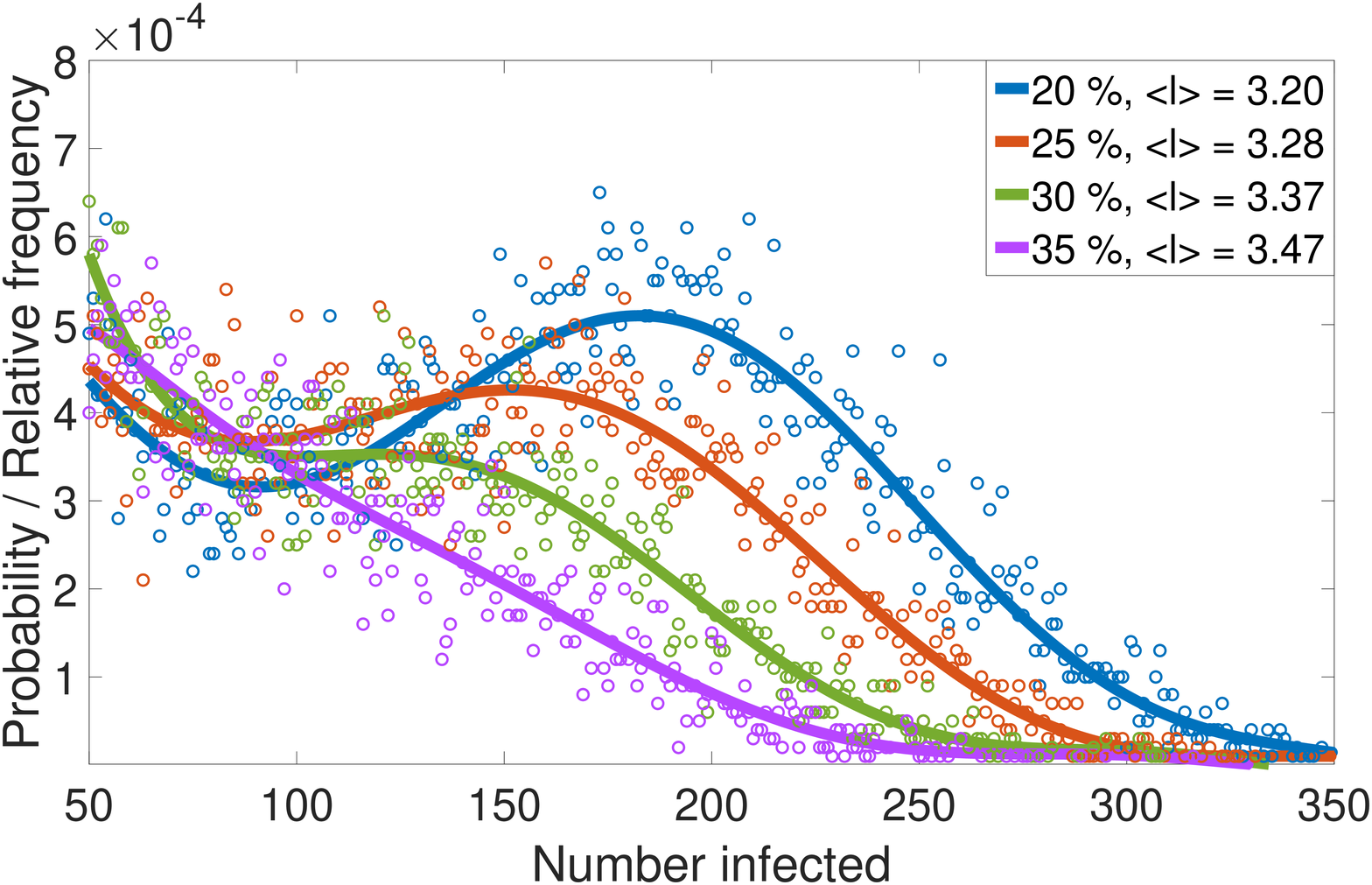}}
	\caption*{random}
\end{minipage}
		\caption{Final outbreak size frequencies given an initial infection of a single network node, over 100,000 simulations for different percentages of deleted edges.  Exact data points from the simulations and regression curves (polynomial of degree 8) are plotted. The results for edge centrality-based removals are depicted in the left figure, and the percentage of critical links is found to be about 14 \%. In contrast, random edge removals are shown in the right figure, and this procedure is clearly less effective: Approximately 30-35\% of edges need to be removed here to eliminate the risk of cyber pandemics. The randomized edge removals are newly conducted for each of the 100,000 simulations.}
		\label{fig:EdgeDel}
	\end{figure}

To illustrate the effectiveness of the proposed procedure, we determine the value $\vert \mathcal{E}_c\vert$ and the average shortest path length $\langle l_c\rangle$ of the resulting network $G_c$ for the Barab\'asi-Albert network depicted in Figure \ref{fig:Deg} with initial functionality of $\langle l\rangle \approx 2.96$ and outbreak size frequencies as shown in Figure \ref{fig:BA5vsER001}. The results after edge deletion are depicted in Figure \ref{fig:EdgeDel}.

\begin{figure}[h]
		\centering
\begin{minipage}[t]{0.495\linewidth}
\centering	
	{\includegraphics[width=1\textwidth]{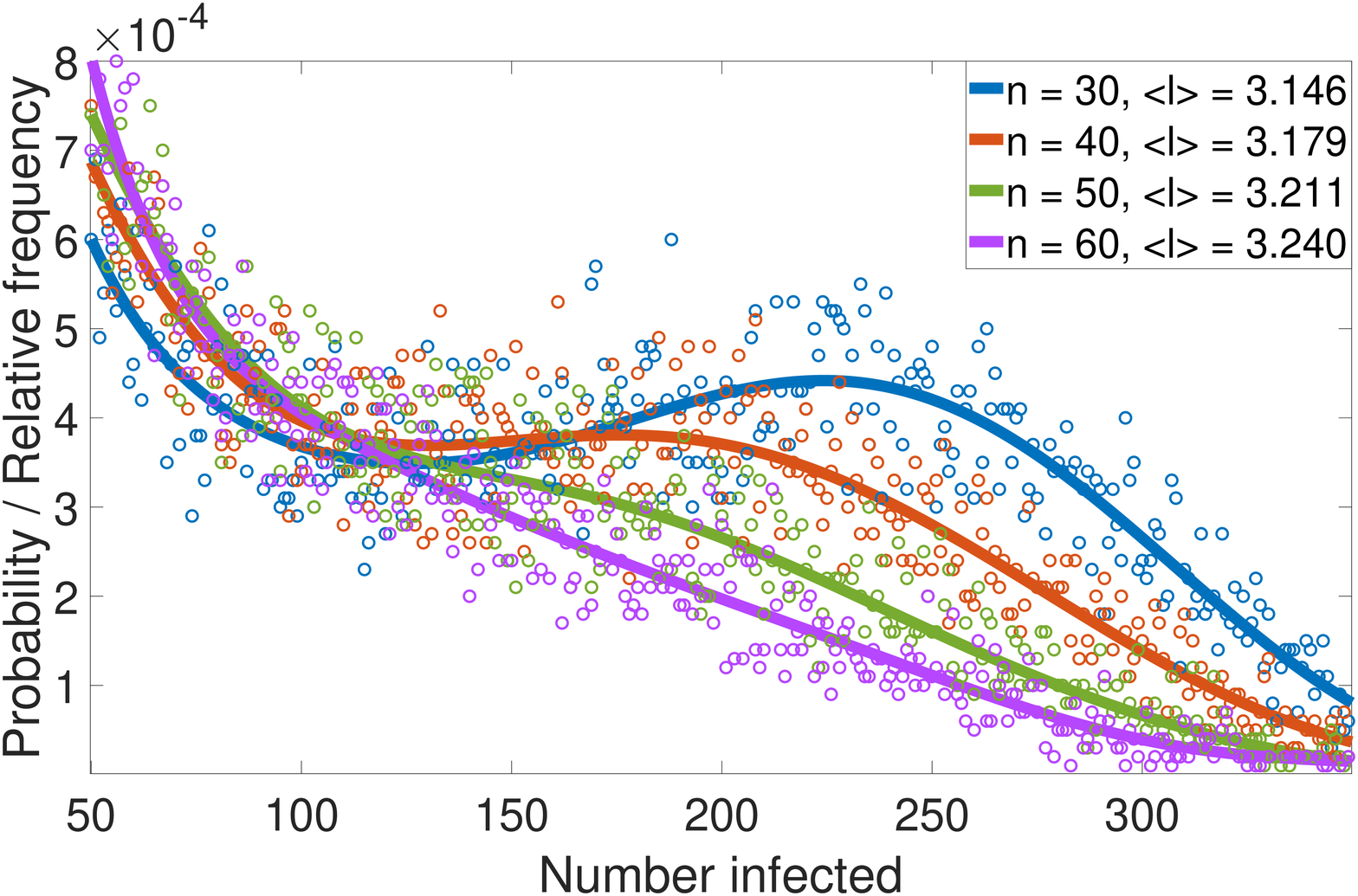}}
	\caption*{degree-based}
\end{minipage}
\begin{minipage}[t]{0.495\linewidth}
\centering
	{\includegraphics[width=1\textwidth]{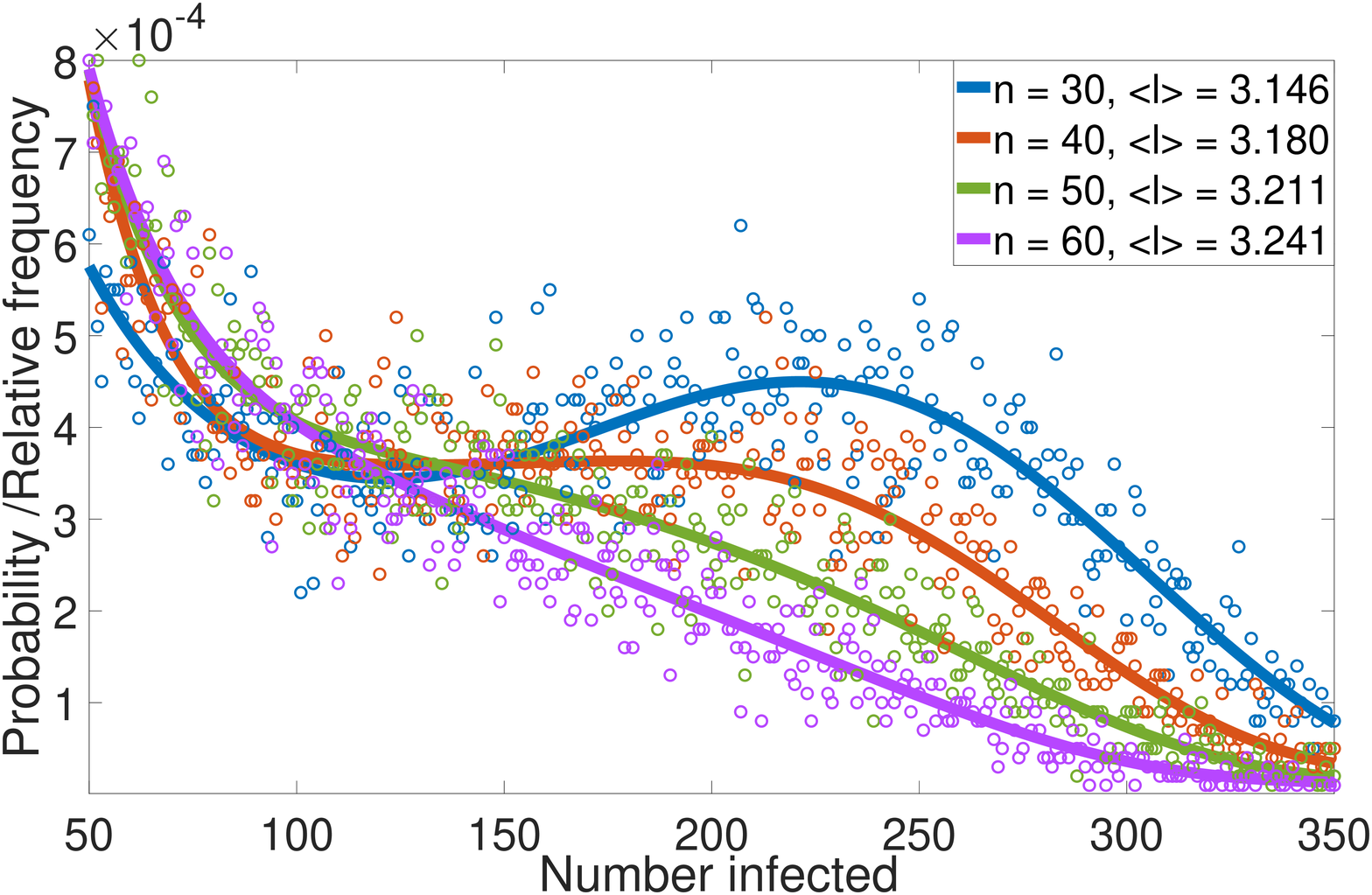}}
	\caption*{betweenness-based}
\end{minipage}
		\caption{Final outbreak size frequencies given an initial infection of a single network node, over 100,000 simulations for different numbers of splitted nodes. Exact data points from the simulations and regression curves (polynomial of degree 8) are plotted. The results for degree-based splittings are depicted in the left figure, the number of critical splits is found to be about $n=60$ which corresponds to $6\%$ of the nodes. Very similar results are found when splitting nodes according to their betweenness centrality, as is shown in the right figure.}  \label{fig:NodeSplit}
	\end{figure}

In comparison to random edge removals, it is clearly observable that the number of necessary edge deletions $\vert \mathcal{E}_c\vert$ can be significantly reduced by following the edge centrality deletion procedure. Moreover, the remaining network possesses a higher functionality represented by a lower average shortest path length $\langle l_c\rangle$ than in the case of random edge removals. 
\paragraph{Node Splitting}\label{app:nodesplit}
In the following, we propose a splitting procedure which is based on the suitable choice of a node centrality measure $\mathcal{C}$.\footnote{A similar algorithm was introduced in \cite{Chernikova2022}.} Nodes with highest centrality are splitted in an iterative manner, i.e., centralities are re-evaluated after each split. Hence, nodes resulting from a split can be splitted again if they still exceed the rest of the network in terms of centrality. 
\begin{algorithm}[Node Splitting] \textit{ }\\
\emph{Input:} Initial network of $N$ nodes, number $n$ of node splits, node centrality measure $\mathcal{C}$
\begin{enumerate}
\item Determine the centrality of all network nodes.
\item Find the node $i$ with highest centrality.
\item Split node $i$ in the following way:
\begin{enumerate}[i)]
\item Add a new node $j$ to the existing network.
\item Create an order of node $i$'s network neighbors where nodes are sorted according to their centrality.
\item For nodes $l$ with an even order rank, delete the edge between $i$ and $l$ and create a new edge between $l$ and $j$.
\end{enumerate}
\item Repeat steps 1) - 3) until $n$ node splits are conducted.
\end{enumerate}
\emph{Output:} Resulting network $G_c$ of $N+n$ nodes
\end{algorithm}
In analogy to the previously conducted analysis of edge removals, we study the effect of node splitting on the epidemic outbreak size distribution and functionality of the  Barabási-Albert network from Figure \ref{fig:Deg} with initial outbreak size frequencies as shown in Figure \ref{fig:BA5vsER001}. The results for degree- and betweenness-based node splitting are depicted in Figure \ref{fig:NodeSplit}, yielding almost identical results.

In comparison to edge removals, we find that node splitting is even more effective: Only about $6 \%$ of the most central nodes need to be splitted in order to control the risk of cyber pandemics. Further, the functionality of $\langle l \rangle \approx 3.24$ of the resulting network is  better than in the case of edge removals ($\langle l \rangle \approx 3.30$). 

\begin{figure}[h]
		\centering
\begin{minipage}[t]{0.495\linewidth}
\centering	
	{\includegraphics[width=1\textwidth]{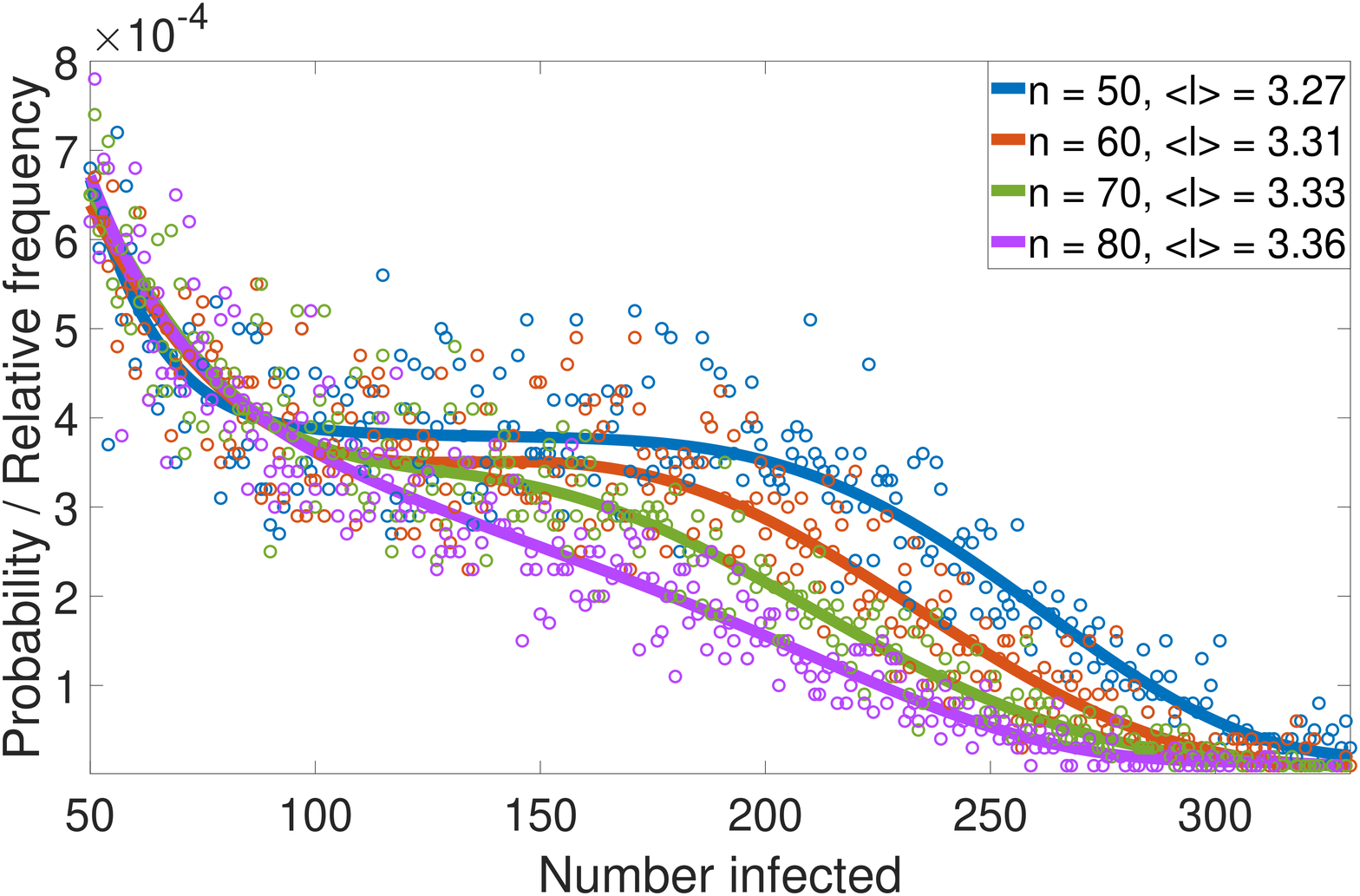}}
\end{minipage}
		\caption{Final outbreak size frequencies given an initial infection of a single network node, over 100,000 simulations for different numbers of split nodes under the modified procedure. Exact data points from the simulations and regression curves (polynomial of degree 8) are plotted. In comparison to the results from Figure \ref{fig:NodeSplit}, we see that the modified rewiring procedure substantially reduces the procedure's efficiency. Indeed, in this case $8\%$ of the nodes need to be splitted and the corresponding network functionality is $\langle l \rangle=3.36$.}  \label{fig:NodeSplit_Bad}
	\end{figure}

In step 3, iii) of the algorithm, rewiring of edges is conducted with the aim of separating critical contagion channels from each other. To study the effectiveness of the procedure, we may modify this step of the algorithm in the following way: Let $k_i$ denote the degree of node $i$. Then, the $\lceil k_i/2\rceil$ neighbors with highest degree remain connected to $i$, and only edges between the $\lfloor k_i/2\rfloor$ lowest degree nodes and $i$ are rewired from node $i$ to $j$. From the outcomes in Figure \ref{fig:NodeSplit_Bad}, we clearly observe that the effectiveness of the node splitting procedure is now remarkably lowered, both in terms of necessary node splits for the prevention of cyber pandemics and network functionality. Hence, the separation of critical contagion channels is essential for the effective implementation of node splitting. 

\subsubsection{Risk Allocation and Design of Contractual Obligations}\label{sec:contact:coeff}
\paragraph{Risk Allocation} Consider an initial graph $G$ and the graph $G_c$ which is obtained by network interventions, either edge removals or node splitting, such that cyber pandemics are sufficiently controlled in $G_c$. 
 The network connections in $G_c$ can be considered \textit{acceptable}, i.e, they should not warrant further regulatory action. 
Instead,  suitable risk allocation schemes and possible obligations should be derived from the set of \textit{deleted} (edge removals) or \textit{rewired} (node splitting) connections. To allocate the cyber pandemic risk to the individual nodes in accordance with their systemic risk contribution, we thus introduce the concept of \textit{contact coefficients}:

\begin{itemize}
    \item \textit{Edge removals:} Let $\epsilon_i = \vert \{ j\mid (i,j)\in \mathcal{E}_c\}\vert$ denote the number of critical connections of node $i$.\footnote{Note that every critical edge $(i,j)\in \mathcal{E}_c$ connects two nodes $i$ and $j$, thus $\sum_{i=1}^N \epsilon_i = 2\vert\mathcal{E}_c\vert$.} To measure the cyber pandemic risk contribution of the single node $i$, we define the contact coefficient $c_i$ of $i$ by
\begin{equation*}
c_i = \frac{\epsilon_i}{2\vert \mathcal{E}_c\vert}, \text{ normalized to} \sum_{i=1}^N c_i = 1.
\end{equation*}
\item \textit{Node splitting:} Let $\mathcal{I}\subseteq \{1,\dots , N\}$ denote the set of nodes from the initial network $G$ which are splitted during the procedure. Then, in analogy to the centrality weights $w_i$ from Case Study I, we choose a node centrality measure $\mathcal{C}$ and define the contact coefficient $c_i$ by
\begin{equation*}
    c_i = \begin{cases}  \Large(\mathcal{C}(i)/\sum_{j\in\mathcal{I}} \mathcal{C}(j)\Large),  &\text{if } i\in\mathcal{I}, \\
        0, &\text{else}.
         \end{cases}
\end{equation*}
\end{itemize}
In the following, we sketch preliminary ideas on how specific topology-based obligations for network nodes $i$ could be established.
\paragraph{Contractual Obligations}
A major problem of (private) regulators such as insurance companies is that they might not be able to directly control or limit connections within cyber networks. In that case, contractual obligations, like surcharges or insurance risk premiums, may incentivize the deletion or protection of critical contagion channels. In the following we briefly discuss such insurance-related obligations. 
\begin{itemize}
    \item \emph{Fixed surcharge:} Given a cyber premium $\pi_i\in\mathbb{R}_+$ for node $i$, not yet accounting for systemic cyber risks, the contact coefficient $c_i$ could serve to determine the fraction of a fixed systemic risk surcharge $f>0$ which has to be borne by node $i$. This means that node $i$'s total premium would equal $$\widetilde{\pi_i}=\pi_i+c_i\cdot f\ge \pi_i,$$ with equality if and only if $c_i=0$, i.e., if and only if node $i$ possesses no critical network connections. For example, these surcharges could be implemented in the context of the insurance backstop mechanism that is discussed in \cite{Lemnitzer2021}.
    \item \emph{Risk premia:} Let $L$ represent the random total loss (over all nodes) in the original network $G$, and let $L_c$ represent the total loss in the new network $G_c$. Then $L_e:=L-L_c$ may be interpreted as the cyber pandemic loss. Consider a risk measure $\rho$ such as the Value at Risk or Expected Shortfall\footnote{For a rigorous introduction to monetary risk measures, we refer the interested reader to Section 4 in \cite{FS}.} and let $\rho(L_e)$ denote the corresponding risk capital. When $\rho(L_e)>0$ we define a topology-based premium $\pi(c_i)$ for each node $i$ by allocating the risk capital $\rho(L_e)$ among the policyholders according to their individual risk contribution. 
    For fixed networks $G$ and $G_c$, the corresponding function $\pi: [0, 1] \to [0, \rho(L_e)]$ should be non-decreasing and satisfy $\sum_{i=1}^N \pi(c_i) = \rho(L_e)$.
    This amounts to a classical risk allocation problem, see, e.g.\ \cite{feinsteinetal2017}. Obviously, the proportional allocation rule  
    \begin{equation*}
   \pi(c_i) = c_i\cdot \rho(L_e) 
    \end{equation*}
    satisfies these constraints. 
\end{itemize}

\begin{figure}[h]
    \centering
    \begin{minipage}{0.3\textwidth}
    \includegraphics[width=\textwidth]{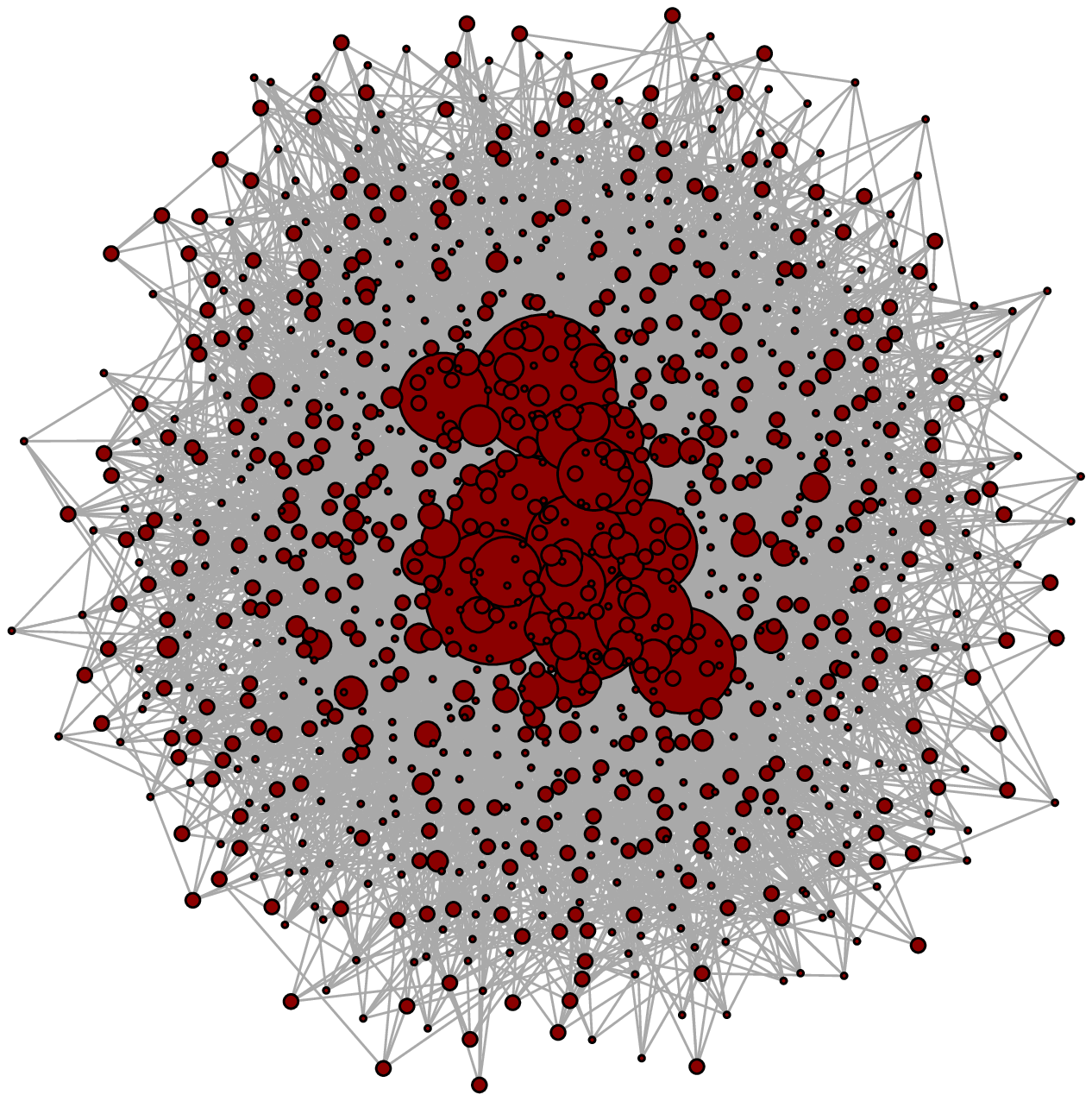}
    \end{minipage}
    \caption{Visualization of contact coefficients based on edge removals in the Barab\'asi-Albert network introduced in Figure \ref{fig:Deg}: Here, node size of node $i$ equals $100\cdot\sqrt{c_i/\sum\nolimits_{j=1}^{1000} c_j}$, an increasing function of the node's importance with respect to its contact coefficient $c_i$.}
    \label{fig:BA1000CP}
\end{figure}

Using edge-removal interventions, we illustrate the effect of these mechanisms in Figure \ref{fig:BA1000CP}  for the Barab\'asi-Albert network from Figure \ref{fig:Deg}. The larger the size of a node in Figure \ref{fig:BA1000CP}, the larger is its underlying contact coefficient $c_i$, and, thus, the higher would be an adequate topology-based obligation. We find that 
 critical network connections are mostly associated to a few central hubs. Comparing Figures \ref{fig:Deg} and \ref{fig:BA1000CP}, these few hubs are even more important than from a degree perspective, and their decisive meaning for the emergence of cyber pandemic risk within the network is clearly observed. Thus, adequate topology-based interventions should target these few central pandemic nodes.

\subsection{Evaluation of Topology-Based Interventions} \label{subsec:CSIIevaluation}

The case study clearly demonstrates that effective manipulations of the network topology can prevent cyber pandemic outbreaks while preserving a reasonable level of network functionality. We obtain the following insights:
\begin{enumerate}[(i)]
    \item In homogeneous networks of large size, connectivity, defined in terms of the sole number of links, plays a major role in the emergence of cyber pandemic risk: 
    A critical connectivity threshold $p_c$ can be identified, below which the frequency of cyber pandemics is negligible. Further, it is possible to prevent cyber pandemics by increasing the overall network security.      
    \item However, 
    many real-world networks are characterized by a more heterogeneous, \textit{scale-free} distribution of node degrees. Examples of networks with a scale-free topology can be modeled using the Barab\'asi-Albert model. Here, we found that highly-connected network participants (hubs) may further \textit{amplify} risk propagation compared to homogeneous networks. Moreover, in the limit of infinite network size, cyber pandemics cannot solely be prevented by strengthening the security of network participants but requires manipulating the degree distribution of the underlying network topology.
    \item  \textit{Centrality} and \textit{contact coefficients} are an effective way to measure an agent's relative topological importance and allocate the cyber pandemic risk of the system to its individual nodes. Regulation taking into account these parameters may significantly reduce the cyber risk and simultaneously preserve a high level of network functionality. However, determining these coefficients requires information on the \textit{full network topology}.
\item In contrast to security-related measures, which should target all large and medium scale entities, topology-based interventions only need to focus on a small group of highly central nodes. Thus, while contact coefficients might be difficult to determine in practice,  it is sufficient to impose obligations, like mandatory backup servers, the protection of data connections, and separation of contagion channels, on a small fraction of highly interconnected network entities. Due to their size and importance, these nodes are more likely to be identified.  
\end{enumerate}	 
We identify the following implications for the cyber resilience measures discussed in Section~\ref{sec:legal}:
\begin{framed}
\begin{itemize}
\item[GOV] 
\begin{itemize}
\item[$\diamond$] \textit{Incident response and reporting:} The implementation of early warning systems and reporting obligations for strongly connected network entities may be an effective way to prevent large-scale events. Immediately disconnecting or otherwise securing these agents after risk arrival may be crucial to prevent the outbreak of a systemic incident. Further, network scanning should evaluate the risk of cyber pandemic outbreaks; in particular, contact coefficients and the analysis of edge removal or node splitting procedures may help to give concrete advice for the design of a more resilient network topology.
\item[$\diamond$]\textit{Critical supply chains:}  Network topology characteristics of industry supply chains should play a major role in risk assessment and resilience building. Highly interconnected entities, cloud service platforms, or frequently used software may pose a severe threat for production chains and industry sectors. 
\end{itemize}
\item[INS]
\begin{itemize}
\item[$\diamond$] \textit{Contact liability premiums:} The systemic risk contribution of a policyholder to the insurers portfolio could be evaluated by means of contact coefficients as introduced in Section~\ref{sec:contact:coeff}.
\item[$\diamond$] \textit{Insurance backstop mechanism:} Our approach provides a reasonable allocation mechanism for mandatory surcharges after the appearance of a systemic cyber risk incident. Further, it may help encourage the deletion or protection of critical network connections and thereby reduce the existing risk potential. 
\end{itemize}
\end{itemize}
\end{framed}

\section{Conclusion and Outlook}\label{sec:conclusion}
As systemic cyber risks such as the well-known WannaCry and NotPetya incidents pose a growing threat to social and economic stability around the world, risk management and resilience building are increasingly becoming the focus of regulators and private actors.
In this context, major issues arise from the limited amount of incident data available and the ever-evolving threat landscape. 

Following the digital twin paradigm, we tackle this issue by introducing the \textit{artificial  cyber lab}: Based on data from virtual counterparts of real-world cyber systems, the artificial cyber lab provides an experimental framework to analyze the impact of both \textit{security-related} and \textit{topology-based} interventions. We find that both types can significantly improve the resilience of interconnected cyber systems -- if they are well-adapted to the topology of the underlying cyber network: In the context of security-related interventions, appropriate obligations can be successfully implemented if, in addition to regulating highly centralized entities, they also apply to
medium-sized network players. Additionally, topology-based measures for preventing cyber pandemic outbreaks in large-scale heterogeneous networks are essential. These constitute a
rather serious regulatory intervention in cyber systems compared to security-related obligations.
However, these interventions may be justified because only a small portion of highly centralized
nodes need to be affected.  Based on our analysis of a virtual counterpart of the real world, digital networks might become more resilient against systemic cyber threats by implementing the discussed cyber resilience measures.

{ Of course, our specific case studies are highly stylized, and the validity of results depends on the appropriateness of the chosen framework. Possible modifications and extensions of the lab environment may be:}

\begin{itemize}
\item {\textit{Attackers and insurers as strategic actors:}} A limitation of our approach is that we have not yet considered in detail the reactions and objectives of the actors involved, e.g., the reaction of malicious actors to the implementation of novel measures, or the impact of information asymmetries in the relationships between insurers and policyholders.
Our approach is a first step toward combining strategic approaches and dynamic cyber risk models. Future research should seek to incorporate these strategic aspects into the modeling framework. 

\item {\textit{Data gathering and model uncertainty:}} Based on artifical lab data, 
our study is able to provide insights on 
critical aspects of  building cyber resilience -- in a \textit{qualitative} sense. However, to determine what \textit{exact} degree of constraints might be appropriate in reality, 
the input parameters of our mathematical cyber risk model need to be fitted to real-world data in order to establish additional data links between the virtual and real-world components of our digital twin. Therefore, gathering data about { network topologies and} cyber incidents remains an important task for regulatory authorities, risk management agencies, and insurance companies.\footnote{{ A brief survey on statistical inference methods for network topologies and/or epidemic model parameters is presented in Appendix E of \cite{Awiszus2021}.} Further, in \cite{Hillairet2021b}, a macroeconomic network model { with weighted edges}  was calibrated from  OECD data on the economic flow between industry sectors.} { Additionally, considering risk management methods under model uncertainty may be necessary to robustify the lab framework.}

\item {\textit{Network size and complexity:} Of course, the computational complexity of algorithms applied within the artificial cyber lab significantly increases with the number of network nodes and edges. However, our studies indicate that an effective risk assessment can be achieved by focusing on the most central parts of the network only. Hence, a possible way to overcome complexity issues could be to artificially reduce the size of the network subject to preserving important characteristics. For instance, large real-world networks could be downsized by merging the peripheral parts to a tractable number of nodes.  The suitability of such approaches is part of future research.}

\item {\textit{Feedback mechanisms in dynamic and adaptive networks:} Over the course of an ongoing contagious cyber incident, nodes may in turn react to the threat evolution dynamics by link activation, shift, or deletion. For example, in response to the downfall of a server, new links may be created to servers which are still operational. Models for dynamic and adaptive networks with link rewiring, activation, and deletion are extensively discussed in \cite{Masuda2017} and Chapter 8 of \cite{Kiss2017}.}
\end{itemize}

This list of future research and modeling perspectives is not exhaustive. Moreover, new aspects of cyber risk will emerge over time as cyber technology evolves. Nevertheless, artificial cyber labs are a promising tool for analyzing and understanding threats -- supporting the evaluation of potential countermeasures when building a more resilient cyber landscape for the future.

\clearpage
\begin{appendices}
\appendix

\section{Markovian SIR Dynamics}\label{app:SIR Dynamics}
\paragraph{Continuous-Time Markov Chains} In Markovian spread models on networks of $N$ nodes, the evolution of the state vector $X(t)$
 \begin{equation*}
 X(t) = (X_1(t),\ldots , X_N(t))\in E^N,
 \end{equation*}
is described by a continuous-time Markov chain on the discrete state space $E^N$. $E$ is the \textit{compartment set} of possible single node states.
We assume that the Markov chain is \textit{time-homogeneous}, i.e., that the probability of changing from state $x\in E^N$ to state $y\in E^N$ within a time window of length $t>0$ does not depend on the current time $u$
\begin{equation*}
    P_{xy}(t) := \mathbb{P}(X(u + t) = y\mid X(u) = x) = \mathbb{P}(X(t) = y\mid X(0) = x),\quad u>0.
\end{equation*}
These probabilities constitute the $|E|^N\times |E|^N$ \textit{transition probability matrix} $P(t)=(P_{xy}(t))$ with $\sum_{y \in E^N} P_{xy}(t) = 1$. For $t=0$, it is consistent to assume that $P(0) = \lim_{t\searrow 0} P(t)$ equals the $|E|^N\times |E|^N$-dimensional identity matrix.
Then $P(t)$ is continuous for all $t\geq 0$ and satisfies the \textit{Chapman-Kolmogorov equation}
\begin{equation}\label{eq: chapkol}
    P(t+u) = P(u) P(t) = P(t) P(u).
\end{equation}
The transition probabilities $P(t)$ fully characterize the evolution of a continuous-time Markov chain. For practical purposes, however, they provide too much information. Hence, we will focus on infinitesimal transition probabilities instead.

The continuity of $P(t)$ implies that the derivative matrix
\begin{equation*}
    Q := P'(0) = \lim_{h\searrow 0}\frac{P(h)- P(0)}{h}
\end{equation*}
exists.\footnote{ see Theorem 2.1 in \cite{Bremaud1999}} $Q$ is called the \textit{infinitesimal generator} of the process, and its entries $q_{xy}$ are called \textit{transition rates} since they describe the probability per unit time of a transition from state $x$ to state $y$. Using the Chapman-Kolmogorov equation (\ref{eq: chapkol}), the evolution of the complete process $(X(t))_{t\geq 0}$ can be described by its infinitesimal generator $Q$ via the \textit{Kolmogorov forward} and \textit{backward equations} 
\begin{equation}\label{eq:kolforback}
    P'(t) = P(t) Q\qquad \text{and}\qquad P'(t) = Q P(t).
\end{equation}
The latter matrix differential equation is solved by the matrix exponential $P(t) = e^{Qt}$, i.e., the transition probabilities can directly be retrieved from the infinitesimal transition rates.
Moreover, this solution implies that the \textit{holding time} $T_x$, i.e., the waiting time for leaving state $x\in E^N$, is exponentially distributed with parameter $q_x := \sum_{y\in E^N,  y\neq x} q_{xy} = - q_{xx} \geq 0$. In addition, the Markov property of $(X(t))_{t\geq 0}$ implies the independence of holding times.\footnote{For details see, e.g.,
Chapter 10 in \cite{Mieghem2014}. We refer to this book for more in-depth reading on stochastic processes and complex networks. 
}

\paragraph{SIR Dynamics} The SIR spread process is determined by $X_i(t)\in E =\{S,I,R\}$.  A transition of $X$ from one state in $E^N$ to another is only possible if exactly one node changes its state $X_i$ in $E$. State changes can occur through infection or recovery: It is assumed that each node may be infected by its infected neighbors, but can be cured independently of all other nodes in the network. Transitions are depicted in Figure \ref{fig:SISSIR}. 
Formally, the entries of the infinitesimal generator $Q$ are given by

\begin{equation}\label{eq:generator}
      q_{xy} = \begin{cases}
        \gamma_i, &\text{if }   x_i = I,  y_i = R,  \text{ and } x_j = y_j \text{ for } j\neq i \\
        \tau \sum_{j=1}^N a_{ij} \mathbbm{1}_{x_j = I}, &\text{if } x_i = S, y_i = I, \text{ and } x_j = y_j \text{ for } j\neq i \\
        -\sum_{z\in E^N, z\neq x} q_{xz}, &\text{if } x=y \\
        0, &\text{otherwise}.
        \end{cases}
    \end{equation}

 Of particular interest are the dynamics of the state probabilities of individual nodes $\mathbb{P}(X_i(t)=x_i),\; t\ge 0$.
They can be derived from Kolmogorov's forward equation and written in general form as ($i=1,\ldots,N$)
\begin{equation}\label{eq: master}
	\frac{d\mathbb{P}(X_i(t) = x_i)}{dt} = \sum_{y: y_i=x_i} \sum_{z\neq y} [\mathbb{P}(X(t)={z}) q_{zy} - \mathbb{P}(X(t)={y}) q_{yz}],
\end{equation}
where $q_{zy}$ denotes the transition rate of the entire process $X$ from $z\to y$.
Using Bernoulli random variables $S_i(t):= \mathbbm{1}_{\{X_i(t) = S\}}$, $I_i(t):= \mathbbm{1}_{\{X_i(t) = I\}}$, and $R_i(t):= \mathbbm{1}_{\{X_i(t) = R\}}$, the dynamics of state probabilities of individual nodes \eqref{eq: master} can conveniently be written via moments:\footnote{The dynamics of the recovery Bernoulli random variable $R_i(t)$ result from the dynamics of $I_i(t)$ and $S_i(t)$ due to $\mathbb{E}[R_i(t)] = 1-\mathbb{E}[S_i(t)] - \mathbb{E}[I_i(t)]$.}
\begin{shaded}
\begin{align}\label{eq:SIR}
\begin{split}
\frac{d\mathbb{E}[S_i(t)]}{dt} &= -\tau\sum_{j=1}^N a_{ij} \mathbb{E}[S_i(t)I_j(t)] , \\
\frac{d\mathbb{E}[I_i(t)]}{dt} &= \tau\sum_{j=1}^N a_{ij} \mathbb{E}[S_i(t)I_j(t)] - \gamma_i \mathbb{E}[I_i(t)] ,\\
\frac{d\mathbb{E}[S_i(t) I_j(t)]}{dt} &= \tau\sum_{k=1, k\neq i}^N a_{jk}\mathbb{E}[S_i(t)S_j(t)I_k(t)] - \tau \sum_{k=1,k\neq j}^N a_{ik}\mathbb{E}[I_k(t)S_i(t)I_j(t)]\\ &\quad- \tau a_{ij}\mathbb{E}[S_i(t)I_j(t)] - \gamma_j \mathbb{E}[S_i(t)I_j(t)] ,\\
\frac{d\mathbb{E}[S_i(t)S_j(t)]}{dt} &= -\tau \sum_{k=1, k\neq j} a_{ik}\mathbb{E}[I_k(t)S_i(t)S_j(t)] - \tau \sum_{k=1, k\neq i}^N a_{jk}\mathbb{E}[S_i(t)S_j(t)I_k(t)] ,
\end{split}
\end{align}
where $i,j = 1,2,\ldots , N$ and $i\neq j$. 
\end{shaded}
Note that system \eqref{eq:SIR} is \textit{not closed}: The dynamics of second order moments depend on third order moments, which, in turn, depend on fourth order moments etc. This dependence structure cascades up to network size $N$. Therefore, in general, solving the exact system of moment equations becomes intractable, especially for larger networks. To deal with this issue, the following two approximation approaches have been proposed:
\begin{enumerate}
\item \textbf{Monte Carlo simulation:}  Monte Carlo simulation using the Gillespie algorithm from \cite{Gillespie1976} and \cite{Gillespie1977} constitutes a powerful tool to obtain various quantity estimates related to the evolution of the epidemic spread.
Pseudocode is given in Appendix \ref{app:Gillespie}, and further explanations of the Gillespie algorithm applied to SIR epidemic network models is, e.g., given in Appendix A.1.1 of \cite{Kiss2017}.
\item \textbf{Moment closures:} 
If a set of nodes $J$ is infected, this increases the probability of other nodes in the network (that are connected to the set $J$ via an existing path) to become infected as well. Hence, node states are to some extent \textit{correlated}.
To break the cascade of equations and to make ODE systems tractable, the moment closure approach consists in assuming independence at a certain order $k$, neglecting any further correlations. This is done by considering the exact moment equations up to this order $k$ and \textit{closing} the system by approximating moments of order $k+1$ in terms of products of lower-order moments using a mean-field function. 
However, a \textit{major problem} with moment closures is that only little is known about rigorous error estimates.
\end{enumerate}

\section{Gillespie Algorithm}\label{app:Gillespie}

\begin{algorithm*}[Gillespie]\label{alg:Gillespie}\,\\
\emph{Input:} Initial state of the system $x^0\in E^N$ and initial time $t_0\ge 0$.
	\begin{enumerate}
		\item\emph{(Initialization)} Set the current state $x\to x^0$, current time $t\to t_0$, and $k \to 0$.
		\item\emph{(Rate Calculation)} For the current state of the system $x$, calculate the sum of rates for all possible transitions $q_x=\sum\nolimits_{i=1}^N q_{x_i}$, where $q_{x_i}$ denotes the rate for a state change of node $i$ according to \eqref{eq:generator}.
		\item\emph{(Generate Next Event Time)} Sample the next event time $t_{\text{new}}$ from an exponential distribution with parameter $q_x$.
		\item\emph{(Choose Next Event)} Sample the node $i_{\text{new}}$ at which the next transition occurs: Each node $i=1,\ldots,N$ is chosen with probability ${q_{x_i}}/{q_x}$. \\
		Change the state $x_{i_{\text{new}}}\to y_{i_{\text{new}}}$ according to $\eqref{eq:generator}$.
		\item Set $t\to t+t_{\text{new}}=:t_{k+1}$, $x\to (x_1,\ldots,x_{i_{\text{new}}-1},y_{i_{\text{new}}},x_{i_{\text{new}}+1},\ldots,x_N)=:x^{t_{k+1}}$, $k\to k+1$, and return to Step 2 until a prespecified stopping criterion is met.
	\end{enumerate}
	\emph{Output:} Trajectory $[t_0, t_{end}]\ni t\to X(t, \omega)$ of the spread process, where $X(t,\omega ) := x^{t_k}$ for $t\in [t_k, t_{k+1}]$; $t_{end}$ denotes the end time of the simulation
\end{algorithm*}

\section{Network Algorithms}\label{app:networks}

\begin{algorithm*}[Erd\H{o}s-R\'enyi] \,\\
\emph{Input:} Number of network nodes $N$, connection probability $p$.
\begin{enumerate}
\item Choose a pair of nodes $(i,j)$ with $i,j\in \{1,\cdots , N\}$, $i\neq j$.
\item Simulate a uniformly distributed number $\tilde{p}$ between 0 and 1.
\item If $\tilde{p}<p$, create an edge between node $i$ and $j$. Else, no edge is created.
\item Repeat steps $1) - 3)$ for all the other possible pairs of nodes.
\end{enumerate}
\emph{Output:} Network $G$ from class $G_p(N)$
\end{algorithm*}

\begin{algorithm*}[Barab\'asi-Albert] \,\\
\emph{Input:} Number of network nodes $N$, small initial network of $n_0$ connected nodes, number $m$ of nodes to which every newly added node is connected.
\begin{enumerate}
\item Add a new node $i$ to the small initial network.
\item Create a new edge for $i$ in the following way:
\begin{enumerate}[i)]
\item Uniformly generate a node number $j$  from the existing network $(i\neq j )$.
\item Simulate a uniformly distributed number $r$ between 0 and 1. 
\item Let $k_l$ denote the current degree of node $l$, $l=1,\cdots N$. If  $r < k_j / \sum_l k_l$, then the edge should be created between $i$ and $j$. Else, go back to step i).
\end{enumerate}
\item Repeat step 2) until $m$ edges are created for the new node $i$.
\item Repeat steps 1) - 3) until a network of $N$ nodes is formed.
\end{enumerate}
\emph{Output:} Network $G$ from class $BA(N;m)$
\end{algorithm*}


\section{$L_i$ as a Strictly Convex Function of $\gamma_i$}\label{sec:lossmodel}

For our cyber loss model 
 \begin{equation*}
        L_i := L_i(\gamma_1,\ldots , \gamma_N) := \mathbb{E}\Large[ \int_0^\infty I_i(t) dt\Large],
    \end{equation*}
    we can derive an elegant expression in terms of $\gamma_i$: Let $A_i:=\{\exists t\in [0,\infty ): I_i(t) =1\}$ be the event that node $i$ will be infected at some moment in time $t$. Then 
        \begin{equation*}
        L_i = \mathbb{E}\Large[ \int_0^\infty I_i(t) dt \mathbbm{1}_{A_i}\Large] + \mathbb{E}\Large[ \int_0^\infty I_i(t) dt \mathbbm{1}_{A_i^c}\Large]
    \end{equation*}
    where $\int_0^\infty I_i(t) dt \mathbbm{1}_{A_i^c} = 0$ by definition.  Note that $\gamma_i$ only affects the recovery process of node $i$, and since reinfection events are ruled out in the SIR modeling framework, the probability of infection $\mathbb{P}(A_i)$ for node $i$ does not depend on the recovery rate of node $i$ but only on the vector $\gamma_{-i}$ of the other node's recovery rates. 
    
    Further, since the initial infections are randomly chosen, we have $\mathbb{P}(A_i) \geq 1/N$. Thus, by using the rules of conditional expectation, $L_i$ can be expressed as
    \begin{equation*}
      L_i = \mathbb{E}\Large[ \int_0^\infty I_i(t) dt \mathbbm{1}_{A_i}\Large] = \mathbb{P}(A_i)\cdot \mathbb{E}\Large[ \int_0^\infty I_i(t) dt\mid A_i\Large].
    \end{equation*}
    By definition,  $\int_0^\infty I_i(t) dt$ is the amount of time  $i$ spends in the infectious state $I$.
    If an infection of node $i$ actually occurs, then this is the time of transition from state $I$ into state $R$. Hence, $\mathbb{E}\Large[ \int_0^\infty I_i(t) dt\mid A_i\Large]$ is the expected waiting time for recovery of node $i$. Since the SIR spread process is assumed to be Markovian, the waiting time is exponentially distributed with parameter $\gamma_i$. Therefore, we obtain
    \begin{equation*}
       \mathbb{E}\Large[ \int_0^\infty I_i(t) dt\mid A_i\Large] = \frac{1}{\gamma_i}.
    \end{equation*}
    Hence,  
    \begin{equation}\label{eq:loss:model}
        L_i(\gamma_1,\ldots , \gamma_N) = \frac{\mathbb{P}(A_i)}{\gamma_i}
    \end{equation}
 where the numerator does not depend on node $i$'s recovery rate.  
    Thus, $L_i$ is a \textit{strictly convex} function of $\gamma_i$. 
   
  \section{Modeling Cyber Losses for the Security Investment Game}\label{sec:lossesgame}

    The decomposition of  loss functions $L_i$ in Appendix \ref{sec:lossmodel} can be used for an efficient stochastic simulation procedure of cyber losses in Algorithm \ref{alg:Security}: To find the recovery rate $\gamma_i(r+1)$ for round $r+1$ in step 2 of the algorithm, we need to determine the minimizer 
    \begin{equation*}
       \gamma_i(r+1) = \argmin_{\gamma_i} \mathcal{E}_i(\gamma_i,\gamma_{-i}(r)) = \argmin_{\gamma_i} [C_i(\gamma_i) + L_i(\gamma_i, \gamma_{-i}(r))].
    \end{equation*}
    Using the aforementioned representation of the loss functions, this means that for every node $i$, we need to determine the infection probabilities $\mathbb{P}(A_i)$ to describe $L_i$ as a function of $\gamma_i$. Now, since the infection probability of every node $i$ is not depending on its own recovery rate, these probabilities can be determined in a joint procedure: 
    \textit{\begin{enumerate}
    \item Choose a sufficiently high number of simulation runs $\mathcal{T}$ to generate trajectories of the SIR process. For example, we chose $\mathcal{T} =10,000,000$ for simulations in Figure \ref{fig:SInetwork} and \ref{fig:effalloc}. 
    \item For every node $i$, let the recovery rate be given by $\gamma_i(r)$.
    \item For every simulation run, initially infect a randomly chosen single node and generate a trajectory of the SIR process on the network. For every node $i$, save whether $i$ was infected during this run. 
    \item After the conduction of the $\mathcal{T}$ simulation runs, for every node $i$, let $\mathcal{T}_i$ be the number of simulation runs where node $i$ was infected. Set
    $\mathbb{P}(A_i) = \mathcal{T}_i/\mathcal{T}$.
    \end{enumerate}}
    Then, for every node $i$ the total expenses $\mathcal{E}_i$ are solely given as a function of $\gamma_i$, and it is straightforward to determine 
    \begin{equation*}
        \gamma_i(r+1) =  \argmin_{\gamma_i} [C_i(\gamma_i) + \mathbb{P}(A_i)/\gamma_i].
    \end{equation*}

\section{Proof of Theorem \ref{theorem}}\label{app:proof}
\begin{proof}
\begin{enumerate}
    \item\textit{Continuity of total expenses:} We prove that 
$  \mathcal{E}_i: (0,\infty)^N \to \mathbb{R}$
is continuous. 
Recall that $$\mathcal{E}_i(\gamma_1,\ldots, \gamma_N)=C_i(\gamma_i)+ L(\gamma_1,\ldots, \gamma_N)$$ where $C_i(\gamma_i)=e^{k\gamma_i}-1$\footnote{In Case Study 1, we choose $k=\frac{1}{3}$.} and $L_i(\gamma_1,\ldots, \gamma_N)= \mathbb{E}\Large[ \int_0^\infty I_i(t) dt\Large] $. Obviously, $C_i$ is continuous. As regards $L_i$ note that 
\begin{equation*}
        L_i(\gamma_1,\ldots, \gamma_N) = \mathbb{E}\Large[ \int_0^\infty I_i(t) dt\Large] = \int^\infty_0 \mathbb{E}[I_i(t)] dt = \int^\infty_0 \mathbb{P}(X_i(t) = I) dt
 \end{equation*}
 by the Fubini-Tonelli Theorem. 
 Therefore, it is sufficient to prove the continuity of $\mathbb{P}(X_i(t) = I)$ in $(\gamma_1,\ldots,\gamma_N)\in (0,\infty)^N$. From equation \eqref{eq:generator}, we see the the generator matrix of the SIR Markov process is continuous  (w.r.t. the Frobenius norm), and therefore, the same applies to the solution $P(t) = e^{Qt}$ of the Kolmogorov backward equation. The continuity (w.r.t. the Euclidean norm) is preserved by the continuous transform
 \begin{equation*}
     \mathbb{P}(X_i(t) = I) = \sum_{x\in E^N} \mathbb{P}(X(0) = x) \sum_{\substack{y\in E^N,\\  y_i=I}} P_{xy}(t)
 \end{equation*}
 of transition probability matrix $P(t)$.
 
 \item Recall the representation  \eqref{eq:loss:model} $L_i(\gamma_1,\ldots , \gamma_N) = \tfrac{\mathbb{P}(A_i)}{\gamma_i}$. Hence, according to 1.\ also $\mathbb{P}(A_i)=\gamma_i L_i(\gamma_1,\ldots , \gamma_N)$ is continuous as a function of $\gamma_{-i}$.
    
\item Note that both $C_i$ and $L_i$, and thus $\mathcal{E}_i$, are strictly convex functions of $\gamma_i$. Recalling that $\mathbb{P}(A_i)$ does not depend on $\gamma_i$, the first order condition for the unique minimum is $$ \frac{\partial}{\partial \gamma_i}\mathcal{E}_i(\gamma_1,\ldots, \gamma_N) = k e^{k\gamma_i}-\frac{\mathbb{P}(A_i)}{\gamma_i^2}=0.$$ Since $\mathbb{P}(A_i)$ is continuous as a function of $\gamma_{-i}$ by 2.,  it also follows that unique minimizer $\gamma^\text{ind}_i(\gamma_{-i})$ is continuous in $\gamma_{-i}$. 
On the one hand, note that $$\frac{\partial}{\partial \gamma_i}\mathcal{E}_i(\gamma_1,\ldots, \gamma_N)=k e^{k\gamma_i}-\frac{\mathbb{P}(A_i)}{\gamma_i^2}\geq k e^{k\gamma_i}-\frac{1}{\gamma_i^2}$$ and the latter expression does not depend on $\gamma_{-i}$ and is positive for, for instance, $\gamma_i> \frac{1}{\sqrt{k}}$. On the other hand,  since the initial infections are randomly chosen, we have $\mathbb{P}(A_i) \geq 1/N$, and thus  
$$k e^{k\gamma_i}-\frac{\mathbb{P}(A_i)}{\gamma_i^2}\leq k e^{k\gamma_i}-\frac{1}{N\gamma_i^2}$$ where again the latter expression does not depend on $\gamma_{-i}$. Now let $\varepsilon(N)>0$ such that $k e^{k\varepsilon(N)}-\frac{1}{N\varepsilon(N)^2}<0$ (which always exists depending  only on $N$). Then it follows that for any $i=1,\ldots, N$ and 
$(\gamma_1,\ldots , \gamma_N)\in (0,\infty)^N$ we have $\gamma^\text{ind}_i(\gamma_{-i})\in [\varepsilon(N), \frac{1}{\sqrt{k}}]$.

\item  In 3.\ we showed that the function \begin{equation*}
    [\varepsilon(N),\frac{1}{\sqrt{k}}]^N \to [\varepsilon(N),\frac{1}{\sqrt{k}}]^N , \quad (\gamma_{1}, \ldots, \gamma_N)\mapsto (  \gamma^\text{ind}_1 (\gamma_{-1}), \ldots, \gamma^\text{ind}_N (\gamma_{-N}))
\end{equation*} is well-defined and continuous. Hence, according to Brouwer's fixed point theorem\footnote{see e.g.\ Corollary 17.56 in \cite{Aliprantis2006}} it has a fixed point, that is \begin{equation*}
    \exists \gamma\in [\varepsilon(N),\frac{1}{\sqrt{k}}]^N\;  \forall i=1,\ldots , N: \quad \gamma^\text{ind}_i (\gamma_{-i}) = \gamma_{i} .
\end{equation*}

  \end{enumerate}
\end{proof}


\section{Security Choices for a Network of Two Nodes}\label{app:2nodes}
\begin{figure}[h]
    \centering
  \begin{tikzpicture}[scale=1] 
\node[draw, shape=circle] (1)[fill=White!85!gray] at (-1.5, 0) {1};
\node[draw, shape=circle] (2) [fill=White!85!gray] at ( 1.5, 0) {2};
\node[draw=none] (4) at (-2.9, 0) {$\gamma_1$};
\node[draw=none] (5) at (2.9, 0) {$\gamma_2$};
\node[draw=none] (6) at (0, 0.3) {$\tau$};
\draw (1) -- (2);
\draw[thick, ->,line width=.3mm] (-1.7,0.3) arc (30:320:0.5cm);
\draw[thick, ->,line width=.3mm] (1.7,0.3) arc (150:-140:0.5cm);
\end{tikzpicture}
\caption{Line network with $N=2$ nodes and the corresponding epidemic transition rates. }
\label{fig:2nodes}
\end{figure}
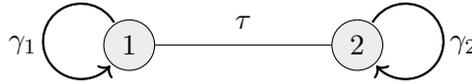

Straightforward exact computations of optimal investment levels are possible for the simple case of two interconnected nodes as illustrated in Figure \ref{fig:2nodes}.

In this special case, the infection probabilities $\mathbb{P}(A_i)$ from the loss model decomposition in Appendix \ref{sec:lossmodel} can be explicitly calculated from the waiting time distributions of the Markov chain: Due to the random uniform choice of the initially infected node, it is $\mathbb{P}(X_i(0) = I) = 1/2$ for $i=1,2$, and thus, we obtain
\begin{align*}
    \mathbb{P}(A_i) &= \mathbb{P}(X_i(0) = I) + \mathbb{P}(X_j(0) = I)\cdot \mathbb{P}( (S_i, I_j)  \to (I_i, I_j)) \\
    &=\frac{1}{2}\cdot (1 +\mathbb{P}( (S_i, I_j)  \to (I_i, I_j))), \qquad i,j\in \{1,2\},\quad i\neq j.
\end{align*}
$\mathbb{P}( (S_i, I_j)  \to (I_i, I_j))$ can be expressed in terms of the waiting time for recovery $T_j^\text{recov}$ of node $j$ and the infection event $T^\text{infec}$, and we can use the fact that waiting times for Markov chains are independent and exponential, yielding
\begin{equation*}
    \mathbb{P}( (S_i, I_j)  \to (I_i, I_j)) = \mathbb{P}(T_j^\text{recov} > T^\text{infec}) = \frac{\tau}{\gamma_j + \tau}.
\end{equation*}
Hence, a closed expression of the cyber losses $L_i$ in terms of epidemic transition rates is given by
\begin{equation*}
    L_i(\gamma_1,\gamma_2) = \frac{\mathbb{P}(A_i)}{\gamma_i} = \frac{1}{2\gamma_i}\cdot \Big(1 + \frac{\tau}{\gamma_j + \tau}\Big),\qquad i,j\in\{1,2\},\quad i\neq j,
\end{equation*}
and this can be inserted into the total expense functions $\mathcal{E}_i(\gamma_1,\gamma_2) = C(\gamma_i) + L_i(\gamma_1,\gamma_2)$.

Thus, the individual optimal security choice $\gamma_i^\text{ind}(r+1)$ in round $r+1$ of the security investment game is given by 
\begin{equation*}
    \gamma_i^\text{ind}(r+1) = \argmin_{\gamma_i} \mathcal{E}_i(\gamma_i,\gamma_j^\text{ind}(r)),
    \qquad i,j\in\{1,2\},\quad i\neq j.
\end{equation*}
Now, in agreement with the chosen parameters in Case Study \ref{subsec:CSI}, we fix $\tau=0.1$ and initialize the security investment game with recovery rates $\gamma_1(0) =\gamma_2(0) = 0.1$. 
\begin{table}[h]
\centering
     \begin{tabular}{|p{1cm} ||p{1.5cm}|}
     \hline
     r  & $\gamma_i^\text{ind}(r)$  \\
    \hline
    \hline
    
     1  & 1.2234  \\
      \hline
    2     & 1.0638  \\
    \hline
    3     & 1.0681  \\
    \hline
     4    & 1.0680 \\
    \hline
     5    & 1.0680 \\
    \hline
    \end{tabular}
    \caption{The security investment game for a line network of $N=2$ nodes}
    \label{fig:twonodestable}
\end{table}
From Table \ref{fig:twonodestable}, we see that the game converges to the security configuration $(\gamma_1^\text{ind},\gamma_2^\text{ind}) = (1.068,1.068)$ after $r=4$ rounds. 

However, from an overall network perspective, the best security configuration $(\gamma_1^\text{soc},\gamma_2^\text{soc})$ would be the one which minimizes the accumulated total expenses $\mathcal{E}$, i.e., from a social welfare perspective, the choice
\begin{equation*}
    (\gamma_1^\text{soc}, \gamma_2^\text{soc}) := \argmin_{(\gamma_1,\gamma_2)} \sum_{k=1,2} \mathcal{E}_k(\gamma_1,\gamma_2)  = (1.0984, 1.0984)
\end{equation*}
would be beneficial. Since $\gamma_i^\text{ind} \neq \gamma_i^\text{soc}$, this simple example illustrates that, in general, individually optimal security choices will \textit{not} correspond to investment levels which minimize the overall network expenses.

\section{Allocation Data}\label{app:upperalloc}
\subsection{Data for Upper and Lower Allocation Strategies}
\begin{table}[h!]
\centering
     \begin{tabular}{|M{2cm} ||M{2.6cm}
     |M{2.6cm}|M{2.6cm}|}
     \hline
       & $c^\text{deg}$ 
       & $c^\text{bet}$ & $c^\text{inv}$  \\
    \hline
    \hline
    
     upper  & {\color{NavyBlue!80!black} 19.363 (0.0030)} {\color{salmon!80!black} 19.444 (0.0030)}  
     & {\color{NavyBlue!80!black} 19.323 (0.0030)} {\color{salmon!80!black} 19.230 (0.0028)} & {\color{NavyBlue!80!black} 19.460 (0.0031)} {\color{salmon!80!black} 19.813 (0.0032)} \\
      \hline
    lower    & {\color{NavyBlue!80!black} 19.891 (0.0033)} {\color{salmon!80!black} 20.450 (0.0036)} 
    & {\color{NavyBlue!80!black} 21.551 (0.0040)} {\color{salmon!80!black} 21.171 (0.0039)} &  {\color{NavyBlue!80!black} 19.601 (0.0032)} {\color{salmon!80!black} 20.096 (0.0034)} \\
    \hline
    untargeted  & \multicolumn{3}{|c|}{{\color{NavyBlue!80!black}19.516 (0.0031)} {\color{salmon!80!black} 19.954 (0.0033)}} \\
    \hline
    \end{tabular}
    \caption{Accumulated total expenses $\mathcal{E}$ after the allocation of the additional budget $\beta = 5$ among all network nodes. The three proposed allocation strategies are evaluated for each of the suggested centrality measures. Entries for the Erd\H{o}s-R\'enyi network are colored in blue (upper entries), and for the Barab\'asi-Albert network in salmon (lower entries), respectively. Reference values without the injection of additional security are 21.66 for the  Erd\H{o}s-R\'enyi, and 21.92 for the Barab\'asi-Albert network.  For each entry, cyber losses were generated from $\mathcal{T}=10,000,000$ simulations of the SIR epidemic process; standard errors are given in brackets.}
\end{table}
\newpage
\subsection{Data for Centralized Upper Allocations}
\begin{table}[h]
\centering
     \begin{tabular}{|M{2cm} ||M{2.6cm}
     |M{2.6cm}|M{2.6cm}|}
     \hline
     targeted  & $c^\text{deg}$ 
     & $c^\text{bet}$ & $c^\text{inv}$  \\
    \hline
    \hline
    
     10 \%  & {\color{NavyBlue!80!black} 21.129 (0.0037)} {\color{salmon!80!black}20.163 (0.0031)} 
     & {\color{NavyBlue!80!black} 21.120 (0.0036)} {\color{salmon!80!black}20.214 (0.0031)} & {\color{NavyBlue!80!black} 21.130 (0.0037)} {\color{salmon!80!black}20.156 (0.0031)}  \\
      \hline
    20 \%    & {\color{NavyBlue!80!black} 20.091 (0.0033)}  {\color{salmon!80!black}19.459 (0.0028)}  
    & {\color{NavyBlue!80!black} 20.152 (0.0033)} {\color{salmon!80!black}19.495 (0.0028)} & {\color{NavyBlue!80!black}20.111 (0.0033)} {\color{salmon!80!black}19.507 (0.0029)} \\
    \hline
    30 \%    & {\color{NavyBlue!80!black} 19.702 (0.0031)} {\color{salmon!80!black}19.302 (0.0028)}
    & {\color{NavyBlue!80!black}19.769 (0.0031)} {\color{salmon!80!black}19.311 (0.0028)}& {\color{NavyBlue!80!black}19.769 (0.0031)}  {\color{salmon!80!black}19.398 (0.0029)}\\
    \hline
     40 \%    & {\color{NavyBlue!80!black}19.523 (0.0030)}  {\color{salmon!80!black}19.265 (0.0028)}
     & {\color{NavyBlue!80!black}19.554 (0.0031)} {\color{salmon!80!black}19.251 (0.0028)}&  {\color{NavyBlue!80!black}19.545 (0.0030)} {\color{salmon!80!black}19.408 (0.0029)} \\
    \hline
     50 \%    & {\color{NavyBlue!80!black} 19.406 (0.0030)} {\color{salmon!80!black}19.271 (0.0028)}    
     & {\color{NavyBlue!80!black}19.454 (0.0030)} {\color{salmon!80!black}19.233 (0.0028)} & {\color{NavyBlue!80!black}19.454 (0.0030)} {\color{salmon!80!black}19.461 (0.0030)} \\
    \hline
     60 \%    & {\color{NavyBlue!80!black}19.347 (0.0030)} {\color{salmon!80!black}19.312 (0.0029)}   
     & {\color{NavyBlue!80!black}19.367 (0.0030)} {\color{salmon!80!black}19.229 (0.0028)} &  {\color{NavyBlue!80!black}19.378 (0.0030)} {\color{salmon!80!black}19.552 (0.0030)} \\
    \hline
     70 \%    & {\color{NavyBlue!80!black}19.328 (0.0030)} {\color{salmon!80!black}19.339 (0.0029)}
     & {\color{NavyBlue!80!black}19.324 (0.0030)} {\color{salmon!80!black}19.227 (0.0028)}& {\color{NavyBlue!80!black}19.364 (0.0030)} {\color{salmon!80!black}19.625 (0.0031)} \\
    \hline
     80 \%    & {\color{NavyBlue!80!black}19.329 (0.0030)} {\color{salmon!80!black}19.372 (0.0029)}   
     & {\color{NavyBlue!80!black}19.319 (0.0030)} {\color{salmon!80!black}19.231 (0.0028)}& {\color{NavyBlue!80!black}19.374 (0.0030)} {\color{salmon!80!black}19.708 (0.0032)} \\
    \hline
     90 \% & {\color{NavyBlue!80!black}19.345 (0.0030)}  {\color{salmon!80!black}19.412 (0.0030)}  
     &{\color{NavyBlue!80!black}19.319 (0.0030)} {\color{salmon!80!black}19.233 (0.0028)}& {\color{NavyBlue!80!black}19.405 (0.0031)} {\color{salmon!80!black}19.758 (0.0032)}  \\
    \hline
    \end{tabular}
    \caption{Accumulated total expenses for different percentages of targeted nodes under the upper allocation strategy for each centrality measure. Again, the total additional security budget is fixed with size $\beta = 5$.  Entries for the Erd\H{o}s-R\'enyi network are colored in blue (upper entries), and for the Barab\'asi-Albert network in salmon (lower entries), respectively. The standard error of the stochastic simulations is given in brackets. For each entry, $\mathcal{T}=10,000,000$ simulations of the epidemic process were generated.}
    \label{fig:uppertable}
\end{table}

\end{appendices}

\clearpage
\printbibliography
\end{document}